\documentclass[a4paper, 12pt]{article}

\usepackage{tikz}
\usetikzlibrary{arrows,decorations.markings,cd}
\usepackage{latexsym,amsmath,amsfonts,amssymb}
\usepackage[latin1]{inputenc}
\usepackage[american]{babel}
\usepackage{bbm}
\usepackage[nosort]{cite}
\usepackage{hyperref}
\hypersetup{linktocpage, colorlinks, citecolor=[rgb]{0,0,0.8}, linkcolor=[rgb]{0,0,0.8}, urlcolor=[rgb]{0.7,0,0.5}}

\renewcommand{\baselinestretch}{1.2} 
\newcommand{\wt}{\widetilde}
\newcommand{\wh}{\widehat}

\newcommand{\matht}[1]{\ensuremath{\boldsymbol{#1}}}

\newcommand{\eg}{\textit{e.g.}}

\newcommand{\ie}{\textit{i.e.}}

\numberwithin{equation}{section}

\newcommand{\nn}{\nonumber}

\newcommand{\be}{\begin{equation}} \newcommand{\ee}{\end{equation}}
\newcommand{\bea}{\begin{equation} \begin{aligned}} \newcommand{\eea}{\end{aligned} \end{equation}}

\newcommand{\cA}{\mathcal{A}}
\newcommand{\cB}{\mathcal{B}}
\newcommand{\cC}{\mathcal{C}}
\newcommand{\cD}{\mathcal{D}}

\newcommand{\cL}{\mathcal{L}}
\newcommand{\cM}{\mathcal{M}}

\newcommand{\cO}{\mathcal{O}}

\newcommand{\cR}{\mathcal{R}}

\newcommand{\cT}{\mathcal{T}}

\newcommand{\bG}{\mathbb{G}}

\newcommand{\bK}{\mathbb{K}}

\newcommand{\bR}{\mathbb{R}}

\newcommand{\bZ}{\mathbb{Z}}

\newcommand{\fg}{\mathfrak{g}}

\newcommand{\fP}{\mathfrak{P}}

\newcommand{\se}{{\sf{e}}}
\newcommand{\sm}{{\sf{m}}}

\newcommand{\bfg}{{\bf g}}
\newcommand{\bfh}{{\bf h}}
\newcommand{\bfk}{{\bf k}}
\newcommand{\bfl}{{\bf l}}
\newcommand{\bfm}{{\bf m}}
\newcommand{\unit}{\mathbbm{1}}

\def\repa{\raise4pt\hbox{$\square$}\mkern-14mu\raise-4pt\hbox{$\square$}}
\def\repab{\overline{\raise4pt\hbox{$\square$}\mkern-14mu\raise-4pt\hbox{$\square$}\mkern-1mu}}

\DeclareMathOperator{\Aut}{Aut}
\DeclareMathOperator{\Hom}{Hom}
\DeclareMathOperator{\id}{id}
\DeclareMathOperator{\Bock}{Bock}

\makeatletter
\def\blfootnote{\gdef\@thefnmark{}\@footnotetext}
\makeatother

\tikzset{
->-/.style={decoration={markings, mark=at position .5 with {\arrow[scale=1.5]{stealth}}}, postaction={decorate}}
}

\begin{document}

\thispagestyle{empty}
\begin{flushright}
SISSA  10/2018/FISI
\end{flushright}
\vspace{10mm}
\begin{center}
{\huge  On 2-Group Global Symmetries \\[.5em] and Their Anomalies} 
\\[15mm]
{Francesco Benini$^{1,2}$, Clay C\'ordova$^2$, Po-Shen Hsin$^3$}\blfootnote{\it E-mails: fbenini@sissa.it, claycordova@ias.edu, phsin@princeton.edu}
\vskip 6mm
 
\bigskip
{\it
$^1$ SISSA, via Bonomea 265 \& INFN - Sezione di Trieste, via Valerio 2, Trieste, Italy \\[.5em]
$^2$ School of Natural Sciences, Institute for Advanced Study, Princeton NJ, USA \\[.5em]
$^3$ Physics Department, Princeton University, Princeton NJ, USA
}
\vskip 6mm

\bigskip
\bigskip

{\bf Abstract}\\[5mm]
{\parbox{14cm}{\hspace{5mm}

In general quantum field theories (QFTs), ordinary (0-form) global symmetries and 1-form symmetries can combine into 2-group global symmetries. We describe this phenomenon in detail using the language of symmetry defects. We exhibit a simple procedure to determine the (possible) 2-group global symmetry of a given QFT, and provide a classification of the related 't~Hooft anomalies (for symmetries not acting on spacetime). We also describe how QFTs can be coupled to extrinsic backgrounds for symmetry groups that differ from the intrinsic symmetry acting faithfully on the theory. Finally, we provide a variety of examples, ranging from TQFTs (gapped systems) to gapless QFTs. Along the way, we stress that the ``obstruction to symmetry fractionalization'' discussed in some condensed matter literature is really an instance of 2-group global symmetry.

}}
\end{center}
\newpage
\pagenumbering{arabic}
\setcounter{page}{1}
\setcounter{footnote}{0}
\renewcommand{\thefootnote}{\arabic{footnote}}

{\renewcommand{\baselinestretch}{1} \parskip=0pt
\setcounter{tocdepth}{2}
\tableofcontents}

\section{Introduction}

Symmetry is one of the most enduring and fruitful tools in the analysis of Quantum Field Theory (QFT).  The most elementary consequence of symmetry is to organize observables into representations and to enforce selection rules on correlation functions.  A more subtle aspect of symmetry is that there may be obstructions ('t~Hooft anomalies) to gauging a symmetry, \ie{} to coupling the system to dynamical gauge fields.  These obstructions are properties of the theory that are inert under renormalization group (RG) flow and are therefore powerful constraints on dynamics. 

The most familiar kind of global symmetry (ordinary or 0-form) acts naturally on local operators. If the symmetry is continuous its implications are encoded in the Ward identities of the associated conserved current.  More generally, it is useful to organize symmetries according to the dimension of the charged objects \cite{Gaiotto:2014kfa}. For instance, 1-form global symmetries act on line operators.  Unlike the case of 0-form symmetries which can be Abelian or non-Abelian, higher-form symmetries are necessarily Abelian.  In the case of continuous higher-form global symmetry, the associated Ward identities are again encoded in the correlation functions of conserved currents, which are differential forms of general degree.

To analyze discrete global symmetries one requires a presentation that does not rely on conserved currents.  This is achieved through the notion of symmetry defects, which are extended operators representing the symmetry transformations.  An ordinary global symmetry labelled by an element $\bfg$ in the 0-form symmetry group $G$ gives rise to a codimension-1 non-local operator $U_{\bfg}$, with the property that as a local operator is dragged through the defect it is acted on by the symmetry transformation associated to $\bfg$.  The fact that this operator is a global symmetry is encoded through a remarkable property of $U_{\bfg}$: it is topological.  Hence its correlation functions do not change under small deformations of the manifold supporting the operator.  Similarly, $n$-form global symmetries are realized by codimension-$(n{+}1)$ operators with topological correlation functions.  The fact that symmetries are implemented through a topological subsector of correlation functions also explains why they are so robust and why they can be analyzed at any energy scale along the RG flow of a field theory.

\begin{figure}[t]
\centering
\begin{tikzpicture}[scale=.6]
\draw [very thick, red!90!black] (0,-2) to (0,2); 
\draw (0,2) to (3,-.5) to (3,-4.5) to (0,-2); 
\draw ([shift=(-130:6)] 0,8) arc (-130:-50:6);
\draw ([shift=(-130:6)] 0,4) arc (-130:-90:6); \draw [dashed, line cap=round] ([shift=(-90:6)] 0,4) arc (-90:-60:6); \draw ([shift=(-60:6)] 0,4) arc (-60:-50:6);
\draw ([shift=(-130:6)] 0,4) to ++(0,4); \draw ([shift=(-50:6)] 0,4) to ++(0,4);
\node at (-3.4,2.2) {$\bfg$}; \node at (3.3,2.2) {$\bfh$}; \node at (2.3,-2.9) {$\bfg\bfh$}; 
\end{tikzpicture}
\caption{A junction (in red) where three 0-form symmetry defects of type $\bfg$, $\bfh$, $\bfg\bfh$ meet in codimension 2. This configuration is generic in spacetime dimension 2 and above.  These junctions encode the group law of the 0-form symmetry.
\label{fig: basic junction}}
\end{figure}
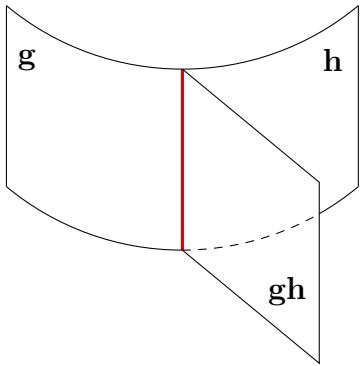

All aspects of the global symmetry of a given QFT can be understood from the properties of the associated topological symmetry defects. Recent investigations include \cite{Bhardwaj:2017xup, Chang:2018iay}. For instance, the most basic property of the 0-form symmetry is that it forms a group $G$.  At the level of defects this is specified by the existence of junctions where three defects meet obeying the multiplication law (as in Figure \ref{fig: basic junction}).  More generally, as we describe below, in the presence of higher-form symmetries there are additional types of junctions (of higher codimension), where 0-form symmetry defects and higher-form symmetry defects meet, and this gives rise to the generalized concepts of symmetry that we explore.

\subsection{2-Group Global Symmetry}

In this paper we focus on the particular case of 0-form and 1-form global symmetries, and we address the question: \emph{What is the most general possible symmetry structure including a 0-form group $G$ and a 1-form group $\mathcal{A}$?}  We show that one general possibility is that $G$ and $\cA$ are combined into a higher-categorical structure known as a 2-group (see \eg{} \cite{baez2004higher, Baez:2004in, Baez:2005qu, Schreiber:2008} and references therein). Including even higher-form global symmetries naturally leads to more general $n$-groups.   Concretely, this means that the symmetries of the theory do not factorize, but rather are fused in a way encoded by the existence of a junction of both 0-form and 1-form symmetry defects discussed below and illustrated in Figure~\ref{fig: 2-group 3D}. When this fusion occurs, we say that a QFT has 2-group global symmetry.

The interplay between 2-groups and QFT has been explored in several contexts in the literature.  In \cite{Kapustin:2013uxa}, symmetry protected phases with discrete 2-group global symmetry were constructed following earlier related work \cite{Gukov:2013zka, Kapustin:2013qsa}.  These topological actions for 2-group gauge fields were subsequently generalized in \cite{Thorngren:2015gtw, Gaiotto:2017zba} and play an essential role in our discussion of 2-group 't~Hooft anomalies in Section~\ref{sec: 't Hooft anomaly}.  Other recent discussions of discrete 2-group symmetry appear in \cite{Bhardwaj:2016clt, Tachikawa:2017gyf, Delcamp:2018wlb}. QFTs with continuous 2-group symmetry have recently been studied in \cite{Cordova:2018cvg} and are briefly reviewed below.%
\footnote{Reference \cite{Sharpe:2015mja} also describes some aspects of continuous 1-form symmetry and 2-group symmetry in two spacetime dimensions.}

One aspect of 2-group global symmetry is simple to describe.  The 0-form symmetry can act on the 1-form charges.  A simple example of this is familiar from three-dimensional Abelian Chern-Simons theory.  In this case the 1-form symmetry group is generated by Wilson lines of various electric charges $q$.  There is also a 0-form symmetry group that includes charge conjugation.  Charge conjugation acts on the Wilson lines by exchanging $q\leftrightarrow -q.$ 

\begin{figure}[t]
\centering
\begin{tikzpicture}
\draw [thick, blue!80!black] (-1,.5) to (1,.5);
\draw [thick, blue!80!black] (2,.5) to (3,.5);
\draw (0,0) to (0,2) to (2,1) to (2,-1) to cycle;
\draw [thick, blue!80!black, dashed, dash phase=1.5] (1,.5) to (2,.5);
\draw [blue!80!black, fill=white] (1,.5) circle [radius=.05];
\node at (.2,1.6) {\small $\bfg$}; \node at (-.7,.7) {\small $\rho_\bfg a$}; \node at (2.7, .75) {\small $a$};
\end{tikzpicture}
\caption{When a symmetry operator of type $a \in \cA$ crosses a codimension-1 symmetry operator of type $\bfg\in G$, it emerges transformed by an automorphism $\rho_\bfg$ of $\cA$.
\label{fig: action on higher symmetry}}
\end{figure}
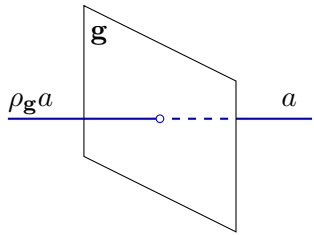

More abstractly, the action of the 0-form symmetry on the 1-form symmetry is encoded by the properties of the symmetry defects described above.  When an operator $a\in \cA$ pierces a codimension-1 symmetry operator for $\bfg\in G$, it emerges as a new 1-form charge denoted $\rho_{\bfg}a$ (see Figure~\ref{fig: action on higher symmetry}).  This transformation must preserve the group structure of the 1-form symmetries and therefore defines a map (a group homomorphism)
\begin{equation}
\rho: G\rightarrow \Aut(\cA) \;,
\end{equation}
where $\Aut(\cA)$ is the group of automorphisms of $\cA$.

The other component of 2-group global symmetry is a Postnikov class $[\beta]$, which is a group cohomology class
\begin{equation}
[\beta]\in H^{3}_{\rho}(BG,\mathcal{A}) \;.
\end{equation} 
Concretely, this means that $\beta$ is a function:
\begin{equation}
\beta: G\times G \times G \rightarrow \mathcal{A} \;,
\end{equation}
which obeys certain identities, and is subject to certain equivalence relations so that only its equivalence class $[\beta]$ is meaningful (see Appendix~\ref{app: cohomology} for a review of group cohomology).  The physical meaning of the $H^3$-class $[\beta]$ is most clearly illustrated using the topological symmetry defects introduced above.  When a configuration of four 0-form symmetry defects is deformed, a 1-form symmetry defect controlled by $\beta$ appears (see Figure~\ref{fig: obstructed F-move}).   The precise representative function $\beta$ of the cohomology class $[\beta]$ can be changed by adding local counterterms to the action.  Hence only the cohomology class is physically meaningful.

The description above presents the class $[\beta]$ from a transformation of 0-form symmetry defects. However, to parallel our discussion above of the group law for 0-form symmetry defects, it is more instructive to view this as arising from a junction of symmetry defects. In spacetime dimension three and above, there are generic intersections of 0-form symmetry defects which occur in codimension-3 in spacetime. The signature of 2-group global symmetry is that at these codimension-3 junctions, a 1-form symmetry defect controlled by $\beta$ is also present.  See Figure~\ref{fig: 2-group 3D}.  Notice that the pattern of rearrangement of the 0-form symmetry defects in that diagram is related to the associativity of the 0-form symmetry defects.  Thus, when the Postnikov class $[\beta]$ is non-trivial the 0-form symmetry defects are non-associative in their action on lines. 

\begin{figure}[t]
$$
\raisebox{-4em}{\begin{tikzpicture}
\draw [thick] (0,-.8) node[below] {$\bf ghk$} to (0,0);
\draw [thick] (0,0) to (-1.2,1.6) node[above] {$\bf g$};
\draw [thick] (0,0) to (1.2,1.6) node[above] {$\bf k$};
\draw [thick] (-.4,.533) to (0,1) to (0,1.6) node[above] {$\bf h$};
\end{tikzpicture}}
\qquad\longleftrightarrow\qquad
\raisebox{-4em}{\begin{tikzpicture}
\draw [thick] (0,-.8) node[below] {$\bf ghk$} to (0,0);
\draw [thick] (0,0) to (-1.2,1.6) node[above] {$\bf g$};
\draw [thick] (0,0) to (1.2,1.6) node[above] {$\bf k$};
\draw [thick] (.4,.533) to (0,1) to (0,1.6) node[above] {$\bf h$};
\filldraw [blue!80!black] (-.3,-.1) node[black, below left] {\small $\beta(\bfg,\bfh,\bfk)$} circle [radius=.07];
\end{tikzpicture}}
$$
\caption{A transformation of symmetry defects can be used to detect 2-group global symmetry.  The lines (labelled $\bfg,\bfh,\bfk, \bfg\bfh\bfk$) are codimension-1 symmetry operators of $G$. Transforming from left to right (also called an $F$-move), a 1-form symmetry defect $\beta(\bfg,\bfh,\bfk)\in \cA$ is created (blue dot). In $d$ dimensions, all objects span the remaining $d-2$ dimensions.  A probe line passing through the 0-form symmetry defects, detects $[\beta]$ and hence sees the non-associativity of the 0-form symmetry defects.
\label{fig: obstructed F-move}}
\end{figure}
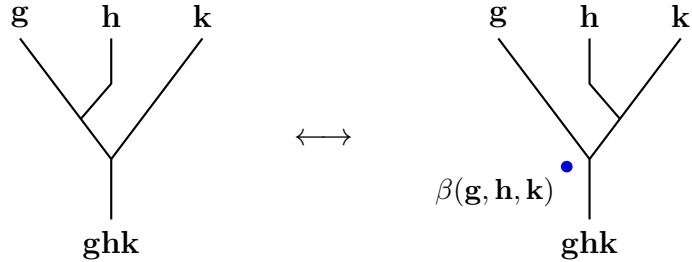     

\begin{figure}[t]
\centering
\begin{tikzpicture}
\draw [thick, red!90!black] (0,-2) to (0,2); 
\draw [thick, red!90!black] (0,0) arc (125:155.3:4.5); 
\draw [thick, red!90!black, dashed, line cap=round] (0,0) arc (-55:-31.7:3.8); 
\draw [thick, red!90!black] (1.5,2.19) arc (-14:-31.5:3.8);
\draw [thick, blue!80!black, decoration = {markings, mark=at position .6 with {\arrow[scale=1.5,rotate=0]{stealth}}}, postaction=decorate] (0,0) arc (-135:-150:6.5); 
\draw (0,2) to (3,-.5) to (3,-4.5) to (0,-2); 
\draw ([shift=(-130:6)] 0,8) arc (-130:-50:6);
\draw ([shift=(-130:6)] 0,4) arc (-130:-90:6); \draw [dashed, line cap=round] ([shift=(-90:6)] 0,4) arc (-90:-60:6); \draw ([shift=(-60:6)] 0,4) arc (-60:-50:6);
\draw ([shift=(-130:6)] 0,4) to ++(0,4); \draw ([shift=(-50:6)] 0,4) to ++(0,4);
\draw [dashed, dash phase = -2, line cap=round] (-1.5,-1.81) to ++(-1,1.8); \draw (1.5,2.19) to ++(-1,1.8); 
\draw [dashed, line cap=round] (-2.5,-0.01) to ([shift=(53.1:2.61)] -2.5,-0.01);
\draw [line cap=round] ([shift=(53.1:2.61)] -2.5,-0.01) to ([shift=(53.1:5)] -2.5,-0.01);
\node at (-3.5,2.5) {$\bfg$}; \node at (.5,3.3) {$\bfh$}; \node at (3.4,2.5) {$\bfk$}; \node at (2.4,-3.3) {$\bfg\bfh\bfk$}; 
\node at (-.7,-1.5) {\small $\bfg\bfh$}; \node at (0.9,1.75) {\small $\bfh\bfk$};
\node at (-1.3,.2) {\footnotesize $\beta(\bfg, \bfh,\bfk)$};
\filldraw (0,0) circle [radius=.05];
\end{tikzpicture}
\caption{A junction where 0-form symmetry defects of type $\bfg$, $\bfh$, $\bfk$, $\bfg\bfh\bfk\in G$ meet in codimension 3. This configuration is generic in spacetime dimension 3 and above.  The junctions of three codimension-1 defects are in red, and their intersection is the black point. At the codimension-3 intersection, a 1-form symmetry defect $\beta(\bfg,\bfh,\bfk)$ emanates, signaling the 2-group symmetry. In $d$ dimensions, all objects span the remaining $d-3$ dimensions.  
\label{fig: 2-group 3D}}
\end{figure}
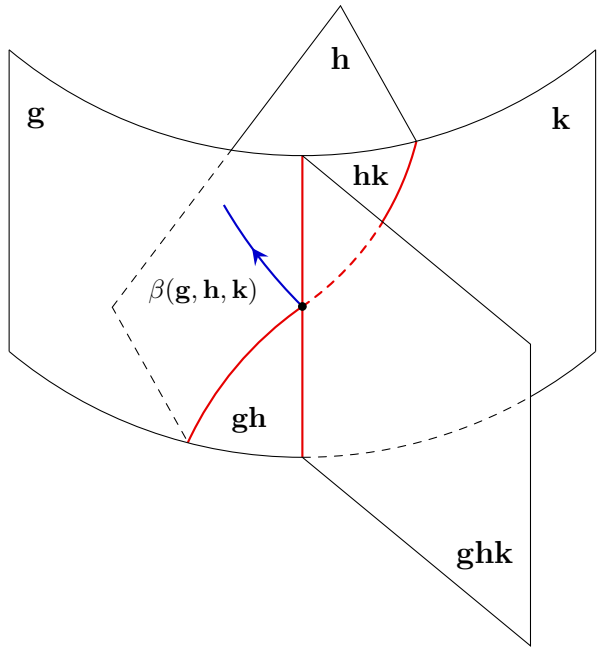

Taking all this data together, we say that a QFT has 2-group global symmetry 
\begin{equation}
\label{2gdataintro}
\bG=\big( G, \cA, \rho, [\beta] \big) \;.
\end{equation}
We stress that this data is intrinsic to the QFT.  If both the 0-form symmetry and the 1-form symmetry are continuous, then 2-group symmetry leads to modified Ward identities that are visible at the level of current correlation functions as discussed in \cite{Cordova:2018cvg}.  In Section~\ref{sec: 2-groups}, on the other hand, we discuss the case of general 0-form and 1-form symmetry groups (including in particular discrete groups) and explain how to extract the defining data \eqref{2gdataintro} from correlation functions of symmetry defects.  

Many of the general phenomena that are possible with standard global symmetries can also occur for 2-group symmetry.  For instance 2-group global symmetry can be spontaneously broken, or alternatively, 2-group global symmetry can be emergent and appear after an RG flow.  Examples in the case of continuous 2-group symmetry are discussed in \cite{Cordova:2018cvg}. Here we describe various examples illustrating aspects of these phenomena in Section~\ref{sec: examples} below.  As we also discuss, discrete 2-group global symmetries can have non-trivial 't~Hooft anomalies.

\subsection{2-Group Background Fields}

A basic way to probe global symmetry in quantum field theory is to couple to background gauge fields.   A 0-form symmetry can be coupled to standard 1-form gauge fields.  A 1-form symmetry can be coupled to 2-form gauge fields---also called Abelian gerbes. (When the 1-form symmetry is $U(1)$, they are also called Deligne-Beilinson 2-cocycles, see \eg{} \cite{Bauer:2004nh, Kapustin:2014zva}.) This concept is familiar in supergravity and string theory, in the $U(1)$ case: a 2-form potential (such as the NS 2-form) is a $U(1)$ gerbe.

The behavior of the theory in the presence of background fields encodes the correlation functions of the symmetry defects discussed above.  Indeed, a codimension-1 0-form symmetry defect can be viewed as a transition function connecting two locally trivial patches in a principal $G$-bundle.  Similarly, codimension-2 1-form symmetry defects define transition functions for an $\mathcal{A}$-gerbe.

For theories with 2-group global symmetry the appropriate background fields are connections forming a 2-group gauge theory \cite{Baez:2005qu, Kapustin:2013uxa}.  These are a 1-form gauge field for $G$, a 2-form gauge field for $\mathcal{A}$, as well as a rule for gauge transformations controlled by $\rho$ and $[\beta]$.  The defect junction illustrated in Figure~\ref{fig: 2-group 3D} has a sharp meaning in this language: at codimension-3 intersections of 0-form symmetry defects, there is a flux for $\mathcal{A}$ described by the Postnikov class $[\beta].$  In particular, this means that when the theory is coupled to $G$ gauge fields, a non-trivial 1-form background for $\mathcal{A}$ is sourced.

Although 2-group symmetry and background gauge fields may sound exotic, there is an important special case where they are familiar from supergravity and string theory. Suppose that both $G$ and $\cA$ are $U(1)$ and $\rho$ is trivial (\ie{} there is no action of $G$ on $\cA$).  In this case a Green-Schwarz mechanism \cite{Green:1984sg} for the associated background fields can naturally be interpreted as saying that the two symmetries are combined into a non-trivial 2-group. This relationship between 2-group global symmetry and the Green-Schwarz mechanism was first observed in \cite{Kapustin:2013uxa}, and this continuous example has been studied in detail from the point of view of 2-group global symmetry in \cite{Cordova:2018cvg}.

In more detail, if $G$ and $\mathcal{A}$ are $U(1)$ there are 1-form and 2-form conserved currents $j^{(1)}$ and $j^{(2)},$ and at the linearized level these couple to a 1-form gauge field $A^{(1)}$ and 2-form gauge field $B^{(2)}$ via terms in the action of the form
\begin{equation}
\int d^dx \Big[ A^{(1)}\wedge *j^{(1)} + B^{(2)} \wedge * j^{(2)} \Big] \;.
\end{equation}
The action is invariant under gauge transformations controlled by a local function $\lambda^{(0)}$ and 1-form $\Lambda^{(1)}$ as well as an integer $\kappa$:
\begin{equation}
\label{2grp}
A^{(1)}\rightarrow A^{(1)}+d\lambda^{(0)} \;, \hspace{.5in} B^{(2)}\rightarrow B^{(2)}+d\Lambda^{(1)} + \frac{\kappa}{2\pi}\lambda^{(0)} dA^{(1)} \;.
\end{equation}
When $\kappa$ vanishes these gauge transformations are standard, while if $\kappa \neq 0$ this theory has continuous 2-group global symmetry.  In particular in this case the cohomology group controlling the Postnikov class is $H^3\big(BU(1),U(1)\big)\cong \mathbb{Z}$ and the integer $\kappa$ can be identified with the class $[\beta]$.  As is standard with the Green-Schwarz mechanism, the modified gauge transformations for $B^{(2)}$ implies that the gauge invariant field strength 3-form $H^{(3)}$ is defined via
\begin{equation}
\label{2-group in SUGRA}
dB^{(2)} = H^{(3)} + \frac{\kappa}{2\pi}\, A^{(1)} \wedge dA^{(1)} \;,
\end{equation}
and moreover implies a modified Bianchi identity for $H^{(3)}$.

As described in \cite{Cordova:2018cvg}, gapless QFTs with continuous 2-group global symmetry are typically IR free. (For instance, the 2-form current $*j^{(2)}$ could be the field strength of an Abelian gauge field \cite{Gaiotto:2014kfa}.)  In this paper we instead focus on the case where the 1-form global symmetry is a finite group.  As we illustrate in the examples below, in this case 2-group global symmetry is compatible with a variety of dynamics ranging from topological theories to gapless interacting systems.  

When the global symmetry is discrete the natural background fields are flat connections, and the associated bundles are described by nets of symmetry defects.  In this case, the analog of \eqref{2-group in SUGRA} is obtained by setting the field strength $H^{(3)}$ to zero and viewing all gauge fields as discrete:
\begin{equation}
d_{A^{(1)}} B^{(2)} = (A^{(1)})^*\beta \;.
\end{equation}
In this equation, $A^{(1)}$ is the (possibly non-Abelian) background field for the 0-form symmetry $G.$  If $G$ is finite, $A^{(1)}$ defines a standard flat connection on a principal $G$-bundle.  By contrast, $B^{(2)}$, the background gauge field for the 1-form symmetry is not flat, but rather has a fixed differential related to the Postnikov class $[\beta]$.  More precisely, we view $A^{(1)}$ as a homotopy class of maps from spacetime to $BG$---the classifying (or Eilenberg-Mac~Lane \cite{Eilenberg:1953}) space of $G$---and $d_{A^{(1)}}$ is a twisted differential.  The right-hand side is the pullback of a representative $\beta$ of the class $[\beta]\in H^3_\rho(BG,\mathcal{A}),$ and hence gives an appropriate 3-cochain on spacetime with values in $\mathcal{A}$. We elaborate on this formalism in Section~\ref{sec: 2-groups}, and also describe how the Postnikov class $[\beta]$ affects the gauge transformations of $B^{(2)}$.

\subsection{2-Group Global Symmetry vs ``Operator-Valued Anomalies'' }

Although aspects of 2-group global symmetry have been described previously, it is sometimes conflated with an anomaly (\eg{} it is referred to as an ``obstruction to symmetry fractionalization'' in the condensed matter literature \cite{Barkeshli:2014cna}).
The origin of this confusion can be understood from Figure~\ref{fig: obstructed F-move}.  There, a junction of 0-form $G$-defects is deformed, and this process may be understood as describing a gauge transformation for a background flat $G$-bundle.  When the $H^{3}$ class $[\beta]$ is non-trivial, this process creates a 1-form symmetry defect.\footnote{We thank Yuji Tachikawa for discussion on this issue.}  

Thus, a gauge transformation of the 0-form background fields modifies the partition function by the insertion of a 1-form symmetry defect, which is a non-trivial operator in the theory.  This gauge non-invariance of the partition function is superficially similar to phenomena usually referred to as 't~Hooft anomalies, where gauge transformations of background fields modify the partition function by $c$-number phases.  However there are important differences.  For instance, a standard $c$-number 't~Hooft anomaly can be cancelled by inflow from a non-dynamical bulk, while in the case of the ``operator-valued anomaly'' described above such a bulk would necessarily be dynamical.

This can also be seen directly in the example of continuous 2-group symmetry (with $G\cong \cA \cong U(1)$) discussed in \cite{Cordova:2018cvg}.  Invariance of the partition function under the 2-group gauge transformations in \eqref{2grp} implies that the 1-form current obeys
\begin{equation}
\label{dj}
d\, {*}j^{(1)}=\frac{\kappa}{2\pi} \, dA^{(1)}\wedge *j^{(2)} \;.
\end{equation}
When the background $A^{(1)}$ vanishes the current is conserved, but when the background is activated the right-hand side is a non-trivial operator.

In fact, it is incorrect to view the phenomena described in Figure~\ref{fig: obstructed F-move} or its continuous analog \eqref{dj} as an anomaly.  Instead, the appearance of the 1-form symmetry defect encodes the non-anomalous Ward identity for 2-group global symmetry.  In particular,
the correlation functions are in fact invariant under 2-group gauge transformations.  More generally in QFT, there are no ``operator-valued anomalies" as we can always add to the partition function a background source for the operator in question and adjust the transformation rules of the source to cancel the hypothetical anomaly by a generalization of the Green-Schwarz mechanism.

Although 2-group global symmetry does not in and of itself constitute an anomaly, examples of theories with 2-group global symmetry can be produced by starting with theories with certain mixed anomalies and gauging global symmetries.  For instance, the non-conservation equation \eqref{dj} can occur in Abelian gauge theories where $*j^{(2)}$ is the field strength of a dynamical gauge field \cite{Cordova:2018cvg}.  A discrete analog of this construction was recently described in \cite{Tachikawa:2017gyf}: starting from theories with only 0-form symmetries and appropriate mixed 't~Hooft anomalies, one can construct theories with 2-group global symmetry by gauging.  We briefly review this construction in Section~\ref{sec: examples} where we use it to construct concrete examples of interacting theories with 2-group global symmetry.

\subsection{2-Group 't Hooft Anomalies}

One reason why it is essential to distinguish 2-group global symmetry from an anomaly is that 2-group Ward identities may themselves be violated by standard $c$-number 't~Hooft anomalies.  Such anomalies are most simply formulated by studying the theory in the presence of background 2-group gauge fields.  An 't~Hooft anomaly then means that the partition function is not exactly invariant under gauge transformations of the background fields, but rather transforms by a local functional of the background fields.  As usual, we are only interested in anomalies that cannot be removed by adjusting local counterterms, and thus classifying anomalies reduces to a cohomological problem.

A useful way to characterize the possible anomalies is via inflow \cite{Callan:1984sa}.  In $d$ dimensions the anomaly is determined by a local bulk action in $d{+}1$ dimensions, which is both gauge invariant and topological.  This means in particular that as the $d$-dimensional system undergoes RG flow, the bulk remains inert and hence illustrates that the anomaly is scale invariant.\footnote{This is similar to the original argument \cite{tHooft:1979rat} for invariance of the anomaly under RG flow, with the non-dynamical bulk playing the role of the spectator fermions.}  These 't Hooft anomalies are therefore robust observables of a QFT that can be used to constrain dynamics.

In the case of 2-group global symmetry, the anomaly is thus characterized in terms of topological 2-group gauge theories.   For the case of continuous 2-groups, the anomalies have been studied in \cite{Cordova:2018cvg}.  In the general case of discrete 2-groups, the appropriate topological actions have been described in \cite{Kapustin:2013uxa} generalizing the analysis of anomalies for discrete 0-form symmetries discussed in \cite{Kapustin:2014lwa, Kapustin:2014zva}.  In Section~\ref{sec: 't Hooft anomaly} we apply these results and give a concrete description of 2-group 't~Hooft anomalies for QFTs in 1,2,3 and 4 spacetime dimensions.  Throughout, we focus on anomalies that do not involve intricacies of the spacetime manifold.  Thus, we neglect possible gravitational or mixed 2-group-gravitational anomalies and consider only bosonic theories.\footnote{In particular, to characterize the anomalies that we study we do not require the more sophisticated cobordism theory discussed in \cite{Kitaev:2013talk, Kapustin:2014tfa, Freed:2016rqq, Gaiotto:2017zba}. }  

Intuitively speaking, an 't~Hooft anomaly for a 2-group consists of an anomaly for the \mbox{0-form} symmetry $G$, an anomaly for the 1-form symmetry $\mathcal{A}$ together with possible mixed $G$-$\mathcal{A}$ anomalies.  However, the 2-group gauge transformations mix the background fields and hence both the possible anomalous variations of the action, and the possible local counterterms must be reclassified.

One simple consequence of this is that the anomaly involving only $G$ background fields is truncated.  In $d$ spacetime dimensions, one generally expects an anomaly for 0-form symmetry to be specified by a group cohomology class $ \omega \in H^{d+1}(BG,\mathbb{R}/\mathbb{Z})$.  However, the 2-group gauge transformations permit the existence of new counterterms, which reduce the possible values of the 0-form anomaly to the quotient group
\begin{equation}
H^{d+1}(BG,\bR/\bZ) \;\text{\Large $/$}\; H^{d-2}_\rho(BG,\wh\cA) \cup [\beta] \;,
\end{equation}
where $\wh\cA$ is the Pontryagin dual to $\mathcal{A},$ (this is the group of homomorphism $\wh\cA = \Hom(\cA, \bR/\bZ)$) and in the denominator we have the image of the natural map from $H^{d-2}_\rho(BG,\wh\cA)$ to $H^{d+1}(BG,\bR/\bZ) $ given by cup product with the Postnikov class $[\beta]$.  This result parallels a similar truncation of the 't Hooft anomaly for theories with continuous 2-group symmetry \cite{Cordova:2018cvg}.

\subsection{Summary of Applications and Explicit Examples}

In Sections \ref{sec: 3D TQFT} and \ref{sec: examples}, we study a variety of examples of theories with 2-group global symmetry.  We mainly focus on theories in three spacetime dimensions.  

We first describe how the correlation functions of a general 3d TQFT can be used to fix the data of a 2-group, the global symmetry of the TQFT, and an associated 2-group 't~Hooft anomaly.  The relationship between 2-groups and 3d TQFTs has been previously noted in \cite{Thorngren:2015gtw, Tachikawa:2017gyf} and our work provides a complete and explicit dictionary. In particular the phenomenon sometimes referred to as an ``obstruction to symmetry fractionalization'' (see \eg{} \cite{Barkeshli:2014cna}) is really an instance of 2-group global symmetry.

We use the formalism of \cite{Barkeshli:2014cna} (see also \cite{EtingofNOM2009, Teo:2015xla} for other treatments and \cite{Barkeshli:2017rzd} for examples) to describe the appropriate $G$-graded modular tensor category characterizing a TQFT with global symmetry.  This framework generalizes the axioms of \cite{Moore:1988qv} to include global symmetries and parallels the results for two-dimensional TQFTs in \cite{Moore:2006dw}.  We mainly focus on the case of bosonic TQFTs that do not require a spin structure. 

Finally, we explore a variety of explicit Chern-Simons-matter theories, and show that simple elementary examples often have 2-group global symmetry.  For instance, one concrete example is $U(1)_{q \ell}$ Chern-Simons theory coupled to $N_{f}$ scalars of charge $q$ (recently discussed in \cite{Komargodski:2017dmc}). This theory has the global symmetry groups 
\begin{equation}
G=U(N_{f})/\mathbb{Z}_{\ell}~, \hspace{.8in}\mathcal{A}=\mathbb{Z}_{q} \;.
\end{equation}
The permutation $\rho$ is trivial, and 
\begin{equation}
[\beta]=\Bock \big( w_2^{(\ell)} \big) \quad \in H^{3}\big( BU(N_{f})/\mathbb{Z}_{\ell} \;,\; \mathbb{Z}_{q} \big) \;,
\end{equation}
where in the above $w_2^{(\ell)}$ is a Stiefel-Whitney class (characterizing an obstruction to lifting the $G$-bundle to a $U(N_{f})$ bundle), and $\Bock$ is a Bockstein homomorphism (see Appendix~\ref{app: Bockstein}).  This symmetry and its anomaly are preserved under renormalization group flow.

While most of our examples involve unitary symmetries, much of the formalism can be generalized to include 2-groups where the 0-form symmetry is time-reversal.  We give several explicit examples of such theories in Section~\ref{sec:timereversal}. Along the way, we also present new examples of Chern-Simons theories with time-reversal symmetry.

We discuss some aspects of RG flows of theories with 2-group global symmetry in Section~\ref{sec: general coupling}.  In particular, it often happens that a theory with global symmetry $\mathbb{G}$ flows at long distances to a theory with global symmetry $\mathbb{K}$.  The 2-group $\mathbb{K}$ can be larger or smaller (both in its 0-form and 1-form global symmetry) than $\mathbb{G}$ because some symmetry in the IR can be accidental and some UV symmetry can decouple if all charged objects are massive.  In addition, the 2-groups $\bG$ and $\bK$ can have different Postnikov class.  However, since the UV theory can couple to $\mathbb{G}$ background fields, the IR theory can also couple to such fields.  We describe a general method for this coupling using homomorphisms of 2-groups,%
\footnote{This idea, in the case of the toric code theory, already appeared in Appendix B of \cite{Bhardwaj:2016clt}.}
and discuss some general implications for emergent symmetry.  We apply this technique to some explicit relevant deformations of QFTs in Section \ref{sec: examples}.  

In the context of three-dimensional TQFTs with global symmetry, our results imply that much of the formalism of \cite{Barkeshli:2014cna}, or its mathematical counterparts \cite{EtingofNOM2009, Teo:2015xla} describing $G$-graded modular tensor categories can be more simply understood in terms of the intrinsic symmetry of the TQFT.%
\footnote{We thank Nathan Seiberg for illuminating conversations regarding this issue.}
This is the 2-group whose 0-form part is the automorphism group of the category and whose 1-form part is the subset of Abelian anyons.  Coupling to more general symmetry groups then proceeds by activating these intrinsic symmetries using the formalism of Section~\ref{sec: general coupling}.


\section{2-Group Global Symmetry and Background Fields}
\label{sec: 2-groups}

\subsection{Symmetry Defects}

It is convenient to organize the symmetries of a QFT into $n$-form symmetries \cite{Gaiotto:2014kfa}. Let us review this classification in Euclidean signature in $d$ dimensions.  Throughout, we assume that spacetime is orientable. 

A 0-form symmetry with group $G$ (that can be discrete or continuous, Abelian or non-Abelian) is realized by unitary operators $U_\bfg$, with $\bfg \in G$, supported along codimension-1 submanifolds $X_{d-1}$. We refer to such operators as symmetry defects. If $G$ is continuous, the component of $G$ connected to the identity is realized by
\be
U_\alpha = e^{i\alpha_c Q^c} \;,\qquad\qquad Q^c = \int_{X_{d-1}} *J^c \;,
\ee
where the 1-forms $J^c$ are the conserved currents, $Q^c$ are the conserved charges, $c$ is an index in the Lie algebra $\fg$ and $\alpha \in \fg^*$ is a parameter. For discrete groups there are no currents, and the symmetry is defined through the operators $U_\bfg$. The operators $U_\bfg$ are topological, in the sense that correlation functions that include them are invariant as we deform $X_{d-1}$ in a continuous way, as long as we do not cross other operators.  In the case of continuous symmetry this is a consequence of charge conservation, but for more general symmetry groups the topological nature of the $U_\bfg$ is part of what we mean when we say that they define a symmetry of a field theory. 

The objects that are charged under 0-form symmetries are the local operators $\cO(x)$.%
\footnote{Here we consider operators that are genuinely local \cite{Kapustin:2014gua, Gaiotto:2014kfa}. In general there also exist operators that live at the end of lines.  The correlators of such operators may  pick up a phase when the attached line crosses $U_\bfg$. As a result, the charges of such ``quasi-local'' operators are ambiguous and not well-defined in terms of the quasi-local operators only.%
\label{foo: genuine}}
Specifically, if $X_{d-1}$ surrounds $\cO(x)$ (and nothing else) then it acts on $\cO(x)$ via a representation
\be
\label{action of U_g}
\big\langle U_\bfg \, \cO(x) \dots \big\rangle = \big\langle \cO'(x) \dots \big\rangle
\ee
where $\cO' = {}^\bfg \cO$ is the transformation of $\cO$ under $\bfg.$

We can turn on a background for $G$, \ie\ we can couple the theory to a $G$-bundle. We will mainly focus on the case of finite $G$, then $G$-bundles are necessarily flat.  To describe flat bundles on an Euclidean spacetime manifold $M_d$, we cover $M_d$ with open contractible patches $V_i$ in such a way that all possible multiple intersections are either empty or contractible. We choose an arbitrary ordering of the patches, and indicate intersections as $V_{ij} = V_i \cap V_j$, $V_{ijk} = V_i \cap V_j \cap V_k$, and so on, always using ordered indices. Then we specify transition functions $A_{ij} \in G$ on $V_{ij}$ satisfying
\be
\label{cocycle condition G-bundle}
A_{ij} A_{jk} = A_{ik} \qquad\text{on } V_{ijk} \;,
\ee
for ordered $\{ijk\}$. It is convenient to represent the patches and their intersections in terms of simplices (see also Appendix~\ref{app: cohomology}). We choose a triangulation of $M_d$, which is dual to the open cover. Thus, the vertices, or $0$-simplices, of the triangulation correspond to the open patches $V_i$. The edges, or $1$-simplices, correspond to the intersections $V_{ij}$. The edges can be given an orientation, from the vertex with lower label to the one with higher label. Simplices of higher dimension correspond to multiple intersections. 

The operators $U_\bfg$ can be constructed by turning on suitable $G$-bundles in which the transition functions are equal to $\bfg$ across the submanifold $X_{d-1}$ and to $\unit$ otherwise. More precisely, in the simplicial formulation the surface $X_{d-1}$ cuts a number of edges $\{ij\}$ and we assign $A_{ij} = \bfg$ if the edge crosses the surface with positive orientation, or $A_{ij} = \bfg^{-1}$ if with negative orientation. Since in each triangle (2-simplex) either none or two edges cross $X_{d-1}$, the cocycle condition (\ref{cocycle condition G-bundle}) is satisfied.%
\footnote{Precisely, we can assume that either none or two edges are crossed after suitably refining the open cover. \label{foo:refine}}

It is easy to see that this bundle realizes the operator $U_\bfg$. Suppose that $X_{d-1}$ is a surface wrapping a local operator $\cO$, which we can imagine be located at one of the vertices of the triangulation. Then we can remove $U_\bfg$ by performing a gauge transformation
\be
A_{ij} \;\to\; A_{ij}^f \,\equiv\, f^{\phantom{1}}_i A_{ij} \, f_j^{-1}
\ee
with $f_i = \bfg^{-1}$ if the vertex $i$ is inside $X_{d-1}$ or $f_i = \unit$ if it is outside. The operator $\cO$ is mapped to ${}^\bfg \cO$, therefore correlation functions satisfy (\ref{action of U_g}). Conversely, any flat $G$-bundle can be described by a net of operators $U_\bfg$. The partition function could be invariant under deformations of the net that describe the same $G$-bundle, or might change by a phase. In the latter case the theory has an 't~Hooft anomaly.

A 1-form symmetry with group $\cA$ is realized by unitary operators $W_a$, with $a \in \cA$, supported along codimension-2 submanifolds $Y_{d-2}$. As there is no covariant ordering of these submanifolds, $\cA$ is necessarily Abelian. As above, the operators $W_a$ are topological. The objects that are charged under $1$-form symmetries are the line operators $L(\ell)$, supported along lines $\ell$: if $Y_{d-2}$ links once with $L(\ell)$ then
\be
\big\langle W_a \, L(\ell) \dots \big\rangle = e^{2\pi i \theta(a)} \, \big\langle L(\ell) \dots \big\rangle
\ee
where $\theta \in \wh\cA$ is the charge of $L$ and $\wh\cA = \Hom(\cA, \bR/\bZ)$ is the Pontryagin dual to $\cA$.

We can turn on a background for $\cA$, namely an Abelian gerbe with fiber $\cA$. We assign elements $B_{ijk} \in \cA$ to triple intersections $U_{ijk}$, such that they satisfy the cocycle condition
\be
\label{cocycle condition A-gerbe}
B_{jkl} - B_{ikl} + B_{ijl} - B_{ijk} \,\equiv\, (dB)_{ijkl} = 0 \qquad \text{on } V_{ijkl}
\ee
for ordered $\{ijkl\}$. Thus $B$ is a cocycle in $Z^2(M_d,\cA)$. Since $\cA$ is Abelian, we have used the additive notation for it (as opposed to the multiplicative notation for $G$).

We can use backgrounds for $\cA$ to construct the operators $W_a$: we take the transition functions to be equal to $a$ on triple intersections along $Y_{d-2}$ and equal to $0$ otherwise. In the simplicial formulation, $Y_{d-2}$ pierces a number of triangles (2-simplices) $\{ijk\}$, and we assign $B_{ijk} = a$ if the defect pierces the triangle with positive orientation or $B_{ijk} = -a$ if with negative orientation. Since for each tetrahedron (3-simplex) the defect must enter across one face and exit across another, the cocycle condition (\ref{cocycle condition A-gerbe}) is satisfied. (See footnote \ref{foo:refine}.) 

Now suppose that the operator $W_a$ is supported on a surface $Y_{d-2}$ winding around a line operator $L(\ell)$ with charge $\theta$, that lies along some edges of the triangulation. We can remove $W_a$ by performing a 1-form gauge transformation $B_{ijk} \to B_{ijk} + \gamma_{jk} - \gamma_{ik} + \gamma_{ij}$ on $V_{ijk}$ for ordered vertices, in other words
\be
B \;\to\; B + d\gamma \;.
\ee
To define $\gamma$, we first draw a codimension-1 surface $\Sigma$ whose boundary is $W_a$, then we set $\gamma_{ij} = a$ if the edge $\{ij\}$ crosses $\Sigma$ with positive orientation or $\gamma_{ij}=-a$ if with negative orientation. Since $\Sigma$ also cuts an edge along $L(\ell)$, correlation functions pick up a phase $e^{2\pi i \theta(a)}$ corresponding to the one-dimensional representation under $\cA$. This is precisely the action of $W_a$.

\subsection{Elements of 2-Group Symmetry}

The previous discussion describes the notion of symmetry defects and background fields appropriate for a theory with global symmetry which is a standard product between a \mbox{0-form} symmetry group $G$ and a 1-form symmetry group $\mathcal{A}$.  In this Section we describe how these ideas must be modified if $G$ and $\mathcal{A}$ are fused into a non-trivial 2-group.

The first generalization away from a product symmetry is as follows. When we have both a 0-form symmetry $G$ and a 1-form symmetry $\cA$, $G$ can act on $\cA$. This is described by a group homomorphism
\be
\rho: G \to \Aut(\cA) \;.
\ee
In particular, the action of $G$ can permute the one-dimensional representations of $\cA$. When a symmetry operator of type $a\in\cA$ crosses $U_\bfg$, on the other side it appears as an operator of type $\rho_\bfg a$ (Figure~\ref{fig: action on higher symmetry}), while when a line operator $L(\ell)$ with charge $\theta$ crosses $U_\bfg$, on the other side it appears as a line operator with charge $\rho_\bfg \theta \equiv \theta \circ \rho_\bfg^{-1}$. This guarantees that $\langle \theta, a \rangle \equiv \theta(a)$ is invariant. 

If the permutation $\rho$ is the only modification of the global symmetry, then the appropriate background fields are a flat $G$-bundle and $\cA$-gerbe, where the latter now satisfies a twisted cocycle condition involving $\rho$:
\be
\label{twisted cocycle condition A-gerbe}
\rho(A_{ij}) B_{jkl} - B_{ikl} + B_{ijl} - B_{ijk} \,\equiv\, (d_A B)_{ijkl} = 0 \qquad \text{on } V_{ijkl} \;,
\ee
(for additional details see Appendix \ref{app: cohomology}) guaranteeing that 1-form symmetry operators are correctly transformed. Notice that, since $G$ permutes the representations of $\cA$, when $\cA$ is finite (and thus the number $n$ of unitary irreducible representations of $\cA$ is finite) $\rho$ induces a group homomorphism $\hat\rho$ from $G$ to the permutation group $S_n$, and the unitary irreducible representations of $\cA$ form a (possibly reducible) representation of $\hat\rho(G) \subseteq S_n$.

In addition to the permutation $\rho$, there is another way in which a 0-form symmetry and a 1-form symmetry can mix.  This is via a Postnikov class $[\beta]$.  The meaning of $[\beta]$ is simply stated in the language of symmetry generators. Consider a flat $G$-bundle described by a net of operators $U_\bfg$ that includes, in some region, a junction between $U_\bfg$, $U_\bfh$, $U_\bfk$ into $U_{\bfg\bfh\bfk}$, as in Figure \ref{fig: obstructed F-move} left. We consider a local transformation of the net, to the configuration in Figure~\ref{fig: obstructed F-move} right, that represents the same bundle and corresponds to a gauge transformation. Then the
correlation functions are invariant only if we include the codimension-2 symmetry operator of type $\beta(\bfg, \bfh, \bfk)$, parallel to the junction.

In $d\geq 3$ dimensions, we can construct a background configuration that is a bordism between the left and right configurations in Figure~\ref{fig: obstructed F-move}: the result is represented in Figure~\ref{fig: 2-group 3D} (in the case $d=3$, while in higher dimensions all objects span the remaining $d-3$ dimensions). The codimension-3 locus where four operators of type $\bfg$, $\bfh$, $\bfk$, $\bfg\bfh\bfk$ meet, acts as a source for a symmetry operator of type $\beta(\bfg,\bfh,\bfk)\in \cA$.   Here $\beta$ is ``normalized'' such that $\beta(\bfg, \bfh,\bfk)=0$ if at least one of the entries is $\unit$ and $\beta$ satisfies a twisted cocycle condition, which for finite $G$ reads
\be
\label{cocycle condition beta}
d_\rho\beta(\bfg,\bfh,\bfk,\bfl) \,\equiv\, \rho_\bfg \beta(\bfh,\bfk,\bfl) - \beta(\bfg\bfh, \bfk, \bfl) + \beta(\bfg, \bfh\bfk, \bfl) - \beta(\bfg, \bfh, \bfk\bfl) + \beta(\bfg,\bfh, \bfk) = 0 \;.
\ee
This condition follows from the equality in Figure~\ref{fig: 3D 0-form anomaly}, where the configuration of symmetry defects on the left is continuously deformed into the one on the right (this is a bordism implementation of the pentagon identity of \cite{Moore:1988uz}): since the fusion of 1-form symmetry defects must be the same on the two sides, the cocycle condition for $\beta$ follows. Moreover, as we will see below, the theory depends only on the function $\beta$ up to the equivalence relation $\beta \sim \beta + d_\rho \alpha$ with $\alpha: G\times G \to \cA.$ Thus the invariant data is the equivalence class $[\beta],$ an element of the third group-cohomology group of $G$ with values in $\cA$, twisted by $\rho$:
\be
[\beta] \in H^3_\rho(BG,\cA) \;,
\ee
where $BG$ is the classifying space of $G$. This cohomological characterization of $[\beta]$ also applies to continuous $G$ and $\mathcal{A}$ provided we use the sheaf cohomology of the classifying space \cite{Segal:1970, Brylinski:2000} (see Appendix \ref{app: group cohomology}).

The Postnikov class modifies the appropriate notion of background fields as follows.  The theory may still be coupled to a standard $G$-connection.  However the appropriate background field $B$ for $\mathcal{A}$ is no longer closed but rather has its coboundary fixed by $\beta$ as $(d_AB)_{ijkl} = \beta(A_{ij}, A_{jk}, A_{kl})$. We can also write this concisely by viewing the connection $A$ as a map from spacetime to the classifying space $BG$.  Then
\be
d_A B = A^* \beta \;,
\ee
expressing that the $G$-bundle acts as a source for $B$.  Note that the left-hand side above is necessarily closed.  Thus we see that the fact that $\beta$ is closed ensures that this equation is consistent.

As we explain in more detail in the following Section, the Postnikov class $[\beta]$ also affects the gauge transformations in such a way as to make $B$ transform under $A$ gauge transformations.  As discussed in \cite{Kapustin:2013uxa} this is a discrete analog of the Green-Schwarz mechanism.  In total, the global symmetry of the theory is described by the quadruplet
\be
\bG = \big(G,\cA, \rho, [\beta] \big) \;.
\ee
This data is intrinsic to the QFT: it describes its global symmetry and the action on gauge-invariant operators, and constitutes the definition of a 2-group.

It is instructive to compare the implications of a non-trivial Postnikov class described above to phenomena that are typically described as anomalies.  One point of view on the above discussion is that $[\beta]$ represents an obstruction to coupling the QFT simultaneously to a $G$-bundle and an $\cA$-gerbe.  Indeed, we see from Figure  \ref{fig: obstructed F-move} that the partition function with flat $G$-bundle is ambiguous up to operators that generate $\cA$.  For this reason, this phenomenon has sometimes be termed the ``$H^3$ anomaly'' or ``obstruction to symmetry fractionalization'' in the literature (see \eg{} \cite{Chen:2015bia, Barkeshli:2014cna, Teo:2015xla, Barkeshli:2017rzd}). 

Ambiguity up to a phase is a standard 't~Hooft anomaly.  It can be parameterized using anomaly inflow in terms of a non-dynamical $(d+1)$-dimensional theory (also called an invertible theory), which depends on the background fields. By contrast, the phenomena described by the Postnikov class cannot be described in terms of a non-dynamical $(d+1)$-dimensional theory.  Instead, the bulk would necessarily be dynamical to account for the fact that $\beta(\bfg, \bfh,\bfk)$ is a non-trivial operator in the theory.  This is a radical and undesirable modification of our theory which began as $d$-dimensional.  Instead, the correct interpretation is to modify the background fields to a 2-connection where $\beta$ is simply a parameter defining the symmetry on par with, say, the structure constants in a non-Abelian Lie algebra.

\subsection[2-Group Bundles and $F$-Moves]{2-Group Bundles and \matht{F}-Moves}

Let us describe 2-group bundles in the simplicial formulation, following \cite{Kapustin:2013uxa}. The transition functions $A_{ij}$ form a $G$-bundle: for each 2-simplex with ordered vertices $i<j<k$ we have
\be
\label{cocycle condition A}
A_{ij} A_{jk} = A_{ik} \qquad\qquad\qquad \raisebox{-2em}{\begin{tikzpicture}
\draw [->-] (0,0) node[left] {$j$} to node[midway, above] {$A_{jk}$} (2,0); \draw [->-] (1,-1) to node[midway, below right] {$A_{ik}$} (2,0) node[right] {$k$}; \draw [->-] (1,-1) node[below] {$i$} to node[midway, below left] {$A_{ij}$} (0,0);
\filldraw (0,0) circle [radius=.05]; \filldraw (2,0) circle [radius=.05]; \filldraw (1,-1) circle [radius=.05];
\end{tikzpicture}}
\ee
The background field $B \in C^2(M_d,\cA)$ is an $\cA$-valued 2-cochain  on $M_d$ satisfying the twisted cocycle constraint
\be
\label{tetrahedron figure}
\rho(A_{ij}) \, B_{jkl} - B_{ikl} + B_{ijl} - B_{ijk} = \beta(A_{ij}, A_{jk}, A_{kl})
\qquad\raisebox{-4em}{\begin{tikzpicture}
\coordinate (3) at (1.5,2); \coordinate (2) at (3,.5); \coordinate (1) at (2.5,-.5); \coordinate (0) at (0,0);
\draw [densely dashed,->-] (2) node[right] {$j$} to (0) node[left] {$l$};
\draw [->-] (2) to node[midway,below right] {$A_{jk}$} (1) node[below] {$k$};
\draw [->-] (1) to node[midway, below] {$A_{kl}$} (0);
\draw [->-] (3) node[above] {$i$} to (0); \draw [->-] (3) to (1); \draw [->-] (3) to node[midway, above right] {$A_{ij}$} (2);
\foreach \n in {0,...,3} \filldraw (\n) circle [radius=.05];
\end{tikzpicture}}
\ee
for each 3-simplex $\{ijkl\}$ with ordered vertices. The LHS is the twisted differential $d_AB$. On the RHS there is a representative $\beta \in Z^3_\rho(BG,\cA)$ of the Postnikov class $[\beta] \in H^3_\rho(BG,\cA)$. If we interpret $A$ as a map from the simplicial complex on $M_d$ to $BG$, then we can write (\ref{tetrahedron figure}) in the compact form
\be
\label{cocycle condition B}
d_A B = A^* \beta \;.
\ee

The modified cocycle condition (\ref{cocycle condition B}) depends on the representative $\beta$ of the class $[\beta]$. If we change the representative $\beta \to \beta' = \beta + d_\rho \alpha$ for some 2-cochain $\alpha$, we can simultaneously redefine $B \to B' = B + A^*\alpha$. Then (\ref{cocycle condition B}) is satisfied by the redefined quantities. This follows from the fact that $d_A A^* = A^* d_\rho$ (as long as the cocycle condition (\ref{cocycle condition A}) is satisfied). A field redefinition of background fields implies a modification of the theory by local counterterms.  Therefore, only the equivalence class $[\beta]$ is meaningful.

There are two types of gauge transformations. The first is given by a 0-cochain $f$ in $G$. This means that to each vertex (0-simplex) we associate $f_i \in G$. The 1-cocycle $A$ transforms as
\be
\label{gauge transformation A}
A_{ij} \;\to\; A^f_{ij} \, \equiv\, f_i^{\phantom{1}} A_{ij} \, f_j^{-1}
\ee
for each 1-simplex $\{ij\}$ with $i<j$. The cocycle condition (\ref{cocycle condition A}) remains invariant.  These are the standard gauge transformations for a $G$-connection.  The above is also accompanied by a transformation of the  2-cochain $B$:
\be
\label{gauge transformation B}
B \;\to\; B^f \,\equiv\, \rho(f)\, B + \zeta(A,f) \;.
\ee
The notation $\rho(f)\, B$ indicates the 2-cochain $\rho(f_i) B_{ijk}$ for each 2-simplex $\{ijk\}$ with ordered vertices, while $\zeta(A,f)$ is a 2-cochain that satisfies
\be
\label{def zeta}
d_{A^f} \zeta(A,f) = A^{f*} \beta - \rho(f)\big( A^*\beta \big)
\ee
and that vanishes when the transformation is trivial ($f = \unit$). This definition guarantees that $B^f$ satisfies the transformed version of the modified cocycle condition (\ref{cocycle condition B}). (To show this, one uses the identity $d_{A^f} \circ \rho(f) = \rho(f) \circ d_A$ which simply follows from the fact that $\rho$ is a homomorphism.)  

The second type of gauge transformations are those that are standard given a background $B$ field.  They are defined by an $\cA$-valued 1-cochain $\gamma$. The 1-cocycle $A$ remains invariant, while $B$ transforms as
\be
\label{bgaugetrans2grp}
B \;\to\; B^\gamma \,\equiv\, B + d_A \gamma \;.
\ee
The cocycle condition (\ref{cocycle condition B}) remains invariant. 

Let us comment on the gauge transformations above.
\begin{itemize}
\item Notice that equation (\ref{def zeta}) always has solutions since a gauge transformation of $A$ corresponds to a homotopy of $A: M_d \to BG$ which does not change the cohomology class of $A^*\beta$, and therefore the RHS of (\ref{def zeta}) vanishes in cohomology.  

\item The definition of the gauge transformation (\ref{def zeta}) requires us to pick a $\zeta$.  This choice is ambiguous.  However the ambiguity can be absorbed by a gauge transformation of the form \eqref{bgaugetrans2grp}.

\item Observe that $\zeta$ is related by descent to the Postnikov class $\beta$.  Therefore, as remarked in \cite{Kapustin:2013uxa}, \eqref{def zeta} should be viewed as a discrete version of the Green-Schwarz mechanism.
\end{itemize}

\begin{figure}[t]
\centering
\raisebox{-4em}{\begin{tikzpicture}
\coordinate [label=left:{$i$}] (0) at (0,0); \coordinate [label={[xshift=-.1cm]below:{$j$}}] (1) at (1.2,.8); \coordinate [label={[xshift=.12cm, yshift=-.1cm]below:{$m$}}] (4) at (2.9,.8); \coordinate [label=right:{$n$}] (5) at (4.1,0);
\coordinate [label=below:{$k$}] (2) at (2,-1.6); \coordinate [label=above:{$l$}] (3) at (2,2.8);
\foreach \n in {2,3,5} \draw [->-] (0) to (\n); \draw [densely dashed,->-] (0) to (1);
\foreach \n in {2,3,4} \draw [densely dashed,->-] (1) to (\n);
\foreach \n in {3,4} \draw [densely dashed,->-] (2) to (\n); \draw [->-] (2) to (5);
\draw [densely dashed,->-] (3) to (4); \draw [->-] (3) to (5);
\draw [densely dashed,->-] (4) to (5);
\foreach \n in {0,...,5} \filldraw (\n) circle [radius=.05];
\end{tikzpicture}}
\hspace{3cm}
\raisebox{-3em}{\begin{tikzpicture}
\draw [thick] (0,-1) node[below] {$\bf ghk$} to (0,0);
\draw [thick] (0,0) to (-1.8,2.4) node[above] {$\bf g$};
\draw [thick] (0,0) to (1.8,2.4) node[above] {$\bf k$};
\draw [thick, densely dashed] (-.6,.8) to (0,1.6); \draw [thick, densely dotted] (.6,.8) to (0,1.6);
\draw [thick] (0,1.6) to (0,2.4) node[above] {$\bf h$};
{\small \node at (-1.2,0.2) {$i$}; \node at (1.2,0.2) {$n$}; \node at (-.6,1.8) {$j$}; \node at (.6,1.8) {$m$}; \node at (0,.8) {$k,l$}; }
\end{tikzpicture}}
\caption{Left: Minimal triangulation necessary to encode the configuration of \mbox{codimension-1} symmetry defects in Figure~\ref{fig: 2-group 3D}. The vertices $\{i,j,k,l,m,n\}$ are ordered; in $d$ dimensions the remaining $d-3$ dimensions are implicit. The configuration is made of an upper ``pyramid'' $\{ijlmn\}$ and a lower upside-down pyramid $\{ijkmn\}$. Right: 2D section of the configuration of symmetry defects, seen from above. The nodes contained in each domain are indicated. The dashed line corresponds to the lower portion of Figure~\ref{fig: 2-group 3D} which contains the node $k$; the dotted line corresponds to the upper portion of Figure~\ref{fig: 2-group 3D} which contains the node $l$. Therefore, the triangulation can encode a bordism between the two configurations in Figure~\ref{fig: obstructed F-move}.
\label{fig: triangulation}}
\end{figure}
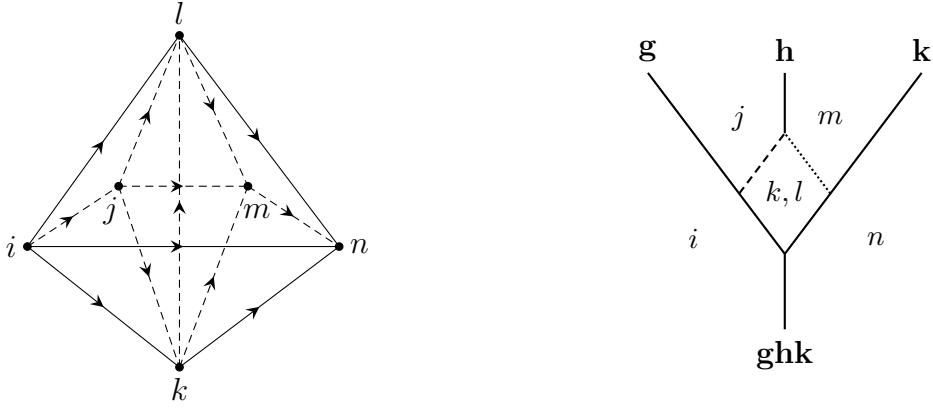

We conclude our discussion of 2-group background fields by demonstrating that the modified cocycle condition (\ref{cocycle condition B}) is equivalent to the statement that the junction point of four codimension-1 symmetry defects $\bfg$, $\bfh$, $\bfk$, $\bfg\bfh\bfk$ is the source of a codimension-2 symmetry defect $\beta(\bfg,\bfh,\bfk)$ as depicted in Figure~\ref{fig: 2-group 3D}. To that purpose, consider the local triangulation in Figure~\ref{fig: triangulation} with ordered vertices $\{i,j,k,l,m,n\}$ (in $d$ dimensions, the remaining $d-3$ dimensions are implicit). The bundle that describes the configuration of codimension-1 defects in Figure~\ref{fig: 2-group 3D} is
\bea
A_{ij} = A_{il} &= \bfg \qquad\qquad & A_{jk} = A_{jm} = A_{lm} &= \bfh \qquad\qquad & A_{kn} = A_{mn} &= \bfk \\
A_{ik} &= \bfg\bfh & A_{jl} = A_{km} &= \unit & A_{ln} &= \bfh\bfk \\
A_{in} &= \bfg\bfh\bfk & A_{kl} &= \bfh^{-1}
\eea
which satisfies the cocycle condition (\ref{cocycle condition A}). There are four 3-simplices in the triangulation, $\{ijkl\}$, $\{jklm\}$, $\{klmn\}$, $\{ikln\}$, and (\ref{cocycle condition B}) gives one equation for each of them. Imposing (with some arbitrariness) that the only external face (2-simplex) on which $B$ could possibly be non-zero is $\{iln\}$, we obtain the following equations:
\bea
\rho_\bfg B_{jkl} - B_{ikl} &= \beta(\bfg, \bfh, \bfh^{-1}) \;,\qquad\quad & B_{kln} - B_{klm} &= \beta(\bfh^{-1}, \bfh, \bfk) \\
\rho_\bfh B_{klm} - B_{jkl} &= \beta(\bfh \bfh^{-1}, \bfh) \;, & \rho_{\bfg\bfh} B_{jkl} - B_{ikl} - B_{iln} &= \beta(\bfg\bfh, \bfh^{-1}, \bfh\bfk) \;.
\eea
Combining them and using that $\beta$ is a (twisted) cocycle, we determine
\be
B_{iln} = \beta(\bfg, \bfh, \bfk) \;.
\ee
We see that $B$ cannot vanish on all external faces: rather the codimension-2 symmetry defect $\beta(\bfg,\bfh,\bfk)$, which pierces the face $\{iln\}$, is sourced. Since the configuration in Figure~\ref{fig: 2-group 3D} is a bordism between the two configurations in Figure~\ref{fig: obstructed F-move}, essentially the same computation shows that the modified gauge transformation (\ref{gauge transformation B})-(\ref{def zeta}) is equivalent to the $F$-move in Figure~\ref{fig: obstructed F-move}.


\section{'t Hooft Anomalies}
\label{sec: 't Hooft anomaly}

In this Section we discuss the classification of 't Hooft anomalies in theories with 2-group global symmetry.  In the special case of continuous 2-group global symmetry anomalies have been analyzed in \cite{Cordova:2018cvg}.  Here, we consider the general case.   As we have described above, a theory with 2-group global symmetry can couple to background fields $A$ and $B$ that describe a connection on a 2-group bundle.  These gauge fields are redundant and can be changed by the gauge transformations \eqref{gauge transformation A}, \eqref{gauge transformation B}, \eqref{bgaugetrans2grp}.  Naively, all correlation functions, thought of as functions of $A$ and $B$, must be invariant under these background gauge transformations.

An 't Hooft anomaly for a theory with 2-group global symmetry is a mild violation of this invariance. Specifically, we allow the (unnormalized) correlators to change by a phase when we change gauge.  This phase is universal in that all correlation functions are modified by the same phase.  If a theory exhibits gauge non-invariance, it is natural to attempt to modify the definition of the theory by adjusting gauge non-invariant counterterms (\ie\ local functions of the background fields) to restore gauge invariance.  The 't Hooft anomalies of interest are those that cannot be removed by adjusting such local terms, and hence they admit a cohomological classification.

A useful way to classify anomalies is via inflow.  In this paradigm the 't Hooft anomalies of a $d$-dimensional QFT are described by a $(d+1)$-dimensional topological action depending on the background fields.  (In the condensed matter literature these are referred to as SPTs, while in the mathematics literature they are invertible field theories.)  When formulated on a closed $(d+1)$-manifold $X$ the action is gauge invariant, but on a manifold with boundary its variation is precisely the 't Hooft anomaly of the boundary $d$-dimensional field theory of interest.

For our problem we are thus led to study topological actions for connections on a 2-group $\big( G,\cA, \rho,[\beta] \big).$    Throughout, we restrict our discussion to anomalies of bosonic theories (\ie\ we do not require a spin structure), and we further ignore any gravitational anomalies. Hence our actions depend only on the 2-group background fields, and not say, on Stiefel-Whitney classes of the manifold.  This class of topological actions has been investigated in detail in \cite{Kapustin:2013uxa} and our analysis below closely follows their treatment. We consider the case of one, two, three and four-dimensional QFTs separately, and describe in explicit terms how the anomaly manifests itself both in the inflow presentation, and via rearrangement rules for symmetry defects.

Topological actions take the generic form $S_\text{anom} = 2\pi \int_X \chi$ and appear in the path-integral as $e^{iS_\text{anom}}$, where $X$ is a closed $(d+1)$-manifold and $\chi$ is a $(d+1)$-cochain with values in $\bR/\bZ$ constructed out of the background fields $A$ and $B$.  On a closed $(d+1)$-manifold $X$,  $\chi$ is automatically closed, $\chi \in Z^{d+1}(X, \bR/\bZ)$, and its integral depends only on the cohomology class $[\chi] \in H^{d+1}(X,\bR/\bZ).$

To ensure that $\chi$ defines a satisfactory action, we must impose a further constraint: we view $\chi$ as a $(d+1)$-cochain on a $(d+2)$-manifold, and demand that the non-trivial equation $d\chi=0$ is satisfied.  This serves two purposes:
\begin{itemize}
\item It ensures that the action does not depend on the choice of triangulation used to present the 2-connection.  Indeed, one can modify the triangulation either by changing the simplicial structure at fixed vertices, or by adding or removing a vertex in the middle of a simplex of maximal dimension. These ``moves'' can be combined, and any two triangulations can be connected by a series of moves.%
\footnote{In two dimensions, the 2-2 move corresponds to flipping the diagonal of a quadrilateral, while the 1-3 move corresponds to adding a vertex inside a triangle and dividing the latter into three. In three dimensions, the 2-3 move corresponds to trading two tetrahedra that share a face for three tetrahedra that share an edge, while the 1-4 move corresponds to adding a vertex inside a tetrahedron and dividing the latter into four. In four dimensions there are 3-3, 2-4 and 1-5 moves.}
A change of triangulation can be made into a cobordism.  This is a $(d+2)$-dimensional triangulated manifold with boundary two copies of $X$ with different triangulations.  Thus, demanding that $d\chi$ vanishes on $(d+2)$-manifolds means that $\int_{X} \chi$ does not depend on the triangulation.

\item It ensures that the action is gauge invariant under the appropriate 2-group gauge transformations.  To see this, observe that gauge transformations can be performed locally on a finite number of vertices.  By modifying the triangulation we can remove those vertices where the gauge transformation is non-trivial.  Since we have concluded that the action is independent of the triangulation, it follows that it is also gauge invariant.  
\end{itemize}

When the topological action is a sum, $S_\text{anom} =2\pi\int_X \chi$ with $\chi=\chi_1+\chi_2 + \ldots,$ we must demand that the sum satisfies $d\chi=0$, while each term in the sum may not be closed. This generalizes the discussion in \cite{Kapustin:2013uxa} and gives a slightly different classification of topological actions. We give explicit examples of anomalies of various theories in Section~\ref{sec: examples}.

We now turn to an explicit construction of these actions.

\subsection{Anomalies in 1d}
\label{1danomsec}

To set our notation, we start with the simple case of a one-dimensional theory. This means that we are studying the partition function of a quantum mechanics model on a circle. We can insert unitary operators $\bfg \in G$, that correspond to gauge transformations for the background gauge field. By inserting a sequence $\bfg, \bfh, \bfk, \dots$ we section the circle into segments, each one in a different gauge, and the ordered product of elements determines the holonomy around the circle. These transition functions define a non-trivial (necessarily flat) background \mbox{$G$-bundle}.

The 't~Hooft anomaly shows up as the fact that correlators pick up a phase when we fuse two symmetry generators $\bfg$, $\bfh$ into one:
\be
\langle \dots \bfg\, \bfh \dots \rangle = e^{2\pi i \omega(\bfg, \bfh)} \, \langle \dots (\bfg\bfh) \dots \rangle \;. \label{projrepeq}
\ee
Graphically this is represented as
\be
\label{1D anomaly graphical}
\raisebox{-1em}{\begin{tikzpicture}
\draw (0,0) -- +(45:1.6);
\filldraw (0,0) ++(45:.4) node[left] {\footnotesize ${\bf g}$} circle (1pt);
\filldraw (0,0) ++(45:1.2) node[left] {\footnotesize ${\bf h}$} circle (1pt);
\end{tikzpicture}}
\; = \; e^{2\pi i \omega({\bf g}, {\bf h})} \;\;
\raisebox{-1em}{\begin{tikzpicture}
\draw (0,0) -- +(45:1.6);
\filldraw (0,0) ++(45:.8) node[left] {\footnotesize ${\bf gh}$} circle (1pt);
\end{tikzpicture}} \;.
\ee
Here $\omega \in C^2(BG,\bR/\bZ)$ is a normalized 2-cochain, and we indicate $U(1)$ as $\bR/\bZ$ in order to adhere to the additive notation. Associativity implies
\be
\omega(\bfh, \bfk) - \omega({\bf gh},{\bf k}) + \omega({\bf g},{\bf hk}) - \omega({\bf g},{\bf h}) = d\omega(\bfg, \bfh, \bfk) = 0
\ee
in $\bR/\bZ$, therefore $\omega$ is closed. The presence of the phases $e^{2\pi i \omega}$ implies a lack of gauge invariance of the partition function under gauge transformations of the background. 

We can modify the anomaly by adding the local counterterm
\be
\label{1D local counterterm}
S_\text{c.t.} = 2\pi \int_\text{1d} A^*\nu \qquad\qquad \nu \in C^1( BG, \bR/\bZ ) \;,
\ee
which is not gauge invariant unless $d\nu=0$. Its effect is to assign an extra phase $e^{2\pi i \nu(\bfg)}$ to each insertion of $\bfg$, which shifts $\omega \to \omega + d\nu$ by an exact term. Therefore the anomaly is parameterized by the cohomology class $[\omega] \in H^2(BG,\bR/\bZ)$. When $[\omega]=0$, we can add a local counterterm to the action to make it gauge invariant.%
\footnote{If we take a closed $\nu \in Z^1( BG, \bR/\bZ)$, then $S_\text{c.t.}$ is gauge invariant (in $\bR/\bZ$) and we can add it to the action without modifying its invariance properties. The effect of this term is to weight bundles with total holonomy $\bfg_\text{hol} \in G$ by $e^{2\pi i \nu(\bfg_\text{hol})}$.}

Notice that $H^2(BG, \bR/\bZ)$ parameterizes projective representations of $G$. In fact we can interpret the anomaly $\omega$ as the fact that the Hilbert space is in a projective, as opposed to regular, representation of $G$ as is manifest from \eqref{projrepeq}.

\paragraph{2d actions and anomaly inflow.}
The 't~Hooft anomaly discussed above is described by the following two-dimensional topological action for $G$ gauge fields:
\be
\label{2D action first term}
S_\text{anom} = 2\pi \int_X A^* \omega
\ee
where $[\omega] \in H^2(BG, \bR/\bZ)$. The integrand is closed because $dA^*\omega = A^*d\omega$.

It is easy to check that the action (\ref{2D action first term}) reproduces the anomaly when $X$ is a two-dimensional surface with one-dimensional boundary. Take a triangle $\{ijk\}$ whose clockwise ordered vertices lie on the boundary. To realize insertions $\bfg, \bfh$ on the boundary, we take $A_{ij} = \bfg$ and $A_{jk} = \bfh$. Then $\int A^*\omega = \omega(A_{ij}, A_{jk}) = \omega(\bfg, \bfh)$. To fuse the two insertions into $\bfg\bfh$, we perform a gauge transformation $f_j = \bfh^{-1}$, such that $A_{ij} \to \bfg\bfh$, $A_{jk} \to \unit$. Now $\int A^*\omega = \omega(\bfg\bfh, \unit) = 0$. We have thus reproduced (\ref{1D anomaly graphical}).

In 1d there cannot be a 1-form symmetry since its generators should have codimension two, and therefore there cannot be a 2-group symmetry either. However, if the 1d system is a defect into a higher-dimensional theory and the latter has a 2-group global symmetry, this can affect the anomalies on the 1d defect worldline. (An example is discussed in Section~\ref{sec: 3D TQFT}.)

In fact, we can write the following action for a two-dimensional 2-group gauge theory:
\be
\label{1Dinflow2grp}
S_\text{anom} = 2\pi \int_X \Big[ \eta(B) + A^* \omega \Big]
\ee
where
\bea
\eta &\in \wh\cA \;,\qquad d_\rho\eta =0 \;,\qquad \omega \in C^2(BG, \bR/\bZ) \;,\qquad \eta(\beta) = -d\omega \;.
\eea
The first two conditions say that $\eta$ is a $G$-invariant homomorphism from $\cA$ to $\bR/\bZ$,%
\footnote{In fact $(d_\rho \eta)(\bfg) = \rho_\bfg \eta - \eta = \eta \circ \rho_\bfg^{-1} - \eta$, see Appendix~\ref{app: cohomology}.}
namely $\eta \in Z^0_\rho(BG, \wh\cA) = H^0_\rho(BG,\wh\cA)$. The last condition implies that $\eta([\beta])=0$ in $H^3(BG, \bR/\bZ)$. To check that the integrand in \eqref{1Dinflow2grp} is closed we use
\be
d\eta(B) = \eta( d_A B) = \eta(A^*\beta) = A^* \eta(\beta) =  A^* (-d\omega) = - d A^*\omega \;.
\ee
Notice that when $\eta\neq 0$, the 0-form anomaly $\omega$ is no longer closed and the fusion of symmetry generators in 1d is no longer associative.%
\footnote{Physically, it can be understood as follows. When the 1d defect is immersed in a higher-dimensional theory, the $G$ defects are intersections of the 1d circle with bulk $G$ defects. To change the order of fusion we have to perform an $F$-move in the bulk first. This generates an $\cA$ symmetry defect of codimension 2 linked to the 1d circle, which induces a phase on the latter.}
In this case the anomaly forms an $H^2(BG, \bR/\bZ)$ torsor (that we also call an affine cohomology class): the differential of $\omega$ is not zero but it is fixed, and $\omega$'s that differ by a coboundary represent the same anomaly. One difference with standard cohomology classes is that there is no natural zero. We will see a concrete example of this in Section \ref{sec: 3D anomaly}.

\subsection{Anomalies in 2d}

In two dimensions the symmetry defects of the 0-form symmetry are unitary line operators of type $\bfg \in G$ while the defects of the 1-form symmetry are unitary local operators of type $a\in\cA$. Here, the 0-form symmetry action on local operators restricts to give the action $\rho$ on the 1-form symmetry generators.  Thus, when a local operator of type $a$ is moved across a line $\bfg$ (in the direction fixed by the orientation), it gets transformed to $\rho_\bfg^{-1} a \in \mathcal{A}$.

The $F$-move is a continuous transformation of the junction of four line operators $\bfg$, $\bfh$, $\bfk$, $\bfg\bfh\bfk$ (Figure~\ref{fig: obstructed F-move}) and the 2-group structure is encoded in the appearance of a local operator of type $\beta(\bfg,\bfh,\bfk)$.
Note that in two dimensions the 2-group condition $d_AB = A^*\beta$ is trivial since there are no 3-simplices.  Nevertheless one can still detect the Postnikov class using a gauge transformation as in Figure \ref{fig: obstructed F-move}.

There are two types of anomalies. First, when a local operator of type $a$ is moved across a line of type $\bfg$, it also acquires a phase $\exp\big( 2\pi i \lambda_a(\bfg)\big)$. We can interpret $\lambda$ as an element of $C^1(BG, \wh\cA)$. If we consider a junction as in Figure \ref{fig: basic junction} of $\bfg$, $\bfh$ and $\bfg\bfh$, we can move $a$ through the 0-form defects in two ways. Demanding equality gives $\lambda_a(\bfg) + \lambda_{\rho_\bfg^{-1} a}(\bfh) = \lambda_a(\bfg\bfh)$ which is the condition
\be
d_\rho\lambda = 0 \;.
\ee
Therefore $\lambda \in Z^1_\rho(BG, \wh\cA)$.

Second, the $F$-move involves a phase $\exp\big(2\pi i \, \omega(\bfg,\bfh,\bfk) \big) \in C^3(BG,\bR/\bZ)$:
\be
\label{2D 2-group and anomaly graphical}
\raisebox{-4em}{\begin{tikzpicture}
\draw [thick] (0,-.8) node[below] {$\bf ghk$} to (0,0);
\draw [thick] (0,0) to (-1.2,1.6) node[above] {$\bf g$};
\draw [thick] (0,0) to (1.2,1.6) node[above] {$\bf k$};
\draw [thick] (-.4,.533) to (0,1) to (0,1.6) node[above] {$\bf h$};
\end{tikzpicture}}
\; = \; e^{2\pi i \omega(\bfg, \bfh, \bfk)} \;\;
\raisebox{-4em}{\begin{tikzpicture}
\draw [thick] (0,-.8) node[below] {$\bf ghk$} to (0,0);
\draw [thick] (0,0) to (-1.2,1.6) node[above] {$\bf g$};
\draw [thick] (0,0) to (1.2,1.6) node[above] {$\bf k$};
\draw [thick] (.4,.533) to (0,1) to (0,1.6) node[above] {$\bf h$};
\filldraw [blue!80!black] (-.8,-.2) circle [radius=.05] node [below, black] {\scriptsize $\beta(\bfg, \bfh, \bfk)$};
\end{tikzpicture}} \;.
\ee
If we consider the junction of five generators $\bfg$, $\bfh$, $\bfk$, $\bfl$, $\bfg\bfh\bfk\bfl$, according to the pentagon identity \cite{Moore:1988qv} we can deform by applying (\ref{2D 2-group and anomaly graphical}) either two or three times and reach the same configuration. Associativity gives two relations:
\bea
0 &= \rho_\bfg \beta(\bfh, \bfk, \bfl) - \beta(\bfg\bfh, \bfk, \bfl) + \beta(\bfg, \bfh\bfk, \bfl) - \beta(\bfg, \bfh, \bfk\bfl) + \beta(\bfg, \bfh, \bfk) \\
\big\langle \lambda(\bfg), \rho_\bfg \beta(\bfh,\bfk,\bfl) \big\rangle  &= \omega(\bfh, \bfk, \bfl) - \omega(\bfg \bfh, \bfk, \bfl) + \omega(\bfg, \bfh\bfk, \bfl) - \omega(\bfg, \bfh, \bfk\bfl) + \omega(\bfg, \bfh, \bfk) \;.
\eea
The first equation above comes from matching the local operator that is generated, while the second constraint arises from matching the phases. These are the cocycle conditions
\be
d_\rho\beta = 0 \;,\qquad\qquad d\omega = \langle \lambda, \cup\, \beta \rangle \;.
\ee
We recognize the constraint $d_\rho\beta = 0$ as one of the defining characteristics of a 2-group.  Meanwhile, the second condition implies that $\omega$ is not closed, but has fixed differential.

We can modify the anomaly by adding the following non-gauge-invariant local counterterms:
\be
\label{2D counterterms}
S_\text{c.t.} = - 2\pi \int_\text{2d} \Big[ \eta(B) + A^*\nu \Big]
\ee
where $\eta \in C^0(BG,\wh\cA) \cong \wh\cA$ while $\nu \in C^2(BG,\bR/\bZ)$. They correspond to assigning extra phases $e^{-2\pi i \eta(a)}$ to the local operators and $e^{-2\pi i \nu(\bfg, \bfh)}$ to the junctions of line operators. The result is
\be
\label{2D anomaly shift}
\lambda \,\to\, \lambda + d_\rho \eta \;,\qquad\qquad \omega \,\to\, \omega + \eta(\beta) + d\nu \;.
\ee
The resulting cohomological content of the anomaly is specified below.

\paragraph{3d actions and anomaly inflow.} To encode the 2d anomaly via inflow, we consider the following 3d action:
\be
\label{3D anomaly action}
S_\text{anom} = - 2\pi \int_X \Big[ \big\langle A^*\lambda, \cup\, B \big\rangle + A^* \omega \Big]
\ee
where
\be
\lambda \in Z^1_\rho(BG, \wh\cA) \;,\qquad \omega \in C^3(BG, \bR/\bZ) \;,\qquad d\omega = \langle \lambda, \cup\,\beta\rangle \;.
\ee
The last condition implies that $\langle \lambda, \cup\,\beta \rangle = 0$ in $H^4(BG,\bR/\bZ)$. The integrand is closed because
\be
d \, \langle A^*\lambda, \cup\, B \rangle = \langle A^* d_\rho \lambda, \cup\, B \rangle - \langle A^*\lambda, \cup\, d_A B \rangle = - A^* \langle \lambda, \cup\,\beta \rangle \;.
\ee
We have discussed the local counterterms in (\ref{2D counterterms}) and their effect on the 3d anomaly action in (\ref{2D anomaly shift}). Since we can shift $\lambda$ by an exact term, the 1-form anomaly is parameterized by 
\begin{equation}
[\lambda] \in H^1_\rho(BG, \wh\cA)~.
\end{equation}
After fixing the representative $\lambda$, we can still shift $\omega$ by an exact term as well as by a term $\eta(\beta) \equiv \langle \eta,\beta\rangle$ for a $G$-invariant character $\eta$ of $\cA$, \ie{} for $\eta \in Z^0_\rho(BG, \wh\cA)$. Therefore the 0-form anomaly is a torsor over
$$
H^3(BG,\bR/\bZ) \;\text{\Large $/$}\; H^0_\rho(BG,\wh\cA) \cup [\beta] \;.
$$
Once again, the 0-form anomaly has no natural zero in general. Moreover, the effect of the Postnikov class $[\beta]$ is to trivialize part of the 0-form anomaly in $H^3(BG, \bR/\bZ)$. This is because one can reabsorb part of the anomalous phases in (\ref{2D 2-group and anomaly graphical}) into a phase redefinition of the local operators $\beta(\bfg,\bfh,\bfk)$.

\subsection{Anomalies in 3d}
\label{sec: 3D anomaly}

In three dimensions the generators of the 0-form symmetry are codimension-1 surface operators of type $\bfg \in G$ while the generators of the 1-form symmetry are line operators of type $a \in \cA$. In the language of TQFT, the latter are Abelian anyons. When a line operator of type $a$ pierces a surface of type $\bfg$ (in the direction fixed by the orientation), on the other side it emerges as a line of type $\rho_\bfg^{-1} a$ (Figure~\ref{fig: action on higher symmetry}). The Postnikov class $[\beta]$ can be seen in gauge transformations, when the junction of four surfaces $\bfg$, $\bfh$, $\bfk$, $\bfg\bfh\bfk$ is continuously deformed ($F$-move) and a line of type $\beta(\bfg,\bfh,\bfk)$ appears as in Figure~\ref{fig: obstructed F-move}. More directly, it can be extracted from fixed configurations: the constraint $d_AB = A^*\beta$ means that when four surfaces $\bfg$, $\bfh$, $\bfk$, $\bfg\bfh\bfk$ meet at a single point, a line $\beta(\bfg,\bfh,\bfk)$ emanates from there as in Figure~\ref{fig: 2-group 3D}. This configuration is a bordism between the two configurations on the left and right of Figure~\ref{fig: obstructed F-move}.

There are three types of 't~Hooft anomalies. First, when two line of type $a$, $b$ are crossed, a phase $M_{ab}$ is generated because of the 1-form anomaly \cite{Gaiotto:2014kfa, Gomis:unpublish}:
\be
\label{crossing of lines in 3D}
\raisebox{-1.5em}{\begin{tikzpicture}
\draw [thick, decoration = {markings, mark=at position .9 with {\arrow[scale=1.5]{stealth}}}, postaction=decorate] (-.6,-.6) to (.6,.6) node[above] {$b$};
\draw [thick] (.6,-.6) to (.1,-.1);
\draw [thick, decoration = {markings, mark=at position .75 with {\arrow[scale=1.5]{stealth}}}, postaction=decorate] (-.1,.1) to (-.6,.6) node[above] {$a$};
\end{tikzpicture}}
\; = M_{ab} \;\;
\raisebox{-1.5em}{\begin{tikzpicture}
\draw [thick, decoration = {markings, mark=at position .9 with {\arrow[scale=1.5]{stealth}}}, postaction=decorate] (.6,-.6) to (-.6,.6) node[above] {$a$};
\draw [thick] (-.6,-.6) to (-.1,-.1);
\draw [thick, decoration = {markings, mark=at position .75 with {\arrow[scale=1.5]{stealth}}}, postaction=decorate] (.1,.1) to (.6,.6) node[above] {$b$};
\end{tikzpicture}}
\;.
\ee
Such a phase defines a symmetric bilinear form $\langle \;,\,\rangle_q:\cA\times\cA \to \bR/\bZ$ by
\be
M_{ab} = e^{2\pi i \langle a,b\rangle_q} \;.
\ee
Since both configurations in (\ref{crossing of lines in 3D}) can be continuously moved across a surface $\bfg$ without producing any extra phase, the bilinear form must be $G$-invariant. As reviewed in detail in Appendix \ref{app: Pontryagin}, the bilinear form follows from a quadratic function $q:\cA \to \bR/\bZ$ such that $q(a) = q(-a)$ and
\be
\langle a, b \rangle_q = q(a+b) - q(a) - q(b) \qquad\qquad\forall\; a,b\in\cA \;.
\ee
Such a function is called a quadratic refinement of $\langle \;,\,\rangle_q$, and it can be interpreted as the ``spin'' of the lines, in the sense that a loop of a line $a$ rotating by 360$^\circ$ produces a phase $\theta_a = e^{2\pi i q(a)}$:
\be
\label{representation of spin}
\raisebox{-1.5em}{\begin{tikzpicture}
\draw [thick, decoration = {markings, mark=at position .6 with {\arrow[scale=1.5]{stealth}}}, postaction=decorate] (-1,-.7) to node[midway, left] {\small $a$} (-1,0);
\draw [thick] (-1,0) arc (180:-130:.25);
\draw [thick] (-1,.25) to (-1,.7);
\node at (0,0) {$=$};
\draw [thick, decoration = {markings, mark=at position .8 with {\arrow[scale=1.5]{stealth}}}, postaction=decorate] (1,0) to node[midway, right] {\small $a$} (1,.7);
\draw [thick] (1,0) arc (360:50:.25);
\draw [thick] (1,-.7) to (1,-.25);
\end{tikzpicture}}
\; = \;\; \theta_a \quad \raisebox{-1.5em}{\begin{tikzpicture}
\draw [thick, decoration = {markings, mark=at position .6 with {\arrow[scale=1.5]{stealth}}}, postaction=decorate] (0,-.7) to node[midway, right] {\small $a$} (0,.7);
\end{tikzpicture}} \;.
\ee
By manipulating a junction of $a$, $b$ and $a+b$ one obtains indeed $M_{ab} = \theta_{a+b}/\theta_a\theta_b$ \cite{Kitaev:2005}, and in particular%
\footnote{Taking the looped configuration on the left of (\ref{representation of spin}) and applying (\ref{crossing of lines in 3D}), we conclude that the existence of a non-trivial anomaly $M_{aa}$ implies that looped configurations necessarily pick up a phase as in (\ref{representation of spin}).}
 $M_{aa} = \theta_a^2$. Another equivalent formulation is in terms of a $G$-invariant group homomorphism \mbox{$\tilde q: \Gamma(\cA) \to \bR/\bZ$} where $\Gamma(\cA)$ is the universal quadratic group of $\cA$. The homomorphism can be interpreted as $\tilde q \in Z^0_\rho\big(BG, \wh{\Gamma(\cA)}\big),$ where the fact that $\tilde{q}$ is closed is equivalent to the fact that it is $G$-invariant. 

Second, when a line of type $a\in\cA$ crosses the junction line of three surfaces%
\footnote{In 3D, the junction of three surfaces is a line operator. However it is not a genuine line operator \cite{Gaiotto:2014kfa}, in the sense that it does not exist in isolation but only at the junction of three surfaces.}
$\bfg$, $\bfh$, $\bfg\bfh$, a phase $\exp\big( 2\pi i \lambda_a(\bfg,\bfh) \big)$ is generated because of the mixed 1-form/0-form anomaly:
\be
\label{def fractionalization phases lambda_a}
\raisebox{-2cm}{\begin{tikzpicture}
\draw [thick, red!90!black] (0,0) to (0,-2); 
\draw (0,0) to (-2,.5) to (-2,-1.5) to (0,-2); 
\draw (0,0) to (1,-1) to (1,-3) to (0,-2);
\draw (0,0) to (1.5,1) to (1.5,-1) to (1,-1.33); \draw [dashed, dash phase=-2] (1,-1.33) to (0,-2);
\draw [thick, blue!80!black, decoration = {markings, mark=at position .7 with {\arrow[scale=1.5,rotate=6]{stealth}}}, postaction=decorate, rotate around={-30:(.3,-1.4)}] (.3,-1.4) arc [x radius=.6, y radius =0.4, start angle=-30, end angle=-170]; 
\draw [thick, blue!80!black, dashed, dash phase=1.5, rotate around={-30:(.3,-1.4)}] (.3,-1.4) arc [x radius=.6, y radius =0.4, start angle=-30, end angle=190];
\draw [blue!80!black, fill=white] (-.6,-.73) circle [radius=.05];
\draw [blue!80!black, fill=white] (.3,-1.4) circle [radius=.05];
\draw [blue!80!black, fill=white] (.26,-.7) circle [radius=.05];
\node at (-1.75,0.1) {\small $\bfg$}; \node at (1.2,0.4) {\small $\bfh$}; \node at (.7,-2.2) {\footnotesize $\bfg\bfh$}; 
\node at (-.8,-1.2) {\small $a$};
\end{tikzpicture}}
\quad = e^{2\pi i \lambda_a(\bfg, \bfh)} \quad
\raisebox{-2cm}{\begin{tikzpicture}
\draw [thick, red!90!black] (0,0) to (0,-2); 
\draw (0,0) to (-2,.5) to (-2,-1.5) to (0,-2); 
\draw (0,0) to (1,-1) to (1,-3) to (0,-2);
\draw (0,0) to (1.5,1) to (1.5,-1) to (1,-1.33); \draw [dashed, dash phase=-2] (1,-1.33) to (0,-2);
\draw [thick, blue!80!black, decoration = {markings, mark=at position .3 with {\arrow[scale=1.5,rotate=6]{stealth}}}, postaction=decorate, rotate around={-30:(-.4,-1.4)}] (-.4,-1.4) arc [x radius=.6, y radius =0.4, start angle=-30, end angle=-390]; 
\node at (-1.75,0.1) {\small $\bfg$}; \node at (1.2,0.4) {\small $\bfh$}; \node at (.7,-2.2) {\footnotesize $\bfg\bfh$}; 
\node at (-1.5,-1.2) {\small $a$};
\end{tikzpicture}}
\ee
We interpret $\lambda$ as an element of $C^2(BG, \wh\cA)$; it is constrained as follows. Wrap a line $a$ around the two red lines in the lower part of Figure \ref{fig: 2-group 3D}, below the point where the line $\beta$ emanates. The line $a$ can be pulled away from the intersection producing a certain phase. Alternatively, slide the line $a$ up, in a continuous way, to the upper part of the figure. Now, pulling the line $a$ through the intersection and through the emanated line $\beta$, produces a different phase. Equating the two we get
\be
\label{2D constraint on lambda}
\lambda_a(\bfg, \bfh) + \lambda_a(\bfg\bfh, \bfk) = \lambda_{\rho_\bfg^{-1}a}(\bfh,\bfk) + \lambda_a(\bfg, \bfh\bfk) - \langle \beta(\bfg,\bfh,\bfk) ,a\rangle_q \;.
\ee
This is the constraint $d_\rho\lambda = \langle \beta, \star \rangle_q$, where the right-hand-side is a function on $\cA$ whose argument is $\star$.  

Interestingly, we could interpret the line $a$ as the worldline of a massive particle. If we restrict to the subgroup $G_a \subset G$ that stabilizes $a$, we can think of $G_a$ as the 0-form symmetry of the quantum mechanics of $a$. Then $\lambda_a \in C^2(BG_a, \bR/\bZ)$ is its 0-form anomaly. However, as apparent in (\ref{2D constraint on lambda}), $\lambda_a$ is in general not closed. This is an example where a ``bulk'' 2-group symmetry induces a quantum mechanical anomaly which is not closed. (See this discussion at the end of Section \ref{1danomsec}.)

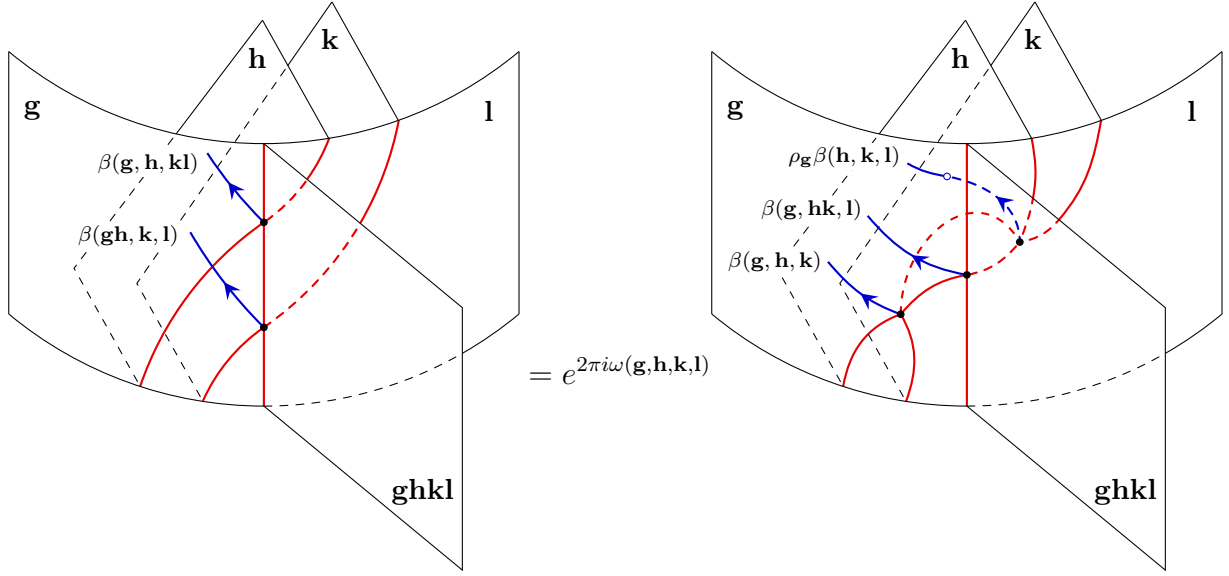
\begin{figure}[t]
$$
\raisebox{-6em}{%
\begin{tikzpicture}[scale=.87]
\draw [thick, red!90!black] (0,-2) to (0,2); 
\draw [thick, red!90!black] (0,.8) arc (125:161.4:5); 
\draw [thick, red!90!black, dashed, dash phase = -1, line cap=round] (0,.8) arc (-55:-35.5:2.7); 
\draw [thick, red!90!black] (.99,2.09) arc (-20.3:-35.5:2.7);
\draw [thick, red!90!black] (0,-.8) arc (125:156.3:2.7); 
\draw [thick, red!90!black, dashed, line cap=round] (0,-.8) arc (-55:-30.2:5);
\draw [thick, red!90!black] (2.04,2.36) arc (-10.7:-30.1:5);
\draw (0,2) to (3,-.5) to (3,-4.5) to (0,-2); 
\draw ([shift=(-130:6)] 0,8) arc (-130:-50:6);
\draw ([shift=(-130:6)] 0,4) arc (-130:-90:6); \draw [dashed, line cap=round] ([shift=(-90:6)] 0,4) arc (-90:-60:6); \draw ([shift=(-60:6)] 0,4) arc (-60:-50:6);
\draw ([shift=(-130:6)] 0,4) to ++(0,4); \draw ([shift=(-50:6)] 0,4) to ++(0,4);
\draw [dashed, dash phase=-2] (-1.87,-1.7) to ++(-1,1.8); \draw (.98,2.08) to ++(-1,1.8); 
\draw [dashed, dash phase=1.5, line cap=round] (-2.87,0.1) to ([shift=(53:2.57)] -2.87,0.1);
\draw [line cap=round] ([shift=(53:2.57)] -2.87,0.1) to ([shift=(53:4.73)] -2.87,0.1);
\draw [dashed, dash phase=-2] (-0.92,-1.93) to ++(-1,1.8); \draw (2.03,2.36) to ++(-1,1.8); 
\draw [dashed, dash phase=1, line cap=round] (-1.92,-0.13) to ([shift=(55.5:4.03)] -1.92,-0.13);
\draw [line cap=round] ([shift=(55.5:4.03)] -1.92,-0.13) to ([shift=(55.5:5.2)] -1.92,-0.13);
\draw [thick, blue!80!black, decoration = {markings, mark=at position .5 with {\arrow[scale=1.5,rotate=0]{stealth}}}, postaction=decorate] (0,-0.8) arc (-135:-150:7); 
\draw [thick, blue!80!black, decoration = {markings, mark=at position .6 with {\arrow[scale=1.5,rotate=0]{stealth}}}, postaction=decorate] (0,0.8) arc (-135:-147:6.5);
\node at (-3.5,2.5) {\small $\bfg$}; \node at (-0.1,3.3) {\small $\bfh$}; \node at (1.0,3.6) {\small $\bfk$}; 
\node at (3.4,2.5) {\small $\bfl$}; \node at (2.4,-3.3) {\small $\bfg\bfh\bfk\bfl$};
\filldraw [white] (-2.5, 1.5) rectangle +(1.5,.4) node[midway, black] {\scriptsize $\beta(\bfg, \bfh,\bfk\bfl)$}; 
\filldraw [white] (-2.8, .4) rectangle +(1.5,.4) node[midway, black] {\scriptsize $\beta(\bfg\bfh,\bfk, \bfl)$};
\filldraw (0,.8) circle [radius=.05]; \filldraw (0,-.8) circle [radius=.05];
\end{tikzpicture}%
} = e^{2\pi i \omega(\bfg, \bfh, \bfk, \bfl)}
\raisebox{-6em}{%
\begin{tikzpicture}[scale=.87]
\draw [thick, red!90!black] (0,-2) to (0,2); 
\draw [thick, red!90!black] (-1.87,-1.7) to[out=80, in=-160] (-1,-.6); 
\draw [thick, red!90!black] (-0.92,-1.93) to[out=70, in=-60] (-1,-.6);
\draw [thick, red!90!black] (-1,-.6) to[out=50, in=-170] (0,0);
\draw [thick, red!90!black, dashed, line cap=round, dash phase=-.8] (0,0) to[out=10, in=-130] (0.8,0.5);
\draw [thick, red!90!black, dashed, line cap=round, dash phase=2.7] (.8,.5) to[out=70, in=-100] (1.0,1.17); \draw [thick, red!90!black] (1.0,1.17) to[out=80, in=-80] (.98,2.08);
\draw [thick, red!90!black, dashed, line cap=round, dash phase=2.2] (.8,.5) to[out=10, in=-130] (1.40,0.85); \draw [thick, red!90!black] (1.40,0.85) to[out=50, in=-100] (2.03,2.36);
\draw [thick, red!90!black, dashed] (-1,-.6) to[out=90, in=120, distance=1.2cm] (.8,.5);
\draw (0,2) to (3,-.5) to (3,-4.5) to (0,-2); 
\draw ([shift=(-130:6)] 0,8) arc (-130:-50:6);
\draw ([shift=(-130:6)] 0,4) arc (-130:-90:6); \draw [dashed, line cap=round] ([shift=(-90:6)] 0,4) arc (-90:-60:6); \draw ([shift=(-60:6)] 0,4) arc (-60:-50:6);
\draw ([shift=(-130:6)] 0,4) to ++(0,4); \draw ([shift=(-50:6)] 0,4) to ++(0,4);
\draw [dashed, dash phase=-2] (-1.87,-1.7) to ++(-1,1.8); \draw (.98,2.08) to ++(-1,1.8); 
\draw [dashed, dash phase=1.5, line cap=round] (-2.87,0.1) to ([shift=(53:2.57)] -2.87,0.1);
\draw [line cap=round] ([shift=(53:2.57)] -2.87,0.1) to ([shift=(53:4.73)] -2.87,0.1);
\draw [dashed, dash phase=-2] (-0.92,-1.93) to ++(-1,1.8); \draw (2.03,2.36) to ++(-1,1.8); 
\draw [dashed, dash phase=1, line cap=round] (-1.92,-0.13) to ([shift=(55.5:4.03)] -1.92,-0.13);
\draw [line cap=round] ([shift=(55.5:4.03)] -1.92,-0.13) to ([shift=(55.5:5.2)] -1.92,-0.13);
\draw [thick, blue!80!black, decoration = {markings, mark=at position .5 with {\arrow[scale=1.5,rotate=0]{stealth}}}, postaction=decorate] (-1,-0.6) to[out=160, in=-50] +(-1.1,.8); 
\filldraw [white] (-3.6, 0) rectangle +(1.4,.4) node[midway, black] {\scriptsize $\beta(\bfg, \bfh,\bfk)$};
\draw [thick, blue!80!black, decoration = {markings, mark=at position .5 with {\arrow[scale=1.5,rotate=0]{stealth}}}, postaction=decorate] (0,0) to[out=170, in=-50] +(-1.5,.9); 
\filldraw [white] (-3.1, .8) rectangle +(1.5,.4) node[midway, black] {\scriptsize $\beta(\bfg, \bfh\bfk, \bfl)$};
\draw [thick, blue!80!black, dashed, line cap=round, dash phase=1, decoration = {markings, mark=at position .5 with {\arrow[scale=1.5,rotate=0]{stealth}}}, postaction=decorate] (.8,.5) to[out=90, in=-10] (-.3,1.5); 
\draw [thick, blue!80!black] (-.3,1.5) to[out=170, in=-30] (-.9,1.7);
\draw [blue!80!black, fill=white] (-.3,1.5) circle [radius=.05];
\filldraw [white] (-2.7, 1.6) rectangle +(1.7,.4) node[midway, black] {\scriptsize $\rho_\bfg \beta(\bfh, \bfk, \bfl)$};
\node at (-3.5,2.5) {\small $\bfg$}; \node at (-0.1,3.3) {\small $\bfh$}; \node at (1.0,3.6) {\small $\bfk$}; 
\node at (3.4,2.5) {\small $\bfl$}; \node at (2.4,-3.3) {\small $\bfg\bfh\bfk\bfl$};
\filldraw (0,0) circle [radius=.05]; \filldraw (-1,-.6) circle [radius=.05]; \filldraw (.8,.5) circle [radius=.05]; 
\end{tikzpicture}%
}
$$
\caption{Following the pentagon identity, the two configurations of five joining surfaces can be modified one into the other. However, because of the 0-form anomaly, they differ by a phase. The red lines are junctions of three surfaces. Where four surfaces meet (two red lines intersect), a line (in blue) emanates according to the 2-group symmetry.
\label{fig: 3D 0-form anomaly}}
\end{figure}

Third, mimicking the pentagon identity, a junction of five surfaces $\bfg$, $\bfh$, $\bfk$, $\bfl$, $\bfg\bfh\bfk\bfl$ can be modified with the application of either two or three $F$-moves, leading to the same final configuration. Interpreting those as two different fixed configurations that realize the two cobordisms, the two fixed configurations differ by a phase $\omega(\bfg, \bfh,\bfk, \bfl)$ because of the 0-form anomaly. This is depicted in Figure~\ref{fig: 3D 0-form anomaly}. Notice that the two emanated lines on the left are equivalent to the three emanated lines on the right because of $d_\rho\beta = 0$.

One can consider the junction of six surfaces $\bfg$, $\bfh$, $\bfk$, $\bfl$, $\bfm$, $\bfg\bfh\bfk\bfl\bfm$, and deform the configuration with successive applications of the anomalous transformation in Figure~\ref{fig: 3D 0-form anomaly}. There exists a ``hexagon relation'' such that the same final configuration can be reached in two different ways, each consisting of three anomalous transformations and some braiding of the emanated lines. Equality of the produced phases gives the equation
\be
\label{3D anomaly: 0-form I}
d\omega(\bfg,\bfh,\bfk,\bfl,\bfm) = \big\langle \lambda(\bfg, \bfh), \rho_{\bfg\bfh} \beta(\bfk, \bfl, \bfm) \big\rangle + \cR \;.
\ee
Here $\cR \in \bR/\bZ$ is the phase coming from braiding, constrained to satisfy
\begin{multline}
\label{3D anomaly: 0-form II}
2\,\cR = \langle \beta, \cup_1 \beta \rangle_q = \big\langle \beta(\bfg\bfh\bfk, \bfl, \bfm), \beta(\bfg, \bfh, \bfk) \big\rangle_q  + {} \\
+ \big\langle \beta(\bfg, \bfh\bfk\bfl, \bfm), \rho_\bfg \beta(\bfh, \bfk, \bfl) \big\rangle_q + \big\langle \beta(\bfg, \bfh, \bfk\bfl\bfm), \rho_{\bfg\bfh} \beta(\bfk, \bfl, \bfm) \big\rangle_q \;.
\end{multline}
Steenrod's cup product $\cup_1$ is reviewed in Appendix \ref{app: cohomology}. In general this equation does not fix $\cR$ completely: in the language of TQFT, this is the freedom in the choice of braiding matrices for fixed monodromy.

We can try to cure the anomaly by adding non-gauge-invariant local counterterms to the action. We will discuss these counterterms below. One corresponds to assigning an extra phase $\exp\big( 2\pi i \eta_a(\bfg)\big)$ to the point where a line $a$ pierces a surface $\bfg$, with $\eta \in C^1(BG, \wh\cA)$. Another one corresponds to assigning an extra phase $\exp\big( -2\pi i \nu(\bfg, \bfh,\bfk)\big)$ to the junction point of four surfaces $\bfg$, $\bfh$, $\bfk$, $\bfg\bfh\bfk$ in Figure~\ref{fig: 2-group 3D}, with $\nu \in C^3(BG,\bR/\bZ)$.

\paragraph{4d actions and anomaly inflow.} The 3d anomaly is described by the following 4d topological action:
\be
\label{4D anomaly action}
S_\text{anom} = 2\pi i \int_X \Big[ \tilde q(\fP B) - \langle A^*\lambda, \cup\, B \rangle + A^* \omega \Big] \;.
\ee
Let us explain the various terms above. As usual, $B \in C^2(X, \cA)$ with $d_A B = A^*\beta$ and $\beta \in Z^3_\rho(BG,\cA)$. Then $\fP B$ is the Pontryagin square of $B$ \cite{Whitehead:1949} (see Appendix \ref{app: Pontryagin}), namely an element of $C^4\big( X, \Gamma(\cA) \big)$ defined modulo coboundaries (with a residual dependence on a choice of ``lift'' for $\beta$), and $\tilde q \in \Hom\big( \Gamma(\cA), \bR/\bZ \big)$. This homomorphism is constrained to be $G$-invariant (in terms of the induced action of $G$ on $\Gamma(\cA)$), which can be phrased as $d_\rho\tilde q= 0$ \ie{} $\tilde q \in H^0_\rho\big( BG, \wh{\Gamma(\cA)} \big)$. The homomorphism is related to a $G$-invariant symmetric bilinear form $\langle \;,\,\rangle_q:\cA \times\cA \to \bR/\bZ$. The differential of the first term in (\ref{4D anomaly action}) is (see Appendix \ref{app: Pontryagin}):
\be
d\, \tilde q(\fP B) = \big\langle d_AB, \cup\,B \rangle_q - \tilde q \big( \fP_1 d_AB \big) = \big\langle A^*\beta, \cup\, B\big\rangle_q - A^* \tilde q(\fP_1\beta) \;.
\ee
Here $\fP_1\beta$ is a higher Pontryagin square: it is an element of $C^5\big( BG, \Gamma(\cA)\big)$ defined modulo coboundaries and satisfying
\be
d\, \tilde q(\fP_1 \beta) = - \langle \beta, \cup\,\beta\rangle_q \;.
\ee
The ambiguity in the addition of exact terms is precisely the residual dependence of $\fP B$ on the choice of lift of $\beta$ and, as we will see, can be reabsorbed in $\omega$. Finally $\lambda \in C^2(BG, \wh\cA)$ and $\omega \in C^4(BG, \bR/\bZ)$, with the two further constraints
\be
\label{3D anomaly: constraints}
d_\rho \lambda = \langle \beta, \star \rangle_q \;,\qquad\qquad d\omega = \langle \lambda, \cup\,\beta\rangle + \tilde q(\fP_1\beta) \;.
\ee
In the first equation, $\langle \beta, \star \rangle_q$ is the element of $C^3(BG,\wh\cA)$ obtained by evaluating on $\cA$ inserted at $\star$. The two equations guarantee that the integrand in (\ref{3D anomaly action}) is closed and thus that the action is independent of the triangulation and gauge invariant. The second equation in (\ref{3D anomaly: constraints}) corresponds exactly to (\ref{3D anomaly: 0-form I})-(\ref{3D anomaly: 0-form II}) and $\cR$ can be identified with $\tilde q(\fP_1\beta)$: the ambiguity by coboundaries of $\tilde q(\fP_1\beta)$ is the freedom in $\cR$---coming from braiding---that we noticed there.

To the 3d action we can add the following non-gauge-invariant local counterterms:
\be
S_\text{c.t.} = 2\pi \int_\text{3d} \Big[ - \langle A^*\eta, \cup\, B\rangle + A^*\nu \Big]
\ee
with $\eta \in C^1(BG, \wh\cA)$ and $\nu \in C^3(BG, \bR/\bZ)$. Their effect is to shift
\be
\lambda \,\to\, \lambda + d_\rho \eta \;,\qquad\qquad \omega \,\to\, \omega + \langle \eta, \cup\,\beta\rangle + d\nu \;.
\ee
Therefore, the 1-form anomaly is directly represented by the bilinear form $\langle\;,\,\rangle_q$ and its quadratic refinement $q$. The mixed anomaly is a torsor over $H^2_\rho(BG, \wh\cA)$. After fixing the representative $\lambda$, we can still shift $\omega$ by an exact term and a term $\langle \eta, \cup\beta\rangle$ for a closed $\eta$. Hence the 0-form anomaly is a torsor over
$$
H^4(BG,\bR/\bZ) \;\text{\Large $/$}\; H^1_\rho(BG,\wh\cA) \cup [\beta] \;.
$$
Again, the Postnikov class $[\beta]$ partially trivializes the 0-form anomaly.

\subsection{Anomalies in 4d}

In four dimensions the generators of the 0-form symmetry are codimension-1 wall operators of type $\bfg \in G$ while the generators of the 1-form symmetry are surface operators of type $a \in \cA$. 

There are three types of 't~Hooft anomalies. When a wall $\bfg$ is pulled through the intersection point%
\footnote{The intersection of two surfaces $a,b$ is a bordism between the two configurations in (\ref{crossing of lines in 3D}).}
of two surfaces $a, b$, a phase $\exp\big( 2\pi i \theta(a,b;\bfg)\big)$ is generated. When a surface $a$ is pulled through the intersection line%
\footnote{Such intersection line corresponds to the special point in Figure~\ref{fig: 2-group 3D} times an extra $\bR$ spanned by all objects. A surface that wraps the intersection line, necessarily intersects the surface $\beta$ at a point.}
of four walls $\bfg$, $\bfh$, $\bfk$, $\bfg\bfh\bfk$, a phase $\exp\big( 2\pi i \lambda_a(\bfg,\bfh,\bfk)\big)$ is generated. Finally, there are two configurations in which six walls $\bfg_1, \dots, \bfg_5, \bfg_1 \cdots \bfg_5$ meet, and the transformation from one to the other introduces a phase $\exp\big( 2\pi i \omega(\bfg_1, \dots, \bfg_5)\big)$. One can try to cure the anomaly by adding local counterterms: an extra phase $e^{2\pi i \langle a,b\rangle_q}$ assigned to the intersection point of two surfaces $a,b$, a phase $e^{2\pi i \eta_a(\bfg,\bfh)}$ assigned to the point where a surface $a$ intersects the intersection surface of three walls $\bfg$, $\bfh$, $\bfg\bfh$, and a phase $e^{2\pi i \nu(\bfg_1, \dots, \bfg_4)}$ assigned to the intersection point of five walls $\bfg_1, \dots, \bfg_4, \bfg_1\cdots \bfg_4$. We will refrain from giving a graphical description of 't~Hooft anomalies in 4d, as this is difficult, and use instead the language of anomaly inflow.

The 2-group anomalies are controlled by the following action:
\be
\label{5D anomaly action}
S_\text{anom} = 2\pi \int_X \bigg[ \langle A^*\theta, \cup\, \fP B \rangle + \langle A^*\lambda, \cup\, B \rangle + A^* \omega \bigg] \;.
\ee
Here
\bea
B &\in C^2(X,\cA) \qquad\qquad\qquad\qquad\qquad & \lambda &\in C^3(BG,\wh\cA) \\
d_A B &= A^*\beta & d_\rho \lambda &= \big\langle \theta, \cup\, \langle \beta, \star\rangle_\gamma \big\rangle \\
\theta &\in Z^1_\rho(BG, \wh{\Gamma(\cA)}) & \omega &\in C^5(BG, \bR/\bZ) \\
\fP B &\in C^4\big( X, \Gamma(\cA)\big) & d\omega &= \langle \lambda, \cup\, \beta\rangle - \langle \theta, \cup\, \fP_1\beta \rangle \;.
\eea
We have used the symmetric bilinear pairing
\be
\langle \,\;,\; \rangle_\gamma: \cA \times \cA \,\to\, \Gamma(\cA) \;,
\ee
reviewed in Appendix \ref{app: Pontryagin}, which is used in the construction of the Pontryagin square. The constraints above guarantee that the integrand in (\ref{5D anomaly action}) is closed.

We can add non-gauge-invariant local counterterms:
\be
S_\text{c.t.} = 2\pi \int_\text{4d} \Big[ \tilde q (\fP B) + \langle A^*\eta, \cup\,B \rangle + A^*\nu \Big] \;.
\ee
Here $\tilde q \in \Hom\big( \Gamma(\cA), \bR/\bZ\big) \cong C^0\big( BG, \wh{\Gamma(\cA)} \big)$ and the first term could be written $\langle A^*\tilde q, \fP B \rangle$. Then $\eta \in C^2(BG, \wh\cA)$ and $\nu \in C^4(BG, \bR/\bZ)$. The effect of these terms is to shift
\bea
\theta &\,\to\, \theta + d_\rho \tilde q \\
\lambda &\,\to\, \lambda + d_\rho \eta + \langle \beta, \star \rangle_q \\
\omega &\,\to\, \omega + d\nu + \langle \eta, \cup\, \beta \rangle - \tilde q(\fP_1 \beta) \;.
\eea
Notice that $\langle \;,\,\rangle_q$ is no longer a $G$-invariant pairing, unless $d_\rho\tilde q = 0$. This is clearer if we write $\langle \;,\,\rangle_q = \big\langle \tilde q, \langle \;,\, \rangle_\gamma \big\rangle$. Therefore the anomaly parameterized by $\theta$ can be viewed as a cohomology class $[\theta] \in H^1_\rho \big(BG, \wh{\Gamma(\cA)}\big).$  The anomaly parameterized by $\lambda$ is a torsor over
$$
H^3_\rho(BG, \wh\cA) \;\text{\Large $/$}\;  H^0_\rho\big( BG, \wh{\Gamma(\cA)} \big) \cup \big[ \langle \beta, \star \rangle_\gamma \big]~,
$$
and the anomaly parameterized by $\omega$ is a torsor over
$$
H^5(BG, \bR/\bZ) \;\text{\Large $/$}\; H^2_\rho(BG, \wh\cA) \cup [\beta] \;.
$$
The two quotients above represent the fact that the Postnikov class $[\beta]$ partially trivializes the mixed 1-form/0-form anomaly described by $\lambda$ and the 0-form anomaly described by $\omega$.

\section{Coupling to General Symmetry Groups}
\label{sec: general coupling}

In previous sections we have discussed the meaning of 2-group global symmetry, background fields, and 't Hooft anomalies.  As we have stressed, 2-group global symmetry is intrinsic to a QFT.  An important caveat in this discussion is that the symmetry that we are interested in acts faithfully on the operators of the theory.  By this we mean that all non-trivial elements of the intrinsic 0-form symmetry $K$ act on or permute some local or extended operator, and similarly all non-trivial elements of the intrinsic 1-form symmetry $\cA$ act on some line operators by non-trivial phases.  (The specific case of 3d TQFTs described by modular tensor categories is discussed in Section \ref{sec: 3D TQFT}.)

It is often the case that, even if a given QFT has some intrinsic 2-group symmetry $\bK = \big( K, \cA, \rho, [\beta]\big)$, we may be interested in coupling to background fields for a different ``extrinsic" 2-group $\bG = \big(G, \cB, \tau, [\Theta] \big)$.  One situation where this arises is in the study of renormalization group flows.  For instance, suppose that a UV QFT has 2-group global symmetry $\bG$.  We assume it to be intrinsic so that it acts faithfully on operators in the UV theory.  It is then natural to couple the theory to background $\bG$ gauge fields.   In the IR we may then arrive at a theory with a different intrinsic symmetry $\bK$.  This symmetry could be larger than $\bG$, because some of the symmetry is accidental in the IR, or it could be smaller because all of the charged objects are massive and have decoupled from the IR field theory (or the flow breaks part of the symmetry).

However, the entire RG flow---and in particular the IR theory---can be coupled to $\bG$ gauge fields.  This is achieved through a homomorphism $\bG\rightarrow \bK$.  In other words, the most general situation in the IR is governed by the coupling to $\bK$ gauge fields, but we can also couple to $\bG$ by using $\bG$-backgrounds to activate intrinsic $\bK$-backgrounds (in the case of the toric code theory, this idea was explored in \cite{Bhardwaj:2016clt}).  In this Section we describe this process in detail.  In particular, this discussion is essential to understand 't~Hooft anomaly matching.  Specifically, if the UV theory has some 't Hooft anomaly for $\bG$, in general it is reproduced in the IR by activating $\bK$-anomalies. 

In the simplest case where the 1-form symmetries $\cA$ and $\cB$ are trivial, the process that we describe below is simply a homomorphism of groups $f_{1}: G\rightarrow K$ that can be used to turn $G$ gauge fields $A$ into $K$ gauge fields $f_1(A)$. It becomes more interesting when the 1-form symmetries are non-trivial.  In that situation we can, for instance, use ordinary $G$ background fields to activate $\cA$ background fields.  A continuous example is illuminating.  Suppose that both $\mathcal{A}$ and $G$ are $U(1)$.  Then $\cA$ can couple to a background 2-form field $B$, and $G$ to a 1-form connection $A$.  We can therefore couple a theory with $\cA$ symmetry to a $G$-background by setting
\begin{equation}
 \label{contback}
B = \alpha \, dA \;,
\end{equation}
where $\alpha\in\mathbb{R}/\mathbb{Z}$ is a coupling constant, and $B$ has the standard 1-form gauge transformation $B\rightarrow B+d\lambda$ with $U(1)$ gauge field $\lambda$ (integer shifts of $\alpha$ are gauge transformations of $B$). The formalism discussed in this Section generalizes this idea both by allowing the symmetries to form a non-trivial 2-group, and by considering the case of discrete symmetries as well.%
\footnote{In general, this formalism can be straightforwardly extended by taking into account the $p$-form symmetry of a model for $p>1$.  }

As above, we consider a theory with  intrinsic 2-group symmetry $\bK = \big( K, \cA, \rho, [\beta]\big)$, where $K$ is the 0-form symmetry, $\cA$ is the total 1-form symmetry, $\rho:K \to \Aut(\cA)$ is a group action of $K$ on $\cA$, and $[\beta] \in H^3_\rho(BK,\cA)$ is the Postnikov class.  We denote the background fields for $(K,\cA)$ by a pair $(B_1, B_2)$.  We wish to couple the theory to background fields $(X_1, X_2)$ for an extrinsic 2-group $\bG = \big(G, \cB, \tau, [\Theta] \big)$. Therefore we would like to understand what 0-form and 1-form groups $G,\cB$ are allowed, what 2-group structures---characterized by $\tau$ and $[\Theta]$---they can form, what freedom we have in the coupling to the QFT, and what is the resulting 't~Hooft anomaly.

The first step is to choose group homomorphism from $G,\cB$ to $K,\cA$ 
\begin{equation}
f_1:~~ G\;\longrightarrow \; K ~, \qquad\qquad\qquad f_2:~~\cB\;\longrightarrow\; \cA \;.
\end{equation}
We will use these homomorphisms, together with other data specified below, to construct background fields.

\paragraph{A Special Case.} Before tackling the most general situation, let us assume that $f_1$ is the trivial map while $f_2$ is an (injective) inclusion. We describe a mechanism to couple to a 2-group bundle for $\bG$ using only the intrinsic 1-form symmetry $\cA$. Since $f_2$ is an inclusion,
\be
\label{short exact sequence of f2}
\begin{tikzcd}[column sep=small]
1 \arrow[r] & \cB \arrow[r, "f_2"] & \cA \arrow[r, "p"] & \cA' \equiv \cA/\cB \arrow[l, dashed, bend left, "\pi"] \arrow[r] & 1
\end{tikzcd}
\ee
is a short exact sequence. Here $p$ is the projection map $\text{mod }\cB$. We have also indicated a lift of $\cA'$, \ie{} an arbitrarily chosen map $\pi:\cA' \to \cA$ such that $p\circ\pi = \id_{\cA'}$. In general, $\cA$ is an extension of $\cA'$ by $\cB$. 

We can use the short exact sequence \eqref{short exact sequence of f2} to construct 2-cocycles for $\cA$, that we use as background fields for the 1-form symmetry, in terms of cochains for $\cB$ and $\cA'$ that satisfy certain conditions. First we need a 2-cocycle for $\cA'$, namely $B_2' \in Z^2(M, \cA')$, such that
\be
\label{condition on B2'}
dB_2' = 0 \;,\qquad\qquad \Bock\big( [B_2'] \big) = 0 \;.
\ee
Here $\Bock:H^i(M, \cA') \to H^{i+1}(M, \cB)$ is the Bockstein homomorphism (App.~\ref{app: Bockstein}) for the exact sequence (\ref{short exact sequence of f2}). Second we need a 2-cochain for $\cB$, namely $X_2 \in C^2(M, \cB)$, such that
\be
\label{condition on X2}
dX_2 = f_2^{-1} \big( d\, \pi(B_2') \big) \;.
\ee
This equation depends on the particular choice of $\pi$ we made. Notice that $f_2^{-1} \circ d \circ \pi$, when acting on cocycles, produces a representative of the Bockstein cohomology class, and thus (\ref{condition on X2}) implies the second eqn. in (\ref{condition on B2'}). Then one constructs the 2-cocycle for $\cA$ as
\be
B_2 = f_2(X_2) - \pi(B_2') \;,
\ee
where $B_2 \in Z^2(M,\cA)$. Note also that this describes the most general $B_{2},$ as we can extract the two components via $B_2' = - p (B_2)$ and $X_2 = f_2^{-1}\circ ( \id_\cA - \pi \circ p)(B_2)$.

On the other hand, we can construct 2-cocycles $B_2'$ out of $G$-backgrounds.   We first choose a class $[q] \in H^2(BG,\cA')$ and construct
\be
\label{possible Postnikov special case}
[\Theta] = \Bock\big( [q]\big) \quad \in H^3(BG,\cB) \;.
\ee
As we show below, these are allowed Postnikov classes for 2-groups $\bG = \big(G,\cB,1,[\Theta]\big)$ the QFT can couple to. We choose a representative $q \in Z^2(BG,\cA')$ for $[q]$ and construct a representative $\Theta = f_2^{-1}d\pi(q)$ for $[\Theta]$. Then, given a $G$-bundle with connection $X_1$, that we can think of as a homotopy class of maps $X_1:M \to BG$, we simply set
\be
\label{setting B2'}
B_2' = X_1^*q \;.
\ee
In other words, from a background $(X_1,X_2)$ for $\bG$ that---combining (\ref{condition on X2}) and (\ref{setting B2'})---satisfies
\be
dX_2 = X_1^*\Theta \;,
\ee
we construct the valid 1-form $\cA$-background
\be
\label{expression B2 special case}
B_2 = f_2(X_2) - X_1^* \pi(q) \;.
\ee
This shows how we can consistently couple a QFT, only through its intrinsic 1-form symmetry $\cA$, to 2-group bundles $\bG = \big(G,\cB,1,[\Theta]\big)$ for all Postnikov classes of the form (\ref{possible Postnikov special case}), \ie{} for all $[\Theta] \in \Bock\big( H^2(BG,\cA')\big)$. We will present a concrete example of this type in Section \ref{sec: example CSM}.%
\footnote{In fact, we could be more general and set $B_2 = f_2(X_2) - X_1^*\big( \pi(q) + \nu\big)$ for some chosen $\nu \in Z^2(BG,\cA)$. The freedom in the choice of $\nu$ corresponds to the so-called fractionalization classes ({\it e.g.} \cite{Essin:2013rca, Chen:2015bia, Barkeshli:2014cna, Teo:2015xla, Tarantino:2016}), and will be discussed in the general case.}

Notice that since $f_1$ is trivial, $G$ maps to the identity in $K$ and so it does not permute the anyons. If the QFT in question is a 3d TQFT, it has been conjectured in \cite{Barkeshli:2014cna} that when the 0-form symmetry does not permute the anyons, the Postnikov class should vanish (in Appendix \ref{app:trivialpermutation} we prove this statement in a special case). This seems in contradiction with the 2-groups $\bG$ we constructed above. However recall that the full 1-form symmetry of the theory is $\cA$, not $\cB$. If we map the Postnikov class $[\Theta] \in H^3(BG,\cB)$ to $H^3(BG,\cA)$ we find
\be\label{ptrivialTQFT}
f_2\big( [\Theta] \big) = 0 \quad\text{in } H^3(BG,\cA) \;,
\ee
\ie{} the class is trivialized in the larger coefficient group $\cA$. In fact, by the extension of (\ref{short exact sequence of f2}) to a long exact sequence in cohomology, $\Bock\big( H^2(BG,\cA')\big) \subset H^3(BG,\cB)$ is precisely the kernel of $f_2$ and hence it is the largest possible set of Postnikov classes when $G$ does not permute the anyons (assuming the conjecture of \cite{Barkeshli:2014cna} is correct).

By substituting (\ref{expression B2 special case}) in the anomaly inflow action of the QFT for the intrinsic 1-form symmetry $\cA$, one can find the implied 't~Hooft anomaly for the 2-group $\bG$. Of course, this anomaly only makes sense up to local counterterms that can be written in terms of $\bG$ gauge fields (we will expand on this point below). The 't~Hooft anomaly for $\bG$ will include a 1-form part, a mixed 1-form/0-form part, and a 0-form part. In particular, a QFT with only 1-form symmetry and no 0-form symmetry can still reproduce anomalous variations for a 0-form $G$-background $X_1$. 

\paragraph{The General Case.} Let us now consider the general problem of coupling a QFT to backgrounds $(X_1,X_2)$ for $\bG = \big(G,\cB,\tau,[\Theta]\big)$. We aim to construct backgrounds $(B_1,B_2)$ for the intrinsic 2-group symmetry $\bK = \big( K, \cA, \rho, [\beta]\big)$ in terms of $(X_1,X_2)$. We set
\bea
\label{expression B1 B2}
B_1 &= f_1(X_1) \\
B_2 &= f_2(X_2) - X_1^*\zeta
\eea
for some $\zeta \in C^2(BG,\cA)$.

The 2-group structure of $\bK$ requires
\be
\label{intrinsic 2-group}
d_{\rho(B_1)} B_2 = B_1^* \beta
\ee
with $\rho:K \to \Aut(\cA)$ and $[\beta] \in H^3_\rho(BK,\cA)$. For the sake of clarity, in this Section we explicitly indicate the group action involved in the twisted differential. Substituting (\ref{expression B1 B2}) into (\ref{intrinsic 2-group}) we get the equation
\be
\label{intermed eqn for general case}
d_{(\rho\circ f_1)(X_1)} \big( f_2(X_2) - X_1^*\zeta \big) = X_1^* f_1^* \beta \;.
\ee
We should determine which $\tau$ and $\Theta$ in the 2-group equation
\be
d_{\tau(X_1)} X_2 = X_1^*\Theta
\ee
with $\tau:G \to \Aut(\cB)$ and $[\Theta] \in H^3_\tau(BG,\cB)$, ensure that (\ref{intrinsic 2-group}) is satisfied. The first condition is that, for every $g\in G$, the following square diagram commutes:
\be
\begin{tikzcd}
G \arrow[d, dashed, "f_1"'] & \cB \arrow[r, "\tau(g)"{name=U}] \arrow[from=1-1, to=U, bend left=40, dashed, "\tau"'] \arrow[d, "f_2"'] & \cB \arrow[d, "f_2"] \\
K & \cA \arrow[r, "(\rho\circ f_1)(g)"'{name=D}] \arrow[from=2-1, to=D, bend right=40, dashed, "\rho"] & \cA
\end{tikzcd}
\ee
(Dashed lines are only included to clarify the origin of the square diagram.) The latter is a constraint on the group action $\tau$,
\be
f_2 \circ \tau(g) = (\rho\circ f_1)(g) \circ f_2
\ee
in $\Hom(\cB,\cA)$, guaranteeing that the action of $G$ on $\cB$ is compatible with the action of $K$ on $\cA$. Using the constraint, we can simplify $d_{(\rho\circ f_1)(X_1)} f_2(X_2) = f_2\big( d_{\tau(X_1)} X_2 \big) = X_1^* f_2(\Theta)$ and $d_{(\rho\circ f_1)(X_1)} X_1^*\zeta = X_1^* d_{\rho\circ f_1} \zeta$. Hence (\ref{intermed eqn for general case}) becomes the pull-back by $X_1^*$ of the equation
\be
\label{constraint general case}
d_{\rho\circ f_1} \zeta = f_2(\Theta) - f_1^*\beta \;.
\ee
If we want (\ref{intermed eqn for general case}) to be satisfied for all backgrounds $X_1$, then (\ref{constraint general case}) should be satisfied. This is a cohomological constraint on the possible Postnikov classes $[\Theta]$ for $\bG$:
\be\label{eqn:constraintsymm}
f_2\big( [\Theta] \big) = f_1^* [\beta] \qquad\text{in } H^3_{\rho\circ f_1}(BG,\cA) \;.
\ee
When this is satisfied and we have chosen representatives $\Theta, \beta$, then there exist $\zeta$'s that solve (\ref{constraint general case}). In fact, (\ref{constraint general case}) leaves us some freedom in the choice of $\zeta$: we can shift
\be
\zeta \;\to\; \zeta + \nu \qquad\text{with}\qquad [\nu] \in H^2_{\rho\circ f_1}(BG,\cA) \;.
\ee
Only the shift by a cohomology class $[\nu]$ matters, because if $\nu$ is a coboundary then $B_2$ in (\ref{expression B1 B2}) in shifted by a gauge transformation. On the contrary, $[\nu]$ parameterizes different allowed couplings of the TQFT to the 2-group $\bG = \big( G,\cB, \tau, [\Theta]\big)$. Those are called ``fractionalization classes'' in the literature (see \eg{} \cite{Essin:2013rca,Chen:2015bia,Barkeshli:2014cna,Teo:2015xla,Tarantino:2016}) and form an $H^2_{\rho\circ f_1}(BG,\cA)$ torsor.

Once again, the 't~Hooft anomaly of the 2-group symmetry $\bG$ can be obtained by substituting the background (\ref{expression B1 B2}) into the 't~Hooft anomaly of the intrinsic 2-group symmetry $\bK$. One should be careful that the resulting anomaly for $\bG$ is defined up to local counterterms that can be written in terms of $\bG$ gauge fields.

\paragraph{An example about possible induced anomalies.} To clarify the meaning of the last paragraph, let us present a simple example. Consider the 3D Chern-Simons TQFT $U(1)_2$, which has a $\bZ_2$ 1-form symmetry with anomaly
\be
\label{example anomaly reduction}
S_\text{anom} = \pi \int_{X_4} \frac12 \fP(B_2)
\ee
where $B_2 \in Z^2(X_4, \bZ_2)$ is the background field. We can couple the theory to an extrinsic $U(1)$ 0-form symmetry, with gauge field $A$, by setting $B_2 = dA/2\pi$. Substituting into (\ref{example anomaly reduction}) and noticing that the Pontryagin square $\fP$ reduces to a standard square, we obtain the action $\frac1{8\pi} \int F \wedge F$. This term is gauge invariant and well-defined on a 4-manifold with boundary, therefore it is an allowed local counterterm%
\footnote{Such a term is gauge invariant in 4D but it depends on the extension. It can be considered as a 3D local but not gauge invariant counterterm.}
that can be removed. The resulting theory has manifestly well-defined 3D Lagrangian
\be
\cL = \frac{2}{4\pi} bdb + \frac1{2\pi} bdA \;.
\ee
On the other hand, we can couple the theory to an extrinsic $SO(3)$ 0-form symmetry, with gauge field $A$, by setting $B_2 = A^* w_2^{SO(3)}$ in terms of the second Stiefel-Whitney class of $SO(3)$. In this case the anomaly
\be
S_\text{anom} = \pi \int_{X_4} \frac{A^* \fP\big(w_2^{SO(3)}\big)}2
\ee
is non-trivial and cannot be removed.

\subsection{Comments on Accidental Symmetries}

The condition \eqref{eqn:constraintsymm} constrains possible RG flows. Suppose that $\bG$ is the UV 2-group symmetry and $\bK$ is the IR 2-group symmetry.  If we assume that $f_1,f_2$ are (injective) inclusion maps, then we can use the construction above to prove that, in certain cases, these maps cannot be isomorphisms and hence that there must be accidental symmetry in the IR.

As an example, suppose that the Postnikov class $[\Theta]$ for the UV 2-group symmetry $\bG$ is non-trivial, while the Postnikov class $[\beta]$ for the IR 2-group symmetry $\bK$ vanishes. Then it follows from \eqref{eqn:constraintsymm} that $f_2$ cannot be an isomorphism, {\it i.e.} the 1-form symmetry must be enhanced in the IR: $\cB \varsubsetneq \cA$. Since this is true for any $f_1$, not necessarily an inclusion, the conclusion remains true even if some local operators decouple during the flow, such as in flows that end up in infrared gapped TQFTs.  A general class of flows where this applies is in 3d QFTs where the UV 0-form symmetry acts on the local operators but does not permute anyons.  When such theories flow to gapped TQFTs at long distances, the 0-form symmetry still does not permute the anyons and hence, (according to a conjecture of  \cite{Barkeshli:2014cna}; see also Appendix \ref{app:trivialpermutation}) the IR Postnikov class is trivial.  Thus the IR TQFT must have emergent 1-form global symmetry. 
See the examples in Sections \ref{sec:eguone} and \ref{sec: example CSM}.

Similarly, if the Postnikov class for the intrinsic 2-group symmetry in the UV is trivial, but in the IR is non-trivial, then $f_1$ cannot be an isomorphism, {\it i.e.} there must be an accidental 0-form symmetry in the IR.


\section{Example: 3d TQFTs with a Global Symmetry}
\label{sec: 3D TQFT}

A general class of examples of theories with 2-group global symmetry is provided by three-dimensional TQFTs---\ie{} gapped systems with topological order---with a global symmetry. TQFTs are particularly important because they describe generic gapped systems, which are a common end of RG flows.\footnote{Below, we do not consider TQFTs with local operators. These are important for describing theories with spontaneously broken 0-form symmetries.}  As we discuss below, for these systems the phenomenon known in the literature (\eg{} \cite{Chen:2015bia, Barkeshli:2014cna, Teo:2015xla, Barkeshli:2017rzd}) as ``obstruction to symmetry fractionalization'' or ``$H^3$ anomaly'' is in fact the signature of a 2-group global symmetry. This relationship was also noted in \cite{Thorngren:2015gtw, Tachikawa:2017gyf}, and here we provide a detailed dictionary. 
   
Let us first review some aspects of the axiomatic construction of 3d TQFTs. Before introducing global symmetry, these theories are described by a unitary modular tensor category $\cC$ (\eg{} \cite{Moore:1988qv, bakalov2001lectures, Kitaev:2005}).  A class of observables is given by line operators.   There is a finite set of such lines $\{a,b,c,\dots\} \in \cC$ and they obey a commutative fusion algebra
\be
\label{fusion algebra}
a \times b = \sum_{c\,\in\,\cC} N_{ab}^c \; c \;.
\ee
In the above, the  $N_{ab}^c = N_{ba}^c$  are non-negative integers, that are equal to the dimensions of vector spaces $V^{ab}_c$ associated to trivalent vertices (or to the sphere $S^2$ with three punctures, in radial quantization). The algebra has an identity $0$, which is the trivial and completely transparent line. We indicate as $\bar a$ the line conjugate to $a$, which has opposite orientation, such that $a \times \bar a = 0$. Associativity requires $\sum_e N_{ab}^e N_{ec}^d = \sum_f N_{af}^d N_{bc}^f$. One can choose bases of vectors $| a,b; c, \mu\rangle$ in these vector spaces, and diagrammatically represent these states as
\be
\bigg( \frac{d_c}{d_a d_b} \bigg)^{1/4} \;
\raisebox{-2em}{\begin{tikzpicture}
\draw [thick, decoration = {markings, mark=at position .5 with {\arrow[scale=1.5]{stealth}}}, postaction=decorate] (0,-.8) node[below] {\small $c$} to (0,0);
\draw [thick, decoration = {markings, mark=at position .7 with {\arrow[scale=1.5]{stealth}}}, postaction=decorate] (0,0) node[below right] {\small $\mu$} to (-.6,.5) node[above] {\small $a$};
\draw [thick, decoration = {markings, mark=at position .7 with {\arrow[scale=1.5]{stealth}}}, postaction=decorate] (0,0)--(.6,.5) node[above] {\small $b$};
\end{tikzpicture}}
\; = |a,b; c,\mu \rangle \in V^{ab}_c \;,\qquad \bigg( \frac{d_c}{d_a d_b} \bigg)^{1/4} \;
\raisebox{-2em}{\begin{tikzpicture}
\draw [thick, decoration = {markings, mark=at position .8 with {\arrow[scale=1.5]{stealth}}}, postaction=decorate] (0,0) to (0,.8) node[above] {\small $c$};
\draw [thick, decoration = {markings, mark=at position .6 with {\arrow[scale=1.5]{stealth}}}, postaction=decorate] (-.6,-.5) node[below] {\small $a$} to (0,0) node[above right] {\small $\mu$};
\draw [thick, decoration = {markings, mark=at position .6 with {\arrow[scale=1.5]{stealth}}}, postaction=decorate] (.6,-.5) node[below] {\small $b$} to (0,0);
\end{tikzpicture}}
\; = \langle a,b; c,\mu| \in V_{ab}^c \;.
\ee
The normalization constants $d_a$ are called quantum dimensions, and will be fixed momentarily. There exist orthogonality and completeness relations for lines:
\be
\label{orthogonality & completeness}
\raisebox{-2.5em}{\begin{tikzpicture}
\draw [thick, decoration = {markings, mark=at position .8 with {\arrow[scale=1.5]{stealth}}}, postaction=decorate] (0,-.5) node[left] {\small $c'$} to (0,0) node[right] {\footnotesize $\mu'$};
\draw [thick, decoration = {markings, mark=at position .56 with {\arrow[scale=1.5, rotate=10]{stealth}}}, postaction=decorate] (0,0) arc (250:110:.5) node[midway, left] {\small $a$};
\draw [thick, decoration = {markings, mark=at position .56 with {\arrow[scale=1.5, rotate=-10]{stealth}}}, postaction=decorate] (0,0) arc (-70:70:.5) node[midway, right] {\small $b$};
\draw [thick, decoration = {markings, mark=at position .8 with {\arrow[scale=1.5]{stealth}}}, postaction=decorate] (0,.93) node[right] {\footnotesize $\mu$} to +(0,.5) node[left] {\small $c$};
\end{tikzpicture}}
= \delta_{cc'} \, \delta_{\mu\mu'} \, \sqrt{ \frac{d_ad_b}{d_c}} \quad \raisebox{-2em}{\begin{tikzpicture}
\draw [thick, decoration = {markings, mark=at position .56 with {\arrow[scale=1.5]{stealth}}}, postaction=decorate] (0,-.5) to node[midway, right] {\small $c$} (0,1.43);
\end{tikzpicture}}
\qquad\qquad\qquad
\raisebox{-2em}{\begin{tikzpicture}
\draw [thick, decoration = {markings, mark=at position .56 with {\arrow[scale=1.5]{stealth}}}, postaction=decorate] (0,-.5) to (0,1.2) node[above] {\small $a$};
\draw [thick, decoration = {markings, mark=at position .56 with {\arrow[scale=1.5]{stealth}}}, postaction=decorate] (.8,-.5) to (.8,1.2) node[above] {\small $b$};
\end{tikzpicture}}
= \sum_{c, \mu}  \sqrt{ \frac{d_c}{d_a d_b}} \quad \raisebox{-3.2em}{\begin{tikzpicture}
\draw [thick, decoration = {markings, mark=at position .7 with {\arrow[scale=1.5]{stealth}}}, postaction=decorate] (-.5, -.5) node[below] {\small $a$} to (0,0) node[right] {\footnotesize $\mu$};
\draw [thick, decoration = {markings, mark=at position .7 with {\arrow[scale=1.5]{stealth}}}, postaction=decorate] (.5, -.5) node[below] {\small $b$} to (0,0);
\draw [thick, decoration = {markings, mark=at position .7 with {\arrow[scale=1.5]{stealth}}}, postaction=decorate] (0,0) to node[midway, left] {\small $c$} (0,.7) node[right] {\footnotesize $\mu$};
\draw [thick, decoration = {markings, mark=at position .7 with {\arrow[scale=1.5]{stealth}}}, postaction=decorate] (0, .7) to (-.5,1.2) node[above] {\small $a$};
\draw [thick, decoration = {markings, mark=at position .7 with {\arrow[scale=1.5]{stealth}}}, postaction=decorate] (0, .7) to (.5,1.2) node[above] {\small $b$};
\end{tikzpicture}} \;.
\ee
On the left, the indices $\mu, \mu'$ run over $N_{ab}^c$ values and so the expression is non-vanishing only if $N_{ab}^c \neq 0$. Taking $c=c'=0$ and $b= \bar a$, one finds a diagrammatic expression for the quantum dimension of $a$:
\be \label{eqqdim}
d_a = d_{\bar a} = \; \raisebox{-1em}{\begin{tikzpicture}
\draw [thick, decoration = {markings, mark=at position .47 with {\arrowreversed[scale=1.5,rotate=10]{stealth}}}, postaction=decorate] (0,0) circle [radius=.5];
\node at (-.8,0) {\small $a$};
\end{tikzpicture}} \;.
\ee
A key identity is $d_a d_b = \sum_c N_{ab}^c d_c$, showing that the quantum dimensions are both eigenvalues (in fact, the largest eigenvalues) and eigenvectors of the matrices $N_{ab}^c$. A line---or anyon---is said to be Abelian if and only if $d_a=1$. We let $\cA$ be the subcategory of Abelian anyons. They have the important property that fusion of a line with an Abelian anyon produces a single line on the right-hand-side of (\ref{fusion algebra}), with multiplicity $1$. In particular $\cA$ is an Abelian group. 

The fusion category is completed by the $F$-matrices that relate the configurations of lines on the two sides of Figure~\ref{fig: obstructed F-move} and satisfy pentagon equations.  We will not use these identities below and refer to \cite{Moore:1988uz, Moore:1988qv} for details.

The line operators also satisfy a variety of braiding relations. The braiding matrices $[R^{ab}_c]_{\mu\nu}$ are defined by
\be
\raisebox{-2.2em}{\begin{tikzpicture}
\draw [thick, decoration = {markings, mark=at position .7 with {\arrow[scale=1.5]{stealth}}}, postaction=decorate] (0,-.5) node[below] {\small $c$} to (0,0) node[right] {\footnotesize $\mu$};
\draw [thick] (0,0) arc [radius=.3, start angle=240, end angle=120];
\draw [thick, decoration = {markings, mark=at position .75 with {\arrow[scale=1.5]{stealth}}}, postaction=decorate] (0,0.52) to +(30:.5) node[right] {\small $b$};
\draw [thick] (0,0) arc [radius=.3, start angle=-60, end angle=42];
\draw [thick, decoration = {markings, mark=at position .8 with {\arrow[scale=1.5]{stealth}}}, postaction=decorate] (0,0.52) ++(150:.1) to +(150:.4) node[left] {\small $a$};
\end{tikzpicture}}
\!\!\! = \sum_\nu [R^{ab}_c]_{\mu\nu} \!\!\!\!\!
\raisebox{-2.2em}{\begin{tikzpicture}
\draw [thick, decoration = {markings, mark=at position .5 with {\arrow[scale=1.5]{stealth}}}, postaction=decorate] (0,-.8) node[below] {\small $c$} to (0,0);
\draw [thick, decoration = {markings, mark=at position .7 with {\arrow[scale=1.5]{stealth}}}, postaction=decorate] (0,0) node[below right] {\footnotesize $\nu$} to (-.6,.5) node[left] {\small $a$};
\draw [thick, decoration = {markings, mark=at position .7 with {\arrow[scale=1.5]{stealth}}}, postaction=decorate] (0,0)--(.6,.5) node[right] {\small $b$};
\end{tikzpicture}}
\quad\;\Rightarrow\quad\;
\raisebox{-1.8em}{\begin{tikzpicture}
\draw [thick, decoration = {markings, mark=at position .9 with {\arrow[scale=1.5]{stealth}}}, postaction=decorate] (-.6,-.6) node[left] {\small $b$} to (.6,.6) node[right] {\small $b$};
\draw [thick] (.6,-.6) node[right] {\small $a$} to (.1,-.1);
\draw [thick, decoration = {markings, mark=at position .75 with {\arrow[scale=1.5]{stealth}}}, postaction=decorate] (-.1,.1) to (-.6,.6) node[left] {\small $a$};
\end{tikzpicture}}
= \sum_{c,\mu,\nu} \sqrt{ \frac{d_c}{d_a d_b}} \; [R^{ab}_c]_{\mu\nu} \!\!\!
\raisebox{-2.4em}{\begin{tikzpicture}
\draw [thick, decoration = {markings, mark=at position .7 with {\arrow[scale=1.5]{stealth}}}, postaction=decorate] (-.5, -.5) node[left] {\small $b$} to (0,0) node[right] {\footnotesize $\mu$};
\draw [thick, decoration = {markings, mark=at position .7 with {\arrow[scale=1.5]{stealth}}}, postaction=decorate] (.5, -.5) node[right] {\small $a$} to (0,0);
\draw [thick, decoration = {markings, mark=at position .7 with {\arrow[scale=1.5]{stealth}}}, postaction=decorate] (0,0) to node[midway, left] {\small $c$} (0,.7) node[right] {\footnotesize $\nu$};
\draw [thick, decoration = {markings, mark=at position .7 with {\arrow[scale=1.5]{stealth}}}, postaction=decorate] (0, .7) to (-.5,1.2) node[left] {\small $a$};
\draw [thick, decoration = {markings, mark=at position .7 with {\arrow[scale=1.5]{stealth}}}, postaction=decorate] (0, .7) to (.5,1.2) node[right] {\small $b$};
\end{tikzpicture}},
\ee
where above we used \eqref{orthogonality & completeness}.
The $R$-matrices satisfy hexagon relations \cite{Moore:1988uz, Moore:1988qv} that correspond to Yang-Baxter equations, and moreover there are hexagon relations between $F$- and $R$-matrices. 

Imposing that the $S$-matrix (defined below) is unitary, one finally defines a unitary modular tensor category (UMTC).

The topological spin of an anyon $a$ was defined in (\ref{representation of spin}). It can be expressed in terms of the $R$-matrices as
\be
\label{spin in TQFT}
\theta_a = \theta_{\bar a} = \sum_{c,\mu} \frac{d_c}{d_a} \, [R^{aa}_c]_{\mu\mu} = \frac1{d_a} \;\; \raisebox{-1.5em}{\begin{tikzpicture}
\draw [thick] (0,0) ++(45:.5) arc (45:315:.5);
\draw [thick] (0,0) ++(-45:.5) to +(45:1);
\draw [thick, decoration = {markings, mark=at position .46 with {\arrowreversed[scale=1.5,rotate=10]{stealth}}}, postaction=decorate] (0,0) ++(-45:1) ++(45:.5) arc (-135:135:.5);
\node at (2.20,0) {\small $a$};
\draw [thick] (0,0) ++(45:.5) to +(-45:.3); \draw [thick] (0,0) ++(45:.5) ++ (-45:.7) to +(-45:.3);
\end{tikzpicture}}
\ee
and it is a root of unity \cite{Vafa:1988ag, Kitaev:2005}. One can show that the $R$-matrices satisfy the ribbon property $\sum_\lambda [R^{ab}_c]_{\mu\lambda} \, [R^{ba}_c]_{\lambda\nu} = \frac{\theta_c}{\theta_a\, \theta_b} \, \delta_{\mu\nu}$. The $S$-matrix is defined as
\be
S_{ab} = \frac1{\cD} \sum_c N^c_{\bar ab} \frac{\theta_c}{\theta_a \theta_b} d_c = \frac1{\cD} \; \raisebox{-1.1em}{\begin{tikzpicture}
\draw [thick, decoration = {markings, mark=at position .33 with {\arrowreversed[scale=1.5,rotate=10]{stealth}}}, postaction=decorate] (0,0) ++(60:.5) arc (60:390:.5);
\node at (-.8,0) {\small $a$};
\draw [thick, decoration = {markings, mark=at position .88 with {\arrowreversed[scale=1.5,rotate=10]{stealth}}}, postaction=decorate] (.7,0) ++(-120:.5) arc (-120:210:.5);
\node at (-.05,0) {\small $b$};
\end{tikzpicture}}
\ee
(the two expressions agree because of the ribbon property) where $\cD = S_{00}^{-1} = \sqrt{ \sum_a d_a^2}$ is the total quantum dimension. The $S$-matrix is unitary and satisfies $S_{ab} = S_{ba} = S^*_{\bar a b}$ and $S_{0a} = d_a/\cD$. It can be used to remove anyons that loop around other anyons:
\be
\label{winding anyon}
\raisebox{-2em}{\begin{tikzpicture}
\draw [thick, decoration = {markings, mark=at position .3 with {\arrowreversed[scale=1.5,rotate=15]{stealth}}}, postaction=decorate] (0,0) ++(100:.5 and .3) arc (100:440:.5 and .3);
\node at (-.7,0) {$a$};
\draw [thick] (0,-.8) to (0,-.4); \draw [thick, ->-] (0,-.2) to (0,.6) node[above] {$b$};
\end{tikzpicture}}
\; = \frac{S_{ab}}{S_{0b}} \quad
\raisebox{-2em}{\begin{tikzpicture}
\draw [thick, decoration = {markings, mark=at position .7 with {\arrow[scale=1.5]{stealth}}}, postaction=decorate] (0,-.8) to (0,.6) node[above] {$b$};
\end{tikzpicture}} \;.
\ee
One can also define the so-called monodromy scalar component
\be
\label{monodromy scalar component}
M_{ab} = \frac{S^*_{ab} \, S_{00}}{S_{0a} \, S_{0b}} = \frac{S_{\bar ab} \, S_{00}}{S_{0a} \, S_{0b}} = \frac1{d_a d_b} \sum_c N^c_{ab} \frac{\theta_c}{\theta_a \theta_b} d_c = \frac1{d_ad_b} \; \raisebox{-1.1em}{\begin{tikzpicture}
\draw [thick, decoration = {markings, mark=at position .94 with {\arrow[scale=1.5,rotate=-10]{stealth}}}, postaction=decorate] (0,0) ++(60:.5) arc (60:390:.5);
\node at (.75,0) {\small $a$};
\draw [thick, decoration = {markings, mark=at position .88 with {\arrowreversed[scale=1.5,rotate=10]{stealth}}}, postaction=decorate] (.7,0) ++(-120:.5) arc (-120:210:.5);
\node at (-.05,0) {\small $b$};
\end{tikzpicture}} \;,
\ee
introduced in (\ref{crossing of lines in 3D}) (for Abelian anyons). If $M_{ab}$ is a phase, $|M_{ab}|=1$, then the braiding of $a$ and $b$ is Abelian. Moreover, when this is true, it follows that $M_{ab} M_{ac} = M_{ad}$ whenever $N_{bc}^d \neq 0$. The converse is also true.  From unitarity of the $S$-matrix it follows that if there exists a set of phase factors $e^{i\phi_a}$, one for each anyon type $a$, satisfying $e^{i\phi_a} \, e^{i\phi_b} = e^{i\phi_c}$ whenever $N_{ab}^c \neq 0$, then
\be
\label{eqn:phasesobeyfusion}
e^{i\phi_a} = M^*_{ae} = M^{-1}_{ae} = \frac{S_{ae}}{S_{0a}}
\ee
for some Abelian anyon $e$.

\subsection{TQFTs with a Global Symmetry}
\label{sec: symmetry}

We now describe some aspects of 3d TQFTs with global symmetry following \cite{Barkeshli:2014cna} (see also \cite{Kitaev:2005, Bernevig:2015fqa}).  Our first task is to identify the intrinsic global symmetry of such a theory as discussed in Section \ref{sec: general coupling}.

We begin with the 0-form symmetry.  First, we define a set of topological symmetries---or auto-equivalences---of a unitary modular tensor category as the set of invertible maps%
\footnote{We restrict here to unitary and parity-preserving maps.}
$\varphi: \cC \to \cC$ that preserve all topological properties, \eg{} $N_{ab}^c$, $d_a$, $\theta_a$, $S_{ab}$ \cite{Barkeshli:2014cna}. They can involve a permutation of the anyons, $\varphi(a) = a'$, and act unitarily on the vector spaces,
\be
\varphi\big( |a,b;c,\mu\rangle \big) = \sum_{\mu'} \big[ u^{a' b'}_{c'} \big]_{\mu\mu'} |a', b'; c',\mu'\rangle \;,
\ee
while preserving the $F$- and $R$-matrices.

Among such transformations, there is a subset, called natural isomorphisms defined as follows:
\be
\Upsilon(a) = a \;,\qquad\qquad \Upsilon \big( |a,b;c,\mu\rangle \big) = \frac{\gamma_a\gamma_b}{\gamma_c} \, |a,b;c,\mu\rangle
\ee
for phases $\gamma_a$. These transformations automatically leave all TQFT data invariant and we do not regard them as symmetries.  Notice in particular that such natural  isomorphisms do not permute the anyons.  More generally, it is conjectured in \cite{Barkeshli:2014cna} that topological symmetries that do not permute anyons are necessarily natural isomorphisms.%
\footnote{This statement is false in non-topological TQFTs.  See Section~\ref{sec: examples} for examples.}
(See Appendix \ref{app:trivialpermutation} for a proof in a special case.)

We thus define the automorphism group of the category, $\Aut(\cC)$, as the group of topological symmetries modulo natural isomorphisms.  In particular, group multiplication in $\Aut(\cC)$ is composition up to natural isomorphisms. If we assume the conjecture above that all non-identity elements of $\Aut(\cC)$ must permute the anyons, we conclude that $\Aut(\cC)$ is a subgroup of the permutation group of the anyons (in particular, it is finite).  We identify this automorphism group with the 0-form symmetry of the theory:
\begin{equation}
\text{Intrinsic 0-form symmetry}=\Aut(\cC) \;.
\end{equation}

Next let us identify the 1-form symmetry of the TQFT.  These are the Abelian anyons in the theory (see the discussion around \eqref{eqqdim}):
\begin{equation}
\text{Intrinsic 1-form symmetry}=\cA=\text{Abelian anyons} \;.
\end{equation}
Since the group of automorphisms $\Aut(\cC)$ acts to permute all the lines, by restriction it also acts to permute only the Abelian anyons $\cA$. This defines the action of the 0-form symmetry on the 1-form symmetry that is part of the defining data of a 2-group.  Finally, we must explain how to extract the Postnikov class $\beta \in H^3\big(B{\Aut(\cC)}, \cA \big)$ from the intrinsic data of the TQFT.  This is done in Section \ref{betaTQFTf} below.

Although the intrinsic symmetry of the TQFT is the 2-group defined above,  it is common in the literature (\eg{} \cite{Barkeshli:2014cna}) to also discuss coupling a TQFT to more general symmetries.  This can be carried out using the formalism of Section \ref{sec: general coupling}.  In particular, we will consider coupling the theory to an extrinsic 2-group with 0-form group $G$ and 1-form group $\cA$. 

To begin the discussion, we consider in general a group homomorphism
\be
[\rho]: G \to \Aut(\cC)
\ee
meaning that $[\rho_{\bf g}]\cdot[\rho_{\bf h}] = [\rho_{\bf gh}]$. The special case that $G = \Aut(\cC)$ (so that we are considering the intrinsic 0-form symmetry of the TQFT) corresponds to $[\rho] = \id$. We represent the classes $[\rho_\bfg]$ by elements of the topological symmetry (and from now on we suppress the index $\mu$):
\be
\label{rho action}
\rho_{\bf g}(a) = a' \equiv {}^{\bf g}a \;,\qquad \rho_{\bf g} |a,b;c\rangle = U_{\bf g}({}^{\bf g}a, {}^{\bf g}b; {}^{\bf g}c) \, | {}^{\bf g}a, {}^{\bf g}b; {}^{\bf g}c\rangle \;,\qquad {\bf g}^{-1} \equiv {\bf \bar g} \;,
\ee
where $U_\bfg$ are unitary matrices. They only have to represent $G$ up to natural isomorphisms:
\be
\label{def kappa}
\kappa_{{\bf g},{\bf h}} \, \rho_{\bf g} \, \rho_{\bf h} = \rho_{\bf gh} \;,
\ee
where $\kappa_{{\bf g},{\bf h}}(a,b;c)$ are natural isomorphisms:
\be
\label{first def eta}
\kappa_{{\bf g}, {\bf h}}(a,b;c)_{\mu\nu} = \frac{\eta_a({\bf g},{\bf h}) \, \eta_b({\bf g},{\bf h})}{\eta_c({\bf g},{\bf h})} \, \delta_{\mu\nu}
\ee
for phases $\eta_a(\bfg,\bfh)$. One easily finds $\kappa_{{\bf g}, {\bf h}}(a,b;c) = U_{\bf g}(a,b;c)^{-1} \; U_{\bf h}({}^{\bf \bar g}a, {}^{\bf \bar g}b; {}^{\bf \bar g}c)^{-1} \; U_{\bf gh}(a,b;c)$. Clearly (\ref{first def eta}) does not uniquely fix the phases $\eta_a(\bfg,\bfh)$, and we will discuss below how they can be unambiguously extracted from the TQFT data. On the other hand, $G$ permutes the anyons and this induces an action of $G$ on the 1-form symmetry group $\cA$, that we indicate with the same symbol $\rho: G \to \Aut(\cA)$.

Decomposing $\rho_{\bf ghk}$ with (\ref{def kappa}) and using associativity, one obtains
\be
\label{cocycle condition kappa}
\kappa_{{\bf g},{\bf hk}} \; \rho_{\bf g} \; \kappa_{{\bf h},{\bf k}} \; \rho_{\bf g}^{-1} = \kappa_{{\bf gh},{\bf k}} \; \kappa_{{\bf g},{\bf h}} \;.
\ee
From the phases $\eta_a({\bf g},{\bf h})$ one can define the phases
\be
\label{def Omega}
\Omega_a({\bf g}, {\bf h}, {\bf k}) = \frac{ \eta_{\rho_{\bf g}^{-1}(a)}({\bf h}, {\bf k}) \, \eta_a({\bf g}, {\bf hk}) }{ \eta_a({\bf gh}, {\bf k})\, \eta_a({\bf g}, {\bf h}) }
\ee
where $\rho_\bfg^{-1}(a) = {}^{\bf\bar g}a$. They satisfy the relations
\be
\frac{\Omega_{\rho_{\bf g}^{-1}(a)}({\bf h}, {\bf k}, {\bf l})\, \Omega_a({\bf g}, {\bf hk}, {\bf l})\, \Omega_a({\bf g}, {\bf h}, {\bf k}) }{ \Omega_a({\bf gh}, {\bf k}, {\bf l})\,  \Omega_a({\bf g}, {\bf h}, {\bf kl}) } = 1 \;.
\ee
From (\ref{first def eta}), (\ref{cocycle condition kappa}) and (\ref{def Omega}) it follows that $\Omega_a({\bf g}, {\bf h}, {\bf k})\, \Omega_b({\bf g}, {\bf h}, {\bf k}) = \Omega_c({\bf g}, {\bf h}, {\bf k})$ whenever $N_{ab}^c \neq 0$. This means that we can write
\be
\label{def beta in TQFT}
\Omega_a({\bf g}, {\bf h}, {\bf k}) = M^*_{a\,\beta({\bf g},{\bf h},{\bf k})}
\ee
for some $\beta({\bf g},{\bf h},{\bf k}) \in \cA$. One can prove that $\beta$ is a cocycle in $Z^3_\rho(BG,\cA)$, and that different solutions to (\ref{first def eta}) for $\eta_a$ lead to $\beta$'s that differ by an exact cocycle. We conclude that the action of the 0-form symmetry $G$ defines a class $[\beta] \in H^3_\rho(BG,\cA)$. In the literature this is called an ``obstruction to symmetry fractionalization'' or ``$H^3$ anomaly''. As we explain in Section \ref{betaTQFTf} this is in fact the Postnikov class of the 2-group global symmetry. Notice that if $G$ does not permute the anyons, then according to a conjecture of  \cite{Barkeshli:2014cna} (see also Appendix~\ref{app:trivialpermutation}) its action on the TQFT can be completely trivialized by a natural isomorphism and thus $[\beta] = 0$.

\paragraph{\matht{G}-crossed braided tensor category.} The $0$-form symmetry $G$ is generated by (two-dimensional) surface operators $\Sigma_\bfg$, one for each element $\bfg \in G$. These are observables of the theory, and we should extend the set of topological correlation functions by including them as well. More generally, we should also include open surfaces ending on defect lines $a_\bfg$. Such defect lines could be considered as new line operators \cite{Barkeshli:2014cna}, even though they are not genuine line operators \cite{Gaiotto:2014kfa}. There must be a background for $G$ with holonomy $\bfg$ around the defect lines that bound $\Sigma_\bfg$---a sort of branch-cut discontinuity along $\Sigma_\bfg$---and correlation functions depend on the topology of the surfaces bounded by the defect lines.

One can then define a $G$-graded tensor category
\be
\cC^\times_G = \bigoplus_{{\bf g}\,\in\, G} \cC_{\bf g} \;,
\ee
called a G-crossed braided tensor category. Each sector $\cC_\bfg$ contains those lines $a_\bfg$ that can bound $\Sigma_\bfg$, and the original tensor category is $\cC \equiv \cC_\unit$. One could generate all the different lines $a_{\bf g} \in \cC_{\bf g}$ by starting with one element in $\cC_{\bf g}$ and then fusing with the elements $a\in \cC$, in other words the action of $\cC$ on $\cC_{\bf g}$ by fusion is transitive.%
\footnote{Such action is not faithful in general, though, in fact the number of lines in $\cC_{\bf g}$ is equal to the number of $\rho_{\bf g}$-invariant lines in $\cC$ \cite{Barkeshli:2014cna}.}
Fusion is compatible with grading, $a_{\bf g} \times b_{\bf h} = \sum_c N^{c_{\bfg\bfh}}_{a_\bfg b_\bfh} \, c_{\bf gh}$. Note that, because of the presence of the surfaces, fusion can be non-commutative.

The surfaces are oriented, and when a line $a_{\bf h}$ crosses $\Sigma_\bfg$, it is mapped to
\be
\rho_{\bf g}(a_{\bf h}) = {}^{\bf g} a_{\bf h} \qquad\text{ of type }\qquad {}^{\bf g}{\bf h} = {\bf g h g}^{-1} \;.
\ee
This generalizes Figure~\ref{fig: action on higher symmetry} to all genuine and non-genuine lines. Besides, one needs to extend $[\rho]: G \to \Aut(\cC)$ to $[\rho]: G \to \Aut(\cC_G^\times)$ such that $[\rho_\bfg]:\cC_{\bf h} \to \cC_{{\bf g h g}^{-1}}$. One also needs to extend fusion and braiding to the full $\cC^\times_G$, including $F$- and $R$-matrices.

In order to implement the presence of surface operators in the diagrammatics, we implicitly place a surface $\Sigma_\bfg$ perpendicular to the page below its respective line $a_\bfg$. Additionally, by a dashed line we represent a 1d section of a surface operator, which may or may not have a boundary defect line. Compatibility between fusion and braiding is expressed by \cite{Barkeshli:2014cna}
\be
\label{diagrammatic matrices U}
\raisebox{-4.5em}{\begin{tikzpicture}
\draw [thick, decoration = {markings, mark=at position .5 with {\arrow[scale=1.5]{stealth}}}, postaction=decorate] (0,-1) node[below] {${}^{\bf \bar k}c_{\bf gh}$} to (0,0) node[below right] {\small $\mu$};
\draw [thick] (0,0)--(-.5,.5);
\draw [thick] (0,0)--(.5,.5) node[right] {${}^{\bf \bar k}b_{\bf h}$};
\draw [thick, decoration = {markings, mark=at position .9 with {\arrow[scale=1.5]{stealth}}}, postaction=decorate] (-.5,.5)--(-.5,1.5) node[above] {$a_{\bf g}$};
\draw [thick, decoration = {markings, mark=at position .9 with {\arrow[scale=1.5]{stealth}}}, postaction=decorate] (.5,.5)--(.5,1.5) node[above] {$b_{\bf h}$};
\fill [white] (-.21,.21) circle [radius = .08];
\fill [white] (.5,.85) circle [radius = .1];
\draw [thick, decoration = {markings, mark=at position .2 with {\arrow[scale=1.5]{stealth}}}, postaction=decorate] (-1,-.5) node[below left] {$x_{\bf k}$} to (1,1.3);
\end{tikzpicture}}
\quad = \sum_\nu U_{\bf k}(a,b;c)_{\mu\nu}
\raisebox{-4.5em}{\begin{tikzpicture}
\draw [thick, decoration = {markings, mark=at position .3 with {\arrow[scale=1.5]{stealth}}}, postaction=decorate] (0,-1) node[below] {${}^{\bf \bar k}c_{\bf gh}$} to (0,1) node[below left] {\small $\nu$};
\draw [thick, decoration = {markings, mark=at position .8 with {\arrow[scale=1.5]{stealth}}}, postaction=decorate] (0,1)--(-.5,1.5) node[above] {$a_{\bf g}$};
\draw [thick, decoration = {markings, mark=at position .8 with {\arrow[scale=1.5]{stealth}}}, postaction=decorate] (0,1)--(.5,1.5) node[above] {$b_{\bf h}$};
\fill [white] (0,.2) circle [radius = .1];
\draw [thick, decoration = {markings, mark=at position .2 with {\arrow[scale=1.5]{stealth}}}, postaction=decorate] (-1,-.7) node[below left] {$x_{\bf k}$} to (1,1.1);
\end{tikzpicture}}
\ee
and
\be
\raisebox{-4.5em}{\begin{tikzpicture}
\draw [thick, decoration = {markings, mark=at position .2 with {\arrow[scale=1.5]{stealth}}}, postaction=decorate] (-1,-.5) node[below left] {$x_{\bf k}$} to (1,1.3) node[above right] {${}^{\bf \bar h \bar g}x_{\bf k}$};
\fill [white] (-.21,.21) circle [radius = .08];
\fill [white] (.5,.85) circle [radius = .1];
\draw [thick, decoration = {markings, mark=at position .5 with {\arrow[scale=1.5]{stealth}}}, postaction=decorate] (0,-1) node[below] {$c_{\bf gh}$} to (0,0) node[below right] {\small $\mu$};
\draw [thick] (0,0)--(-.5,.5);
\draw [thick] (0,0)--(.5,.5);
\draw [thick, decoration = {markings, mark=at position .9 with {\arrow[scale=1.5]{stealth}}}, postaction=decorate] (-.5,.5)--(-.5,1.5) node[above] {$a_{\bf g}$};
\draw [thick, decoration = {markings, mark=at position .9 with {\arrow[scale=1.5]{stealth}}}, postaction=decorate] (.5,.5)--(.5,1.5) node[above] {$b_{\bf h}$};
\end{tikzpicture}}
= \eta_x({\bf g},{\bf h})
\raisebox{-4.5em}{\begin{tikzpicture}
\draw [thick, decoration = {markings, mark=at position .2 with {\arrow[scale=1.5]{stealth}}}, postaction=decorate] (-1,-.7) node[below left] {$x_{\bf k}$} to (1,1.1) node[above right] {${}^{\bf \bar h \bar g}x_{\bf k}$};
\fill [white] (0,.2) circle [radius = .12];
\draw [thick, decoration = {markings, mark=at position .3 with {\arrow[scale=1.5]{stealth}}}, postaction=decorate] (0,-1) node[below] {$c_{\bf gh}$} to (0,1) node[below left] {\small $\mu$};
\draw [thick, decoration = {markings, mark=at position .8 with {\arrow[scale=1.5]{stealth}}}, postaction=decorate] (0,1)--(-.5,1.5) node[above] {$a_{\bf g}$};
\draw [thick, decoration = {markings, mark=at position .8 with {\arrow[scale=1.5]{stealth}}}, postaction=decorate] (0,1)--(.5,1.5) node[above] {$b_{\bf h}$};
\end{tikzpicture}}
\;.
\ee
If we move the lines that are displaced above the page away leaving only the attached surface operators, we find simplified diagrams. The first one implies
\be
\raisebox{-4.5em}{\begin{tikzpicture}
\draw [thick, decoration = {markings, mark=at position .3 with {\arrow[scale=1.5]{stealth}}}, postaction=decorate] (0,-1.5) node[below] {$c$} to (0,0);
\draw [thick, decoration = {markings, mark=at position .9 with {\arrow[scale=1.5]{stealth}}}, postaction=decorate] (0,0) to (-1,1.5) node[above] {$a$};
\draw [thick, decoration = {markings, mark=at position .9 with {\arrow[scale=1.5]{stealth}}}, postaction=decorate] (0,0) to (1,1.5) node[above] {$b$};
\node at (-.55,.4) {${}^{\bf \bar k}a$}; \node at (.67,.4) {${}^{\bf \bar k}b$}; \node at (-.3,-.5) {${}^{\bf \bar k}c$}; \node at (.2,-.1) {\small $\mu$};
\draw [thick, densely dashed, decoration = {markings, mark=at position .49 with {\arrowreversed[scale=1.5,rotate=5]{stealth}}}, postaction=decorate] (0,.1) circle [radius = 1];
\node at (-1.3,0) {$\bf k$};
\end{tikzpicture}}
\quad = \sum_\nu U_{\bf k}(a,b;c)_{\mu\nu}
\raisebox{-4.5em}{\begin{tikzpicture}
\draw [thick, decoration = {markings, mark=at position .5 with {\arrow[scale=1.5]{stealth}}}, postaction=decorate] (0,-1.5) node[below] {$c$} to (0,0);
\draw [thick, decoration = {markings, mark=at position .7 with {\arrow[scale=1.5]{stealth}}}, postaction=decorate] (0,0) to (-1,1.5) node[above] {$a$};
\draw [thick, decoration = {markings, mark=at position .7 with {\arrow[scale=1.5]{stealth}}}, postaction=decorate] (0,0) to (1,1.5) node[above] {$b$};
\node at (.2,-.1) {\small $\nu$};
\end{tikzpicture}}
\;.
\ee
This is precisely the action of $\rho_{\bf k}$ on the basis vectors $|a,b;c,\mu\rangle$, as in (\ref{rho action}). The second one implies
\be
\label{def eta}
\raisebox{-4.5em}{\begin{tikzpicture}
\draw [thick, densely dashed, decoration = {markings, mark=at position .3 with {\arrow[scale=1.5]{stealth}}}, postaction=decorate] (0,-1.5) node[below] {$\bf gh$} to (0,0);
\draw [thick, densely dashed, decoration = {markings, mark=at position .9 with {\arrow[scale=1.5]{stealth}}}, postaction=decorate] (0,0) to (-1,1.5) node[above] {$\bf g$};
\draw [thick, densely dashed, decoration = {markings, mark=at position .9 with {\arrow[scale=1.5]{stealth}}}, postaction=decorate] (0,0) to (1,1.5) node[above] {$\bf h$};
\draw [thick, decoration = {markings, mark=at position .49 with {\arrowreversed[scale=1.5,rotate=5]{stealth}}}, postaction=decorate] (0,.1) circle [radius = .7];
\node at (-1,0) {$x$}; \node at (0,1.2) {${}^{\bf \bar g}x$}; \node at (1,-.8) {${}^{\bf \bar h \bar g}x$};
\end{tikzpicture}}
= \; \eta_x(\bfg, \bfh) \;
\raisebox{-4.5em}{\begin{tikzpicture}
\draw [thick, densely dashed, decoration = {markings, mark=at position .5 with {\arrow[scale=1.5]{stealth}}}, postaction=decorate] (0,-1.5) node[below] {$\bf gh$} to (0,0);
\draw [thick, densely dashed, decoration = {markings, mark=at position .7 with {\arrow[scale=1.5]{stealth}}}, postaction=decorate] (0,0) to (-1,1.5) node[above] {$\bf g$};
\draw [thick, densely dashed, decoration = {markings, mark=at position .7 with {\arrow[scale=1.5]{stealth}}}, postaction=decorate] (0,0) to (1,1.5) node[above] {$\bf h$};
\draw [thick, decoration = {markings, mark=at position .48 with {\arrowreversed[scale=1.5,rotate=8]{stealth}}}, postaction=decorate] (-.9,-.5) circle [radius = .5];
\node at (-1.7,-.55) {$x$};
\end{tikzpicture}}
\quad = \; \eta_x({\bf g}, {\bf h})\; d_x 
\raisebox{-4.5em}{\begin{tikzpicture}
\draw [thick, densely dashed, decoration = {markings, mark=at position .5 with {\arrow[scale=1.5]{stealth}}}, postaction=decorate] (0,-1.5) node[below] {$\bf gh$} to (0,0);
\draw [thick, densely dashed, decoration = {markings, mark=at position .7 with {\arrow[scale=1.5]{stealth}}}, postaction=decorate] (0,0) to (-1,1.5) node[above] {$\bf g$};
\draw [thick, densely dashed, decoration = {markings, mark=at position .7 with {\arrow[scale=1.5]{stealth}}}, postaction=decorate] (0,0) to (1,1.5) node[above] {$\bf h$};
\end{tikzpicture}}
\;,
\ee
where $\eta_a({\bf g},{\bf h})$ are phases, one for each anyon $a \in \cC$. Hence, the phases $\eta_a(\bfg,\bfh)$ generalize the phases $e^{2\pi i \lambda_a(\bfg,\bfh)}$ that we defined in (\ref{def fractionalization phases lambda_a}) to all line operators in the TQFT.

By considering two trivalent junctions that slide one on top of the other in two different ways and requiring that the two operations give the same result, one obtains the relation (\ref{first def eta}) between $\kappa_{\bfg,\bfh}(a,b;c)_{\mu\nu}$ and $\eta_a(\bfg,\bfh)$ (see Figure 8 of \cite{Barkeshli:2014cna} and the related discussion). Therefore from the correlators (\ref{def eta}) we can unambiguously extract the phases $\eta_a(\bfg,\bfh)$.

\subsection{Diagrammatics of the Postnikov Class}
\label{betaTQFTf}

To understand the relation between the ``obstruction to fractionalization'' and the Postnikov class of 2-group symmetry, both of which we have indicated as $[\beta] \in H^3_\rho(BG,\cA)$, take the two configurations on the left and right of Figure~\ref{fig: obstructed F-move} (recalling that solid lines in that figure are codimension-1 symmetry defects, which we represent in this section by dashed lines) and wrap an anyon $a$ around both. Performing a contour deformation and implementing (\ref{def eta}) we find
\be
\label{H3 diagram I}
\raisebox{-4.5em}{\begin{tikzpicture}
\draw [thick, densely dashed, decoration = {markings, mark=at position .14 with {\arrow[scale=1.5]{stealth}}, mark=at position .55 with {\arrow[scale=1.5]{stealth}}, mark=at position .93 with {\arrow[scale=1.5]{stealth}}}, postaction=decorate] (.1,-.2) node[below] {$\bf ghk$} to (-1.3,2.6) node[above] {$\bf g$};
\draw [thick, densely dashed, decoration = {markings, mark=at position .85 with {\arrow[scale=1.5]{stealth}}}, postaction=decorate] (-.4,.8) to (.5,2.6) node[above] {$\bf k$};
\draw [thick, densely dashed, decoration = {markings, mark=at position .9 with {\arrow[scale=1.5]{stealth}}}, postaction=decorate] (-.8,1.6) to (-.3,2.6) node[above] {$\bf h$};
\draw [thick, decoration = {markings, mark=at position .49 with {\arrowreversed[scale=1.5,rotate=5]{stealth}}}, postaction=decorate] (-.6,1.2) circle [radius = .9];
\node at (-1.8,1.15) {$a$};
\end{tikzpicture}}
= \eta_a({\bf g},{\bf h}) \, \eta_a({\bf gh}, {\bf k}) \;
\raisebox{-4.5em}{\begin{tikzpicture}
\draw [thick, densely dashed, decoration = {markings, mark=at position .14 with {\arrow[scale=1.5]{stealth}}, mark=at position .55 with {\arrow[scale=1.5]{stealth}}, mark=at position .93 with {\arrow[scale=1.5]{stealth}}}, postaction=decorate] (.1,-.2) node[below] {$\bf ghk$} to (-1.3,2.6) node[above] {$\bf g$};
\draw [thick, densely dashed, decoration = {markings, mark=at position .85 with {\arrow[scale=1.5]{stealth}}}, postaction=decorate] (-.4,.8) to (.5,2.6) node[above] {$\bf k$};
\draw [thick, densely dashed, decoration = {markings, mark=at position .9 with {\arrow[scale=1.5]{stealth}}}, postaction=decorate] (-.8,1.6) to (-.3,2.6) node[above] {$\bf h$};
\draw [thick, decoration = {markings, mark=at position .47 with {\arrowreversed[scale=1.5,rotate=12]{stealth}}}, postaction=decorate] (-1.1,.6) circle [radius = .4];
\node at (-1.8,.6) {$a$};
\end{tikzpicture}}
\ee
and
\be
\label{H3 diagram II}
\raisebox{-4.5em}{\begin{tikzpicture}
\draw [thick, densely dashed, decoration = {markings, mark=at position .14 with {\arrow[scale=1.5]{stealth}}, mark=at position .55 with {\arrow[scale=1.5]{stealth}}, mark=at position .93 with {\arrow[scale=1.5]{stealth}}}, postaction=decorate] (-.1,-.2) node[below] {$\bf ghk$} to (1.3,2.6) node[above] {$\bf k$};
\draw [thick, densely dashed, decoration = {markings, mark=at position .85 with {\arrow[scale=1.5]{stealth}}}, postaction=decorate] (.4,.8) to (-.5,2.6) node[above] {$\bf g$};
\draw [thick, densely dashed, decoration = {markings, mark=at position .9 with {\arrow[scale=1.5]{stealth}}}, postaction=decorate] (.8,1.6) to (.3,2.6) node[above] {$\bf h$};
\draw [thick, decoration = {markings, mark=at position .49 with {\arrowreversed[scale=1.5,rotate=10]{stealth}}}, postaction=decorate] (.0,1.2) circle [x radius = 1.5, y radius = .8];
\node at (-1.8,1.15) {$a$};
\filldraw [blue!80!black] (-.5,1.3) node[black, below] {\scriptsize $\beta(\bfg,\bfh,\bfk)$} circle [radius=.07];
\end{tikzpicture}}
\quad = \; \eta_{{}^{\bf \bar g}a}({\bf h},{\bf k}) \, \eta_a({\bf g}, {\bf hk}) \;
\raisebox{-4.5em}{\begin{tikzpicture}
\draw [thick, densely dashed, decoration = {markings, mark=at position .14 with {\arrow[scale=1.5]{stealth}}, mark=at position .55 with {\arrow[scale=1.5]{stealth}}, mark=at position .93 with {\arrow[scale=1.5]{stealth}}}, postaction=decorate] (-.1,-.2) node[below] {$\bf ghk$} to (1.3,2.6) node[above] {$\bf k$};
\draw [thick, densely dashed, decoration = {markings, mark=at position .85 with {\arrow[scale=1.5]{stealth}}}, postaction=decorate] (.4,.8) to (-.5,2.6) node[above] {$\bf g$};
\draw [thick, densely dashed, decoration = {markings, mark=at position .9 with {\arrow[scale=1.5]{stealth}}}, postaction=decorate] (.8,1.6) to (.3,2.6) node[above] {$\bf h$};
\filldraw [blue!80!black] (-.8,1.0) node[black, below] {\scriptsize $\beta(\bfg,\bfh,\bfk)$} circle [radius=.07];
\draw [thick, decoration = {markings, mark=at position .48 with {\arrowreversed[scale=1.5,rotate=17]{stealth}}}, postaction=decorate] (-.8,.8) circle [x radius = .9, y radius = .5];
\node at (-2.0,.75) {$a$};
\end{tikzpicture}}
\;.
\ee
The standard $F$-move of 3d TQFTs that transforms the configuration of lines on the left of Figure~\ref{fig: obstructed F-move} to the one on the right, does not involve the creation of any other line $\beta$. If we insist that this remains true for the non-genuine lines $a_\bfg$, and thus also for the surface symmetry defects $\Sigma_\bfg$, we obtain the relation $\eta_{{}^{\bf \bar g}a}({\bf h},{\bf k}) \, \eta_a({\bf g}, {\bf hk}) = \eta_a({\bf g},{\bf h}) \, \eta_a({\bf gh}, {\bf k})$. This is the statement that the class $[\beta] = 0$ in $H^3_\rho(BG,\cA)$. In other words, if the class $[\beta]$ defined in (\ref{def beta in TQFT}) does not vanish, we cannot consistently couple the TQFT to global 0-form symmetry $G$ and 1-form symmetry $\cA$ as \emph{independent} symmetries. This is why the class $[\beta]$ is sometimes termed an ``anomaly'' in the literature.

On the other hand, including the Abelian line $\beta(\bfg, \bfh, \bfk)$ in the $F$-move of surface symmetry defects (and accordingly in the $F$-move of non-genuine lines), we obtain the equation
\be
\frac{ \eta_{{}^{\bf \bar g}a}({\bf h},{\bf k}) \, \eta_a({\bf g},{\bf hk}) }{ \eta_a({\bf gh}, {\bf k}) \, \eta_a({\bf g},{\bf h}) } = \bigg( \frac{S_{a\beta}}{S_{0\beta}} \bigg)^{-1} = M^{-1}_{a\, \beta({\bf g},{\bf h},{\bf k})} = \Omega_a({\bf g},{\bf h},{\bf k}) \;,
\ee
which is precisely the desired relation. We conclude that it is consistent to couple such a TQFT to a 2-group $\bG$-bundle---up to the 't~Hooft anomalies discussed in Section~\ref{sec: 't Hooft anomaly}.

\subsection{Abelian TQFTs}

To illustrate some of the methods, consider the special case of an Abelian TQFT where all anyons are Abelian, \ie{} any two anyons $a,a'$ fuse into a unique anyon $aa'$ (equivalently, all anyons have quantum dimension one). Take two anyons $a,a'$ to encircle the same defect junction as in (\ref{def eta}). Using (\ref{def eta}) subsequently for the anyons $a,a'$ gives the product of phases $\eta_a(\mathbf{g},\mathbf{h}) \, \eta_{a'}(\mathbf{g},\mathbf{h})$, while using (\ref{def eta}) for the anyon $aa'$ gives the phase $\eta_{aa'}(\mathbf{g},\mathbf{h})$. This corresponds to the equation%
\footnote{This argument should not be considered as a proof that $\eta_\star$ is a homomorphism from $\cA$ to $U(1)$. Rather, the fact that $\eta_\star$ is linear in $\cA$ was taken as an assumption in (\ref{def fractionalization phases lambda_a}). (Notice that $\eta$ and $\lambda$ are exactly the same object in an Abelian TQFT.) Linearity is natural from anomaly inflow, since $\lambda$ should take values in $\wh\cA$ in order for (\ref{4D anomaly action}) to be well-defined. It would be nice to prove linearity from the axioms of TQFT. We thank Yuji Tachikawa for pointing this out.}
\begin{equation}
\eta_a(\mathbf{g},\mathbf{h}) \, \eta_{a'}(\mathbf{g},\mathbf{h}) = \eta_{aa'}(\mathbf{g},\mathbf{h}) \;.
\end{equation}
Namely, $\eta_\star(\mathbf{g},\mathbf{h})$ is a homomorphism from the intrinsic 1-form symmetry ${\cal A}$ to $U(1)$.
Since this is true for all anyons in the Abelian TQFT, the phase $\eta_a(\mathbf{g},\mathbf{h})$ can be expressed as (\ref{eqn:phasesobeyfusion}) for some Abelian anyon $e=e(\mathbf{g},\mathbf{h})$.
Substituting the expression for $\eta_a(\mathbf{g},\mathbf{h})$ into (\ref{def Omega}), using the fact that the monodromy matrix $M_{ab}$ is $G$-invariant and (in the case of an Abelian TQFT) multiplicatively linear in the two entries, and comparing with (\ref{def beta in TQFT}), we find that $\beta$ is a coboundary. We conclude that in Abelian unitary TQFTs, the 2-group symmetry has trivial Postnikov class:
\begin{equation}
[\beta]=0 \quad\text{ in }\quad H_\rho^3(BG,{\cal A}) \;.
\end{equation}

\section{More Examples}
\label{sec: examples}

In this Section we discuss many more examples of theories with 2-group global symmetry, or that can be coupled to 2-group backgrounds. We also illustrate the procedure of Section~\ref{sec: general coupling}.

In the first example we discuss how to couple the Abelian $\bZ_N$ Chern-Simons theory to 2-group backgrounds using its 1-form symmetry. The second example is $U(1)_K$ Abelian Chern-Simons-matter theory with matter fields of charge $q>1$, that can have 2-group global symmetry. In the third example we show how a certain mixed anomaly for discrete global symmetries can give rise to 2-group global symmetry after gauging. The fourth example is $Spin(N)_K$ and $O(N)_K$ non-Abelian Chern-Simons-matter theory with matter in the vector representation, that can have 2-group global symmetry. The fifth example is the Chern-Simons theory $Spin(k)_2$ which, for certain values of $k$, can have $\bZ_2^\cT$ time-reversal symmetry forming a 2-group with the $\bZ_2$ 1-form symmetry. We also present an infinite list of $U(1)_k$ and $SU(k)_1$ Chern-Simons theories with time-reversal symmetry. The last example illustrates the method of Section~\ref{sec: general coupling}, by coupling the simplest $\bZ_N$ gauge theory in general spacetime dimension to various global symmetries using its intrinsic higher-form symmetries.

\subsection[Abelian $\bZ_N$ Chern-Simons Theory]{Abelian \matht{\bZ_N} Chern-Simons Theory}
\label{sec:eggauging}

Let us first present the example, discussed in \cite{Kapustin:2014lwa, Kapustin:2014zva, Thorngren:2015gtw}, of a TQFT coupled to 2-group bundles through its 1-form symmetry.

Consider the Abelian $\mathbb{Z}_N$ Chern-Simons theory, that can be described by two $U(1)$ gauge fields $u,w$ \cite{Maldacena:2001ss, Banks:2010zn, Kapustin:2014gua}:
\be
S_\text{CS} = \frac{K}{4\pi}\int u\,du +\frac{N}{2\pi} \int u\,dw ~,
\ee
where $w$ constrains $u$ to be a $\mathbb{Z}_N$ gauge field. Denote $r\equiv \gcd(N,K)$.
The theory has 1-form symmetry $\bZ_{N^2/r} \times \bZ_r$ generated by the lines $\oint w$ and $\frac Kr \oint u + \frac Nr \oint w$, respectively. We will focus on the part $\bZ_{N^2/r} \equiv \cA$, which can be written as an extension of $\bZ_{N/r} \equiv \cA'$ by $\bZ_N \equiv \cB$. We will restrict to the case $r\neq N$.
Using (\ref{expression B2 special case}) and for any group $G$, the theory can be coupled to the background $(X_1,X_2)$ for a 2-group global symmetry $\bG = \big( G, \bZ_N, 1, \Bock( [q] ) \big)$ with $[q] \in H^2(BG, \bZ_{N/r})$.
For a related discussion of this theory coupled to 2-group bundles on the lattice, see \cite{Kapustin:2014zva} (where the Postnikov class is interpreted as an obstruction for ordinary symmetries). 
The 't~Hooft anomaly for the 2-group background can be computed from the anomaly of the $\mathbb{Z}_{N^2/r}$ 1-form symmetry generated by $\oint w$ of spin $-\frac{K}{2N^2}$ mod 1. Using the short-hand notation $\pi(q)= \varsigma$, the anomaly is
\bea
\label{eqn:anomegznk}
S_\text{anom} &= 2\pi \int_X \left( - \frac K{2N^2} \right) \, \fP (B_2) = - \frac{2\pi K}{N^2} \int_X \frac12 \, \fP\left( \frac{N}{r} X_2 - X_1^* \, \varsigma \right) \\
&= 2\pi\int_X \left(-\frac{K}{2r^2} \, \fP(X_2) + \frac{K}{Nr} X_1^* \, \varsigma \cup X_2 - \frac{K}{N^2} \, X_1^* \Big( \tfrac12 \fP(\varsigma) + \varsigma \cup_1 d\varsigma \Big) \right)
\eea
mod $2\pi \bZ$, where $B_2$ is as in (\ref{expression B2 special case}) and $X$ is a closed spin 4-manifold. 
The anomaly agrees with the general structure in (\ref{4D anomaly action})-(\ref{3D anomaly: constraints}) where the $\mathbb{Z}_N\subset \mathbb{Z}_{N^2/r}$ subgroup 1-form symmetry is generated by the line $\frac{N}{r}\oint w$ of spin $-\frac{K}{2r^2}$ mod 1, the permutation $\rho$ is trivial, $\lambda=-\frac{K}{Nr} \varsigma$, $\omega=-\frac{K}{N^2} \big( \frac12 \fP(\varsigma) + \varsigma \cup_1 d\varsigma\big)$ and $\beta=\frac{r}{N} d \varsigma$.

The case $N=3$, $K=2$ formulated on the lattice was discussed in \cite{Thorngren:2015gtw}. In this case $r=1$, and the theory can couple to the 2-group background as above with $[q] \in H^2(BG,\mathbb{Z}_3)$. The 't~Hooft anomaly for the 2-group background can be computed from the anomaly of the $\bZ_9$ 1-form symmetry generated by $\oint w$ of spin $-\frac19 \mod 1$ as in (\ref{eqn:anomegznk}):
\begin{equation}
\label{eqn:anomegzthree}
S_\text{anom} = 2\pi \int_X \left[ \frac{2}{3} X_1^* \varsigma \cup X_2 
- X_1^* \Big( \frac{1}{9}\varsigma^2+\frac{2}{3}\varsigma\cup_1 \frac{d\varsigma}{3} \Big) \right] \quad \text{mod }2\pi\mathbb{Z} \;,
\end{equation}
which agrees with the anomaly computed in \cite{Thorngren:2015gtw}.%
\footnote{Here, however, we interpret $X_2$ as the classical background for the 1-form global symmetry $\cA$, as opposed to what is called $G_\text{gauge}^*$ in \cite{Thorngren:2015gtw}.}

\subsection[$U(1)_K$ with Matter of General Charges]{\matht{U(1)_K} with Matter of General Charges}
\label{sec:eguone}

Consider $U(1)_K$ Chern-Simons theory coupled to $N_f$ scalars of charge $q$.
We study the case that $K=q \ell$ is a multiple of $q$. 
The gauge-invariant unit monopole operator is dressed with $\ell$ scalar fields, and thus the theory appears to have ordinary global symmetry%
\footnote{The group $U(N_f)/\mathbb{Z}_\ell$ can be described as the elements $(x,y)$ of $SU(N_f) \times U(1)$ with the identifications $(x,y) \sim \big( e^{-2\pi i / N_f}x \,,\, e^{2\pi i/ N_f} y \big) \sim \big( x \,,\, e^{2\pi i / \ell}y \big)$.}
$G=U(N_f)/\mathbb{Z}_\ell$ (we neglect charge conjugation symmetry in the following discussion). The theory also has ${\cal A}=\mathbb{Z}_q$ 1-form symmetry generated by the element $e^{2\pi i/q}$ in the center of the gauge group, which assigns the phase $e^{2\pi i Q/q}$ to the Wilson line of charge $Q$.
As we show, the two symmetries combine into a 2-group.

Denote the matter fields by $\phi_{I}$ with flavor index $I=1,\dots, N_f$. The $\mathbb{Z}_\ell$ quotient on $U(N_f)$ is generated by $\phi_I\rightarrow e^{2\pi i/\ell}\phi_I$. 
To see this, note it acts trivially on the perturbative local operators formed by gauge invariant polynomials of $\phi_I$ since it can be absorbed by a $U(1)$ gauge rotation $e^{2\pi i/(q\ell)}$ (recall $\phi_I$ has charge $q$).
It also acts trivially on the local operators with magnetic charge since the basic monopole is dressed with $\ell$ matter-field zero-modes.
Since the $\mathbb{Z}_{\ell}$ transformation is identified with a $e^{2\pi i/(q\ell)}$ gauge rotation on the matter fields, if we activate a $G=U(N_f)/\mathbb{Z}_\ell$ background field that is not a $U(N_f)$ gauge field, we find the dynamical $U(1)$ gauge field is modified by a $\mathbb{Z}_{q\ell}$ quotient.
The $\mathbb{Z}_{q\ell}$ quotient changes the quantization of the dynamical $U(1)$ gauge field from integral periods to $\mathbb{Z}_{q\ell}$ fractional periods, specified by a $\mathbb{Z}_{q\ell}$ 2-cocycle that depends on the background fields of the global symmetry.
Since the $\ell$-th power of the gauge rotation $e^{2\pi i/(q\ell)}$ acts trivially on the matter fields but non-trivially on the Wilson lines that are charged under the 1-form symmetry, such background field for the 0-form symmetry also activates the background for the $\mathbb{Z}_q$ 1-form symmetry.%
\footnote{More precisely, denote $r=\text{gcd}(q,\ell)$: since $(q/r)^{-1}$ can be defined in $\mathbb{Z}_\ell$, the $\mathbb{Z}_\ell$ transformation on the charge-$q$ matter fields can be identified with the gauge rotation $e^{2\pi i (q/r)^{-1}/r\ell}$ (with a lift of $(q/r)^{-1}$ in $\mathbb{Z}_{r\ell}$) which generates a $\mathbb{Z}_{r\ell}$ quotient instead of $\mathbb{Z}_{q\ell}$. Consequently, the background for the 0-form symmetry only activates the background for the $\mathbb{Z}_{r}\subset \mathbb{Z}_q$ subgroup 1-form symmetry. However, since we will couple the theory to the background for the entire $\mathbb{Z}_q$ 1-form symmetry, this extends the $\mathbb{Z}_{r\ell}$ quotient to a $\mathbb{Z}_{q\ell}$ quotient. The special case with only 0-form global symmetry is discussed in \cite{Benini:2017dus}.
\label{foot:exampleuone}}
The $\mathbb{Z}_{q\ell}$ 2-cocycle can be expressed as a background $\mathbb{Z}_q$ 2-cochain $X_2$ for the $\mathbb{Z}_q$ 1-form symmetry, and the $\mathbb{Z}_\ell$ 2-cocycle
\begin{equation}\label{eqn:cocycleeg}
B_2' = X_1^*w_2^{(\ell)}~,
\end{equation}
where $X_1$ is the background $G$ gauge field, and $w_2^{(\ell)}$ is the obstruction to lifting the $U(N_f)/\mathbb{Z}_\ell$ bundle to a $U(N_f)$ bundle. More precisely, the $\mathbb{Z}_{q\ell}$ 2-cocycle is described by $X_2,B_2'$ with the constraint
\begin{equation}
\label{eqn:uone}
d X_2 = \Bock (B_2')= X_1^* \Bock\big( w_2^{(\ell)} \big) \;,
\end{equation}
in terms of the Bockstein homomorphism for the exact sequence $1\rightarrow\mathbb{Z}_q\rightarrow\mathbb{Z}_{q\ell}\rightarrow\mathbb{Z}_\ell\rightarrow 1$ that describes $\mathbb{Z}_{q\ell}$ as the extension of $\mathbb{Z}_\ell$ by $\mathbb{Z}_q$. Namely, the $G=U(N_f)/\mathbb{Z}_\ell$ ordinary global symmetry and the $\mathbb{Z}_q$ 1-form symmetry combine into a 2-group
\be
\bG = \Big( U(N_f)/\bZ_\ell \;,\; \bZ_q \;,\; 1 \;,\; \Bock\big( w_2^{(\ell)}\big) \Big) \;,
\ee
with Postnikov class $\Bock \big( w_2^{(\ell)} \big)\in H^3(BG,\mathbb{Z}_q)$.%
\footnote{In the special case gcd$(q,\ell)=1$, the Bockstein homomorphism is trivial since $\mathbb{Z}_{q\ell}= \mathbb{Z}_q\times\mathbb{Z}_{\ell}$, and the 2-group symmetry has trivial Postnikov class \ie{} it factorizes into a 0-form and a 1-form symmetry. From Footnote \ref{foot:exampleuone} we find that the background field for $G=U(N_f)/\mathbb{Z}_\ell$ with non-trivial $w_2^{(\ell)}$ modifies the ordinary gauge and global symmetry bundle into a $[U(1)_\text{dyn}\times U(N_f)_\text{global}]/\mathbb{Z}_{\ell}$ bundle, where the $\mathbb{Z}_{\ell}$ quotient is generated by the element $(e^{2\pi i (q^{-1})/\ell},e^{-2\pi i/\ell})$. We will focus on the case gcd$(q,\ell)>1$. In particular, if we gauge the $\mathbb{Z}_{\text{gcd}(q,\ell)}$ subgroup of the $\mathbb{Z}_q$ 1-form symmetry to change $(q,\ell)$ into $(q/\text{gcd}(q,\ell),\ell/\text{gcd}(q,\ell))$, the resulting UV 2-group symmetry has trivial Postnikov class.
}
The permutation $\rho$ in the 2-group symmetry is trivial, since the 0-form flavor symmetry $G$ acts on the matter fields without changing their gauge charges.%
\footnote{If we include the charge conjugation symmetry, which we neglected, the permutation would be $\mathbb{Z}_2$.}

The class $w_2^{(\ell)}\in H^2(BG,\mathbb{Z}_\ell)$ for $G=U(N_f)/\mathbb{Z}_\ell$ can be described in more detail as follows.
The $G$-bundle can be described by a $U(1)\times PSU(N_f)$ bundle with the following correlation \cite{Benini:2017dus}:
\begin{equation}
\label{eqn:unfquotient}
\frac{F}{2\pi} =\frac{N_f}{d} w_2^{(\ell)}+\frac{\ell}{d}w_2^{(N_f)}\quad\text{mod }\frac{N_f \ell}{d} ~,
\end{equation}
where $F/(2\pi)$ is the first Chern class of the $U(1)$ bundle, $d=\text{gcd}(N_f, \ell)$ and $w_2^{(N_f)}$ is the obstruction to lifting the $PSU(N_f)$ bundle to an $SU(N_f)$ bundle.
In the special case $N_f=1$ the condition (\ref{eqn:unfquotient}) implies $w_2^{(\ell)}$ can be lifted to an integral cocycle, and in particular a $\mathbb{Z}_{q\ell}$ cocycle, therefore the Postnikov class $\Bock \big( w_2^{(\ell)} \big)$ vanishes.
In this case $G$ is the $U(1)$ magnetic symmetry, and the vanishing Postnikov class reproduces the fact that the theory can couple to the magnetic symmetry without activating any background for the 1-form symmetry.
We will consider the case $N_f\geq 2$.

We can also give the matter field a mass term singlet under the global symmetry, and integrate out the matter fields to find $U(1)_K$ Chern-Simons theory coupled to the background fields. Since the symmetry acts on the matter fields without changing the gauge charge, the symmetry does not permute the anyons in the resulting Chern-Simons theory, and it couples to the theory by the 1-form symmetry.
Note that in the IR the 1-form symmetry is enhanced from $\mathbb{Z}_q$ to $\mathbb{Z}_{q\ell}$.
As a consistency condition for the flow, we can couple the theory to background fields using the formalism of Section~\ref{sec: general coupling} with $f_1$ the trivial homomorphism and $f_2:\mathbb{Z}_q\rightarrow\mathbb{Z}_{q\ell}$ the inclusion given by multiplication by $\ell$.
This reproduces (\ref{eqn:uone}).
Equivalently, the backgrounds $X_1,X_2$ activate the 2-form background for the $\mathbb{Z}_{q\ell}$ total 1-form symmetry
\begin{equation}\label{eqn:uoneanombg}
B_2 = \ell \wt X_2 - \wt B_2' 
\end{equation}
with $B_2'=X_1^*w_2^{(\ell)}$ as in (\ref{eqn:cocycleeg}), where tilde denotes the lift to $\mathbb{Z}_{q\ell}$ cochains while preserving (\ref{eqn:uone}), and thus $B_2$ is a $\mathbb{Z}_{q\ell}$ cocycle independent of the lift. This is the same $\mathbb{Z}_{q\ell}$ 2-cocycle that describes the $\mathbb{Z}_{q\ell}$ quotient on the dynamical $U(1)$ gauge field in the UV.
The 't~Hooft anomaly of the 2-group symmetry can then be computed from the 't~Hooft anomaly of the $\mathbb{Z}_{q\ell}$ 1-form symmetry:
\begin{equation}
\label{eqn:uoneanomeg}
S_\text{anom} = \frac{2\pi}{q\ell}\int_X \frac{\fP B_2}{2} \;,
\end{equation}
where $B_2$ is given by (\ref{eqn:uoneanombg}) and $X$ is a closed spin 4-manifold. The same anomaly is reproduced in the UV theory using the $\mathbb{Z}_{q\ell}$ quotient on the dynamical $U(1)$ gauge field described by the same 2-cocycle (\ref{eqn:uoneanombg}), where the quotient makes the Chern-Simons term of the dynamical gauge field no longer properly quantized.

The discussion can be repeated with the scalars replaced by $N_f$ massless fermions of charge $q$, and $K$ replaced by the bare Chern-Simons level (we consider the case $K_\text{bare}=q\ell$), namely in the theory $U(1)_{q\ell - q^2N_f/2}$ with $N_f$ fermions of charge $q$.%
\footnote{What one means by ``bare CS level'' depends on the scheme used to regularize the fermions. In a different scheme, the bare CS level would be $K_\text{bare} = q\ell - q^2 N_f$. This would not affect the physical result, since the groups $U(N_f)/\mathbb{Z}_\ell$ and $U(N_f)/\mathbb{Z}_{\ell - qN_f}$ are isomorphic, and the corresponding Postnikov classes $\Bock \big( w_2^{(\ell)} \big) \in H^3(BG,\mathbb{Z}_q)$ are the same.}

In the special case $q=1$, the 1-form symmetry is trivial and the ordinary symmetry $G$ does not participate in a non-trivial 2-group symmetry. The 't~Hooft anomaly (\ref{eqn:uoneanomeg}) of the ordinary symmetry $G$ agrees with the computation in \cite{Benini:2017dus} (up to counterterms for the $G$ background gauge field).

Another special case is QED$_3$ with $N_f$ fermions of charge $q$, where $N_f q$ needs to be even to avoid the standard parity anomaly. Following the previous notation, $\ell = qN_f/2$ and the bare Chern-Simons level is $q^2N_f/2$, thus from the previous discussion the theory has 2-group symmetry with ordinary symmetry $G=U(N_f)/\mathbb{Z}_{qN_f/2}$, 1-form symmetry $\mathbb{Z}_q$, trivial permutation and Postnikov class $\Bock \big( w_2^{( qN_f/2 )} \big)$.

\subsection{Gauging a Symmetry with Mixed 't~Hooft Anomaly}
\label{sec: gauging to 2-group}

Consider a theory with an ordinary 0-form global symmetry $\wh\cA \times G$, where $\wh\cA$ is an Abelian group while $G$ is generic, and a mixed 't~Hooft anomaly
\be
\label{eqn:egmixanom}
S_\text{anom} = 2\pi  \int_X C \cup A^*\beta \;.
\ee
Here $X$ is a closed four-manifold, $C$ is a gauge field for $\wh\cA$, $A$ is a gauge field for $G$, the class $[\beta]\in H^3(BG,\cA)$ parameterizes the mixed 't~Hooft anomaly and $\cA$ is the Pontryagin dual to $\wh\cA$. The 't~Hooft anomaly is an obstruction to gauging the entire $\wh\cA \times G$ symmetry. However, following the discussion in \cite{Tachikawa:2017gyf}, we show that gauging only the subgroup $\wh\cA$ produces a theory with 2-group symmetry.

For trivial $G$-background, namely $A=0$, the anomaly (\ref{eqn:egmixanom}) vanishes: we can gauge the subgroup $\wh\cA$ and make $C$ dynamical in the path-integral. The resulting theory has a 1-form center symmetry $\cA$ (isomorphic to $\wh\cA$), which can be coupled to a background 2-form gauge field $B$ through the term
\be
S_\text{1-form} = 2\pi  \int_\text{3d} C \cup B \;.
\ee
In the absence of the mixed anomaly (\ref{eqn:egmixanom}), invariance under $\wh\cA$ gauge transformations would require $B$ to be a cocycle. With (\ref{eqn:egmixanom}), instead, gauge invariance requires that the background fields satisfy
\be
dB = A^* \beta \;.
\ee
We conclude that $G$ and $\cA$ form a 2-group global symmetry $\bG = \big(G, \cA, 1, [\beta]\big)$ in which $G$ does not act on $\cA$.

Summarizing, one can produce examples of 3d theories with 2-group global symmetry by starting with a theory with ordinary global symmetry $\wh\cA \times G$ and 't~Hooft anomaly of the form (\ref{eqn:egmixanom}), and then dynamically gauging $\wh\cA$ \cite{Tachikawa:2017gyf}. Notice that, naively, one could have expected that because of the mixed anomaly (\ref{eqn:egmixanom}), promoting $\wh\cA$ to a dynamical gauge symmetry completely spoils the global symmetry $G$. Instead $G$ survives and becomes part of a 2-group global symmetry.%
\footnote{In fact, all 3d examples with 2-group symmetry discussed in Section \ref{sec: examples} have such a parent theory (or a generalization where the 0-form symmetry is not the product ${\widehat{\cal A}}\times G$ but an extension), which can be obtained by gauging a subgroup of the 1-form symmetry.}

\subsection[$Spin(N)$ and $O(N)$ Chern-Simons-Matter Theories]{\matht{Spin(N)} and \matht{O(N)} Chern-Simons-Matter Theories}
\label{sec: example CSM}

Two interesting examples of the general strategy highlighted above are provided by $Spin(N)_K$ and $O(N)_K$ Chern-Simons theories with $N_f$ massless scalars in the vector representation. Those two theories can have 2-group global symmetry. To explain those examples, let us proceed step by step.

We start considering $SO(N)_K$ Chern-Simons theory with $N_f$ massless scalars in the vector representation.%
\footnote{We can add an $O(N_f)$-invariant potential for the scalar fields and tune the mass term to zero: depending on $N$, $K$, $N_f$ the theory at long distances has been conjectured to flow to a critical point or a symmetry-breaking phase \cite{Aharony:2016jvv, Komargodski:2017keh}. In this discussion we will focus on the microscopic theory.}
We take $N=2 \mod 4$, $K$ even, and $N_f=0$ mod 4. The theory has charge conjugation symmetry $\cC$, magnetic symmetry $\cM$, as well as flavor symmetry $O(N_f)$. More precisely, the action of the $\mathbb{Z}_2$ center of $O(N_f)$ on the matter fields can be identified with the $\mathbb{Z}_2$ center of the $SO(N)$ gauge group, and besides, such a center flavor rotation does not act on the gauge-invariant operators with magnetic charge (which, for $K$ even, need an even number of matter fields to be gauge invariant, and thus they carry an even number of flavor indices). Therefore the faithful flavor symmetry is $PO(N_f)$. We would like to determine the 't~Hooft anomaly for $\cC$, $\cM$ and the connected component $PSO(N_f) = SO(N_f)/\bZ_2$ in this theory.

We can turn on a background gauge field $A$ for a $PSO(N_f)$-bundle with non-trivial second Stiefel-Whitney class, by which we mean the obstruction to lifting the bundle to an $SO(N_f)$ bundle. Let us denote by $[w_2^{PSO(N_f)}]$ the group cohomology class that represents the second Stiefel-Whitney class in the classifying space of $PSO(N_f)$:
\be
\big[ w_2^{PSO(N_f)} \big] \,\in\, H^2\big(BPSO(N_f), \bZ_2 \big) \;.
\ee
This class has the property that for every $PSO(N_f)$-bundle with connection $A$, its second Stiefel-Whitney class is $\big[ A^*w_2^{PSO(N_f)} \big]$. We then choose a representative $w_2^{PSO(N_f)}$.

Because of the non-trivial $PSO(N_f)$ flavor bundle, also the bundle for the dynamical gauge field is forced to be a non-trivial $PSO(N)$-bundle, as the second Stiefel-Whitney classes of the two bundles are constrained to be the same \cite{Benini:2017dus}. Namely, the gauge fields live in
\be
\frac{ SO(N)_\text{dyn} \times SO(N_f)_\text{global} }{ \mathbb{Z}_2} \;.
\ee
From the point of view of the dynamical gauge sector, the constraint to non-trivial $PSO(N)$-bundles is enforced by an effective coupling $\int A_\text{dyn} \cup B_2$ (obtained, for instance, by integrating out the massive scalar fields) to a 2-cochain
\be
\label{eqn:cssotwoform}
B_2 = A^* w_2^{PSO(N_f)} \;.
\ee
In fact, the Chern-Simons term of $SO(N)_K$ has a $\bZ_2$ 1-form symmetry and $B_2$ acts as a source for the latter, enforcing the constraint. Let us stress that in the full theory with matter, such a would-be 1-form symmetry is explicitly broken.

On the other hand, consider for a moment the pure $SO(N)_K$ Chern-Simons theory. Such a TQFT---as we said---has a $\bZ_2$ 1-form symmetry related to the center of $SO(N)$, as well as magnetic and charge conjugation 0-form $\bZ_2$ symmetries $\cM$, $\cC$, respectively. From the analysis in Section 2.4 of \cite{Cordova:2017vab}, the theory has the following 't~Hooft anomaly:
\be
\label{eqn:cssoanom}
S_\text{anom} = 2\pi  \int_X \bigg[ \frac{NK}8 \; \frac{\fP B_2}2 + \Bock(B_2) \cup \bigg( \frac N4 B^\cM + \frac K4 B^\cC \bigg) + \frac12 B_2 \cup B^\cC \cup B^\cM \bigg] \;.
\ee
Here $X$ is a closed spin four-manifold, $\fP$ is the Pontryagin square (Appendix \ref{app: Pontryagin}), $\Bock$ is the $\mathbb{Z}_2$ Bockstein homomorphism (Appendix \ref{app: Bockstein}), while $B^\cM$, $B^\cC$ are background gauge fields for $\cM$, $\cC$, respectively. Substituting the effective value (\ref{eqn:cssotwoform}) for $B_2$ in the theory with matter, we find the 't~Hooft anomaly of $SO(N)_K$ with $N_f$ scalars. Here we are only interested in backgrounds for $PSO(N_f)$ and $\cM$, hence let us set $B^\cC=0$. Recalling that we chose $N=2 \mod 4$ and $K$ even, we find the anomaly
\be
\label{intermediate anomaly}
S_\text{anom} = \pi  \int_X \bigg[ (K/2) \, \frac{A^* \fP w_2^{PSO(N_f)} }2 + A^* \Bock\big( w_2^{PSO(N_f)} \big) \cup B^\cM \bigg] \;.
\ee
Notice that $\Bock\big( w_2^{PSO(N_f)}\big)$ would be trivial for $N_f = 2\mod 4$.%
\footnote{For $N_f = 2\mod4$, $PSO(N_f)$ has a $\bZ_4$-valued Stiefel-Whitney class $\wt w_2$ representing the obstruction to lifting $PSO(N_f)$ bundles to $Spin(N_f)$ bundles, and $w_2^{PSO(N_f)} = \wt w_2 \mod 2$. If follows that there is no obstruction to lifting the $\bZ_2$-valued class $w_2^{PSO(N_f)}$ to a $\bZ_4$-valued class, and therefore the Bockstein homomorphism associated to the exact sequence $1\to\bZ_2 \to \bZ_4 \to\bZ_2\to1$ maps $w_2^{PSO(N_f)}$ to zero.}

Such an anomaly has the same form as in (\ref{eqn:egmixanom}), therefore we can produce a theory with 2-group global symmetry by employing the procedure of \cite{Tachikawa:2017gyf} reviewed in Section \ref{sec: gauging to 2-group}. We promote $B^\cM$ to a dynamical gauge field, which enlarges the gauge group to $Spin(N)_K$. The new theory has a $\bZ_2$ center 1-form symmetry (recall that $Spin(N)_K$ has a $\bZ_4$ center 1-form symmetry, however the coupling to matter in the vector representation breaks it to $\bZ_2$) which we can couple to a background field $X_2 \in C^2(M, \bZ_2)$ by the coupling $\pi \int_\text{3d} B^\cM \cup X_2$. Invariance of the action under gauge transformations of $B^\cM$ requires the background fields $A,X_2$ to satisfy%
\footnote{With some abuse of notation, by $\Bock\big( w_2^{PSO(N_f)}\big)$ we mean a representative of the class $\Bock\big( \big[ w_2^{PSO(N_f)}\big] \big)$. Moreover, we used that $\Bock\big( w_2^{PSO(N_f)}\big)$ is a $\bZ_2$-valued cochain and so it is equal to its opposite.}
\be
d X_2 = A^* \Bock\big( w_2^{PSO(N_f)} \big) \;.
\ee
We conclude that in $Spin(N)_K$ Chern-Simons theory with $N_f$ scalars in the vector representation ($N=2\mod4$, $K$ even, $N_f= 0\mod 4$), the $PSO(N_f)$ flavor symmetry and the $\bZ_2$ 1-form symmetry form a 2-group global symmetry
\be
\label{eqn:cssotwogroup}
\bG = \Big( PSO(N_f) \,,\, \bZ_2 \,,\, 1 \,,\, \Bock\big(\big[ w_2^{PSO(N_f)}\big]\big) \Big) \;.
\ee
The non-trivial Postnikov class $\Bock\big( \big[ w_2^{PSO(N_f)}\big]\big)$ is an element of $H^3\big( BPSO(N_f), \bZ_2 \big)$. We can easily write the 't~Hooft anomaly for the 2-group symmetry $\bG$, as the non-dynamical part of (\ref{intermediate anomaly}):
\be
\label{'t Hooft anomaly 2-group CSM}
S_\text{anom} = \pi  \int_X (K/2) \, \frac{A^* \fP w_2^{PSO(N_f)} }2 \;.
\ee
This anomaly has only a pure 0-form part (non-vanishing for $K=2\mod4$).

An alternative heuristic explanation for the 2-group global symmetry is as follows. 
The action of the $\mathbb{Z}_2$ center of $SO(N_f)$ on the elementary fields is identified with the action of a transformation in the center of $Spin(N)$ of order four.
Thus in the presence of a $PSO(N_f)$ background with nontrivial $A^*w_2^{PSO(N_f)}$, the $Spin(N)$ gauge field is modified by a background for the 1-form symmetry. 
If both transformations were of order two, one could simply set the background to be $A^*w_2^{PSO(N_f)}$. Instead, since the square of the transformation is nontrivial in $Spin(N)$, we need to introduce an additional $\mathbb{Z}_2$ 2-form background that correlates with the $PSO(N_f)$ gauge fields as in (\ref{expression B2 special case}).

Instead of making $B^{\cal M}$ dynamical, we could take $K=2\mod4$, $N$ even, and repeat the discussion making $B^\cC$ dynamical (and setting $B^\cM$ to zero). Hence we find that in $O(N)^0_K$ Chern-Simons theory with $N_f$ scalars in the vector representation \cite{Cordova:2017vab}, the flavor symmetry $PSO(N_f)$ and the $\bZ_2$ center 1-form symmetry form a 2-group global symmetry (\ref{eqn:cssotwogroup}). Or, we could gauge the diagonal combination ${\cal CM}$ by making $B^{\cal C}=B^{\cal M}$ dynamical producing the $O(N)^1_K$ theory: then for $N,K$ even that satisfy $N+K=0\mod 4$ and $N_f=0\mod4$, the theory has 2-group symmetry.

Now, suppose we deform the $Spin(N)_K$ theory with matter by a large $SO(N_f)$-invariant mass for the matter fields: for a suitable sign of the mass term, the theory flows to the pure $Spin(N)_K$ Chern-Simons TQFT in the IR. The $\bZ_2$ 1-form symmetry we had in the UV is enhanced to $\bZ_4$ in the IR. On the other hand, the $PSO(N_f)$ 0-form symmetry does not permute the anyons in the IR: in the UV the symmetry does not change the representation of the matter fields under the gauge group, and also since it is connected and continuous, it cannot be mapped to the permutation group but in the trivial way. In other words, $PSO(N_f)$ is not an intrinsic symmetry of the TQFT. Yet, since the UV theory has 2-group global symmetry $\bG$ as in (\ref{eqn:cssotwogroup}) and the flow preserves the symmetry, it should be possible to couple the TQFT to $\bG$ and the 't~Hooft anomaly (\ref{'t Hooft anomaly 2-group CSM}) should be reproduced. Indeed, in the IR the 2-group $\bG$ is coupled to the TQFT through the $\bZ_2 \equiv \cB$ subgroup of the $\bZ_4 \equiv \cA$ 1-form symmetry, according to the short exact sequence (\ref{short exact sequence of f2}), and $q = w_2^{PSO(N_f)}$ as in (\ref{possible Postnikov special case}) and (\ref{expression B2 special case}). As noted in Section \ref{sec: general coupling}, the non-trivial Postnikov class $\Bock\big( \big[ w_2^{PSO(N_f)}\big]\big) \in H^3 \big( BPSO(N_f),\bZ_2 \big)$ vanishes once it is mapped to $H^3 \big( BPSO(N_f), \bZ_4 \big)$, consistently with the fact that $PSO(N_f)$ cannot permute the anyons. 
To compute the 't~Hooft anomaly of $\bG$ in the IR, we take the anomaly for the 1-form symmetry of $Spin(N)_K$ with $N=2\mod 4$ and $K$ even, namely $\pi  \int_X (K/2) \, \fP B_2/2$, and substitute (\ref{expression B2 special case}). Noticing that $f_2:\bZ_2 \to \bZ_4$ is here multiplication by 2, the only term that survives is precisely (\ref{'t Hooft anomaly 2-group CSM}).

In the special case that $N=2$, we find the theory $U(1)_{4K}$ with $N_f$ scalars of charge $q=2$ ($K$ even and $N_f=0\mod4$), and conclude that it has 2-group symmetry. This is consistent with Section \ref{sec:eguone} by turning off the background for the $U(1)$ magnetic symmetry and restricting the background for the flavor symmetry to be in $PSO(N_f)$. 

The discussion can be repeated with scalars replaced by fermions, and the level replaced by the bare level (for $O(N)$ gauge theory there is also the bare $\mathbb{Z}_2$ level). In particular we take $N=2\mod4$, $K$ even and $N_f= 0\mod4$, and claim that the resulting theories have 2-group global symmetry. This also provides a consistency check for the dualities \cite{Cordova:2017vab}:
\bea
Spin(N)_K\quad\text{with }N_f\;\phi &\quad\longleftrightarrow\quad O(K)^0_{-N+ \frac{N_f}2,-N+\frac{N_f}2}\quad\text{with }N_f\;\psi \\
O(N)^1_{K,K-1+L}\quad\text{with }N_f\;\phi &\quad\longleftrightarrow\quad O(K)^1_{-N+\frac{N_f}2,-N + \frac{N_f}2 +1+L}\quad\text{with }N_f\;\psi~,
\eea
where the scalars $\phi$ and the fermions $\psi$ are in the vector representation.
In the first duality, for $N=2\mod4$, $K$ even, $N_f=0\mod4$, both sides have the same 2-group symmetry and the same 't~Hooft anomaly.
The $\mathbb{Z}_2$ levels are 3d local counterterms added when gauging ${\cal C}$ or ${\cal CM}$, and thus they do not affect the 2-group symmetry and its anomaly.
Similarly in the second duality, for $N,K$ even, $N+K=0\mod4$, $N_f=0\mod4$, both sides have the same 2-group symmetry and the same 't~Hooft anomaly.
In fact, this simply follows from 't~Hooft anomaly matching (\ref{eqn:cssoanom}) in the $SO(N)$ Chern-Simons-matter dualities and the gauging strategy discussed in Section~\ref{sec:eggauging}.

\subsection{Chern-Simons Theories with Time-Reversal Symmetry}
\label{sec:timereversal}

Our next family of examples is given by Chern-Simons theories with a $\mathbb{Z}_2$ 1-form symmetry that forms a 2-group with time-reversal symmetry.

We start with $U(1)_k$ Chern-Simons theory. For special values of $k$ the theory is time-reversal invariant as a spin-TQFT (up to a mixed gravitational anomaly): those are the integer solutions to the negative Pell equation
\be
\label{cond}
kp^2-q^2=1
\ee
for some integers $p,q$ \cite{witten:unpublished}. The first few values are $k=1,2,5,10,13,17,26$. In order to prove time-reversal invariance, we start from the equality of the following two theories:
\begin{multline}
\label{tinvgen}
\frac k{4\pi} bdb - \frac1{4\pi} cdc + \frac1{2\pi} bd(B+kA) + \frac1{2\pi}cdA \qquad\longleftrightarrow \\
- \frac k{4\pi} {\tilde b}d{\tilde b} + \frac1{4\pi}{\tilde c}d{\tilde c} + \frac1{2\pi} {\tilde b}d\big( qB+k(q-p)A\big) + \frac1{2\pi} {\tilde c}d\big( pB+(kp-q)A \big) \;,
\end{multline}
where $b,c,{\tilde b},{\tilde c}$ are dynamical $U(1)$ gauge fields, $B$ is a background $U(1)$ gauge field and $A$ is a background spin$_c$ connection. Equality follows from the field redefinition $b=q {\tilde b}+p {\tilde c}$, $c=-kp {\tilde b}-q {\tilde c}$ which has unit Jacobian because of \eqref{cond}. Integrating $c, \tilde c$ out and setting $A=0$ (which is consistent on spin manifolds) gives
\be
\label{uonek}
\frac k{4\pi} bdb +  \frac1{2\pi}bd B \qquad\longleftrightarrow\qquad
- \frac k{4\pi}{\tilde b}d{\tilde b} + \frac q{2\pi} {\tilde b}dB - \frac{p^2}{4\pi}BdB - 4{\rm CS_{grav}} \;.
\ee
(Note that by the redefinition $\tilde b\rightarrow \tilde b + B$, the coupling $\frac q{2\pi} \tilde bdB$ could be shifted by $-\frac k{2\pi} \tilde bdB$.) Written more simply, this is the duality of spin-TQFTs
\be
U(1)_k \quad\longleftrightarrow\quad U(1)_{-k} \;.
\ee
Thus, the theories are time-reversal invariant.

Taking into account the coupling to $B$, time reversal should act as $\cT(B) = -qB$. Yet, the theory is not time-reversal invariant in the presence of a background $B$, due to the anomalous shift on the right-hand side. To achieve invariance we should take
\begin{multline}
\frac k{4\pi} bdb + \frac1{2\pi} bdB + \frac{1/k}{4\pi} BdB + 2{\rm CS_{grav}} \quad\stackrel{\cT}{\longrightarrow}\quad - \frac k{4\pi}bdb + \frac q{2\pi} bdB - \frac{q^2/k}{4\pi} BdB - 2{\rm CS_{grav}} \\
\stackrel{\text{dual}}{\longleftrightarrow}\quad \frac k{4\pi} bdb + \frac1{2\pi}bdB + \frac{1/k}{4\pi} BdB + 2{\rm CS_{grav}} \;.
\end{multline}
The term $\frac{1/k}{4\pi} BdB$ is not properly quantized and thus not well-defined in 3d. We can realize it---and thus the theory can be made time-reversal invariant---by placing the system on the surface of a bulk with theta term $\theta=2\pi/k$ for $U(1)_B$ (normalized as $\theta\sim \theta+2\pi$), which characterizes the mixed 't~Hooft anomaly.

The mapping of lines can be deduced from their coupling to the background fields $B,A$, or from the change of variables. We find
\begin{equation}
\label{mapuonek}
Q_1 \oint b + Q_2\oint c \quad\longleftrightarrow\quad 
\big( qQ_1-kpQ_2 \big) \oint \tilde{b} + \big( pQ_1-qQ_2 \big) \oint\tilde{c} \;,
\end{equation}
with the identifications $k\oint b+k\oint c\sim 2\oint c\sim 0$ and similarly for $\tilde b,\tilde c$.
Namely, the topological charges satisfy $(Q_1,Q_2)\sim (Q_1+k,Q_2+k)\sim(Q_1,Q_2+2)$.
Applying the duality map twice we get
\begin{equation}
{\cal T}^2(Q_1,Q_2)=(-Q_1,-Q_2)\sim (-Q_1,Q_2) \;.
\end{equation}
Thus the time-reversal symmetry is a $\mathbb{Z}_4$ anyon permutation symmetry for $k>2$ (while for $k=2$ we have $Q_1\sim Q_1+2$).%
\footnote{Examples with this property were already observed in \cite{Barkeshli:2017rzd, Benini:2017aed, Cordova:2017kue}.}
It satisfies
\be
\cT^2 = \cC \;,
\ee
where $\cC$ is charge conjugation symmetry acting as $\cC: B \mapsto - B$ and with $\cC^2= \unit$.

Using the level-rank duality $U(1)_k \;\leftrightarrow\; SU(k)_{-1}$, we also obtain
\be
\label{lrsu}
SU(k)_1 \quad\longleftrightarrow\quad SU(k)_{-1} \;.
\ee
Thus also $SU(k)_1$ is time-reversal invariant as a spin-TQFT for $k$ that satisfies \eqref{cond}. 
We will focus on the case that $k$ is odd, where the theory is also time-reversal invariant as a non-spin TQFT, since the difference in the framing anomaly between the two sides is $2(k-1)\in 8\mathbb{Z}$.%
\footnote{This uses the following property: a duality between two non-spin 3d TQFTs each tensored with the invertible spin-TQFT $\{1,\psi\}$ \cite{Seiberg:2016rsg} implies that the non-spin TQFTs themselves are dual, if and only if their framing anomalies differ by a multiple of 8 \cite{hsin:unpublish}.}
In fact, taking $\text{mod }4$ and $\text{mod }8$ on both sides of (\ref{cond}), one finds that $k=1 \text{ mod } 4$ for $q$ even and $k=2\text{ mod }8$ for $q$ odd (while $p$ is always odd). The anyons of $SU(k)_1$ are labelled by Young diagrams with a single column of height $Q\text{ mod }k$, and the time-reversal symmetry acts as
\be
\label{tdual}
{\cal T}: \;\; Q \;\to\; q Q \;.
\ee
Note that, once again, ${\cal T}^2={\cal C}$ where ${\cal C}(Q)=-Q$ mod $k$. Time-reversal symmetry is a $\mathbb{Z}_4$ anyon permutation symmetry.

Consider gauging charge conjugation ${\cal C}$ in $SU(k)_1$. The 1-form symmetry is $\mathbb{Z}_k$, and since for odd $k$ there is no Abelian anyon stabilized by ${\cal C}$, there is only one way to couple to the $\mathbb{Z}_2$ charge conjugation symmetry. There are two possible non-spin $\mathbb{Z}_2$ counterterms: $(\mathbb{Z}_2)_0$ and $(\mathbb{Z}_2)_4$, where in our convention the $\mathbb{Z}_2$ level is defined mod 8 (see \cite{Cordova:2017vab} for explanations). Such counterterms preserve time-reversal invariance. The duality $SU(k)_1\leftrightarrow SO(k)_2$ maps charge conjugation to the magnetic symmetry. Thus, depending on the choice of $\bZ_2$ counterterm, the gauging produces either $Spin(k)_2$ or $\widetilde{Spin}(k)_{2,4}=[Spin(k)_2\times (\mathbb{Z}_2)_4]/\mathbb{Z}_2$ (where the quotient is generated by the product of the line in the two-index symmetric tensor representation of $Spin(k)$ and the Wilson line of the $\mathbb{Z}_2$ theory, see \cite{Cordova:2017vab}). 

The anyons in the new theory are as follows. After gauging, the state $Q=0$ splits into two states $1, \epsilon$, where $\epsilon$ generates the new $\bZ_2$ 1-form symmetry. Denote $k=4m+1$ for some integer $m$. There are $2m+4$ lines, with two lines $W_1,W_2$ in the spinor representation of $Spin(k)$ of spin $\frac m4$, $\frac m4 + \frac12$ for zero counterterm and spin $\frac{m-1}4$, $\frac{m-1}4+ \frac12$ for the non-trivial counterterm. The two spinor lines are related by fusing with $\epsilon$ and they have the same quantum dimension $\sqrt{k}$.

Let us show that, depending on $k$, the new theories $Spin(k)_2$ and $\wt{Spin}(k)_{2,4}$ can have 2-group symmetry. This discussion follows the one in \cite{Barkeshli:2017rzd}. The new theories have $\mathbb{Z}_2$ anyon-permutation time-reversal symmetry as follows.
It leaves invariant $1,\epsilon$, and maps the spinor lines among themselves (otherwise braiding with $\epsilon$ would be inconsistent).
From the spins of the spinor lines $W_1,W_2$ one can determine the map in the case of trivial counterterm:
\bea
\label{Tspinor}
\text{even $m$:}\qquad &{\cal T}(W_1)=W_1\;,\qquad {\cal T}(W_2)= W_2 \\
\text{odd $m$:}\qquad & {\cal T}(W_1)=W_2 \;,\qquad {\cal T}(W_2)= W_1 \;.
\eea
In the case of non-trivial counterterm, $m$ is replaced by $m-1$ and the two cases are exchanged.

Suppose that the $\mathbb{Z}_2$ anyon-permutation time-reversal symmetry does not form a 2-group with the 1-form symmetry.
Then we can compute the 't~Hooft anomaly of time-reversal symmetry using the anomaly indicator formula in \cite{Wang:2016qkb, Barkeshli:2016mew}, and it must take values in $\{\pm1\}$. If the anomaly is not in $\{\pm 1\}$, then the anyon-permutation symmetry must form a 2-group with the 1-form symmetry.
The 't Hooft anomaly of time-reversal symmetry in a non-spin 3d TQFT can be parameterized by $\pi\int w_2^2$ and $\pi\int w_1^4$ \cite{Kapustin:2014tfa, Barkeshli:2016mew} where $w_1,w_2$ are the Stiefel-Whitney classes of the bulk manifold. The presence of the first term is detected by the framing anomaly $c$ as $e^{2\pi i c/8}$.
For the theory $Spin(k)_2$ with $k=4m+1$, the framing anomaly is $c=4m$ and thus the first anomaly is $(-1)^m\in \{\pm1\}$. The second anomaly can be computed as
\bea
\label{partitionwoct}
\text{even $m$:} \qquad & Z_{\rm anom} = \frac1{2\sqrt{k}} \left[1+\eta_\epsilon^{\cal T} + \sqrt{k}(-1)^{m/2} \left(\eta_{W_1}^{\cal T}-\eta_{W_2}^{\cal T}\right)\right] \\
\text{odd $m$:} \qquad & Z_{\rm anom} = \frac1{2\sqrt{k}} \left(1+\eta_\epsilon^{\cal T}\right) \; \neq \pm 1 \;.
\eea
Here $\eta_x^\cT$ are the phases (\ref{def eta}) associated to the generator of $\cT$ and to the lines $x$. It turns out that for even $m$ we have $\eta^\cT_\epsilon = -1$, while $\eta^\cT_{W_1} = - \eta^\cT_{W_2} = \pm1$ depending on how we couple to the $\bZ_2$ symmetry (there are two fractionalization classes): we see that $Z_\text{anom}$ is in $\{\pm1\}$. On the other hand, for odd $m$ no value of $\eta^\cT_\epsilon = \pm1$ can lead to $Z_\text{anom} = \pm1$. We conclude that for $k=5$ mod 8 satisfying \eqref{cond}, the theory $Spin(k)_2$ has $\mathbb{Z}_2$ anyon-permutation time-reversal symmetry that combines with the $\mathbb{Z}_2$ 1-form symmetry to form a 2-group. In a similar way, we can conclude that for $k=1$ mod 8 satisfying \eqref{cond}, the theory $\widetilde{Spin}(k)_{2,4}$ has $\mathbb{Z}_2$ anyon-permutation time-reversal symmetry that combines with the $\mathbb{Z}_2$ 1-form symmetry to form a 2-group. This is consistent with the conclusion in \cite{Barkeshli:2017rzd}.

In both cases, the 2-group symmetry has 0-form part given by $\bZ_2^\cT$ time-reversal symmetry, 1-form part $\bZ_2$, trivial permutation, and Postnikov class $[\beta]$ given by the unique non-trivial element in $H^3(B\bZ_2, \bZ_2) = \bZ_2$. Such an element is represented, for instance, by the 3-cocycle $(B_1)^3=B_1\cup \Bock (B_1)$, where $B_1$ is a $\bZ_2$ 1-cocycle and the $\bZ_2$ Bockstein homomorphism is for the exact sequence $1 \to \bZ_2 \to \bZ_4 \to \bZ_2 \to 1$.

We could instead couple the theories $Spin(k)_2$ or $\wt{Spin}(k)_{2,4}$, that have intrinsic 2-group symmetry, to an external $\mathbb{Z}_4$ 0-form symmetry and the $\mathbb{Z}_2$ 1-form symmetry using the method of Section \ref{sec: general coupling}, with the projection map $f_1:\mathbb{Z}_4\rightarrow \mathbb{Z}_2$ and the identity map $f_2$. From the discussion in Section~\ref{sec: general coupling}, the Postnikov class for the extrinsic 2-group symmetry is trivial by the property of the Bockstein homomorphism. Note that this applies to any example with such intrinsic 2-group symmetry. This agrees with the discussion in \cite{Barkeshli:2017rzd}.


\subsection[$\mathbb{Z}_N$ Gauge Theory in General Dimension]{\matht{\bZ_N} Gauge Theory in General Dimension}

Consider the simplest $\mathbb{Z}_N$ gauge theory in spacetime dimension $d>1$ \cite{Maldacena:2001ss, Banks:2010zn, Kapustin:2014gua}:
\begin{equation}
S = \frac{N}{2\pi}\int b \, d\phi_{(d-2)} \;,
\end{equation}
where $b$ is a $U(1)$ 1-form gauge field, while $\phi_{(d-2)}$ is a $U(1)$ $(d{-}2)$-form gauge field that constrains $b$ to be a $\mathbb{Z}_N$ gauge field. Likewise, $b$ constrains $\phi_{(d-2)}$ to be a $\mathbb{Z}_N$ gauge field.

The theory has an intrinsic $\mathbb{Z}_N$ $(d{-}2)$-form symmetry generated by the line $e^{i\oint b}$, and an intrinsic $\mathbb{Z}_N$ 1-form symmetry generated by the operator%
\footnote{When $d=2$, the generator is a local operator $e^{i\phi_{(0)}}$ from the periodic scalar field $\phi_{(0)} \sim \phi_{(0)} +2\pi$.}
$e^{i\oint\phi_{(d-2)}}$. The corresponding charged objects are the 't~Hooft operators $e^{i n_\sm \oint\phi_{(d-2)}}$ and the Wilson lines $e^{i n_\se\oint b}$ with integers $n_\sm,n_\se \in\mathbb{Z}_N$.
We can turn on $\mathbb{Z}_N$ background gauge fields $B^\sm_{d-1}$ or $B^\se_2$ for these symmetries (in this discussion they are normalized as $\oint B^\sm_{d-1}$, $\oint B^\se_2 \in \frac{2\pi}N \mathbb{Z}$), and this introduces one of the two couplings
\begin{equation}
\label{eqn:ZNegD}
\frac{N}{2\pi}\int b \, B^\sm_{d-1} \qquad\text{or}\qquad \frac{N}{2\pi}\int \phi_{(d-2)} \, B^\se_2 \;.
\end{equation}
On the other hand, introducing both couplings (\ref{eqn:ZNegD}) produces a mixed 't~Hooft anomaly: the equation of motion of $b$ implies $N \, d\phi_{(d-2)}=-NB^\sm_{d-1}$, and thus the other coupling is not well-defined but has the anomaly
\begin{align}
\label{eqn:ZNegDanom}
S_\text{anom} = -\frac{N}{2\pi}\int_X B^\sm_{d-1} \, B^\se_2 \;,
\end{align}
where $X$ is a closed bulk $(d{+}1)$-manifold.
We can construct several interesting examples using this theory.

As a first example, take $N=pq$ with gcd$(p,q)>1$. Using the method of Section~\ref{sec: general coupling}, we can couple the theory to an external 2-group symmetry with 0-form part $G$, 1-form part $\bZ_q \subset \bZ_N$, trivial action of $G$ on $\bZ_q$, and Postnikov class $\Bock (\zeta)$ for $\zeta\in H^2(BG,\mathbb{Z}_p)$ (the Bockstein homomorphism is for the short exact sequence $1 \to \bZ_q \to \bZ_N \to \bZ_p \to 1$). Denote the \mbox{2-group} background by a $G$ gauge field $X_1$ and a $\mathbb{Z}_q$ 2-cochain $X_2$. The coupling to the external 2-group symmetry is realized by constructing a $\mathbb{Z}_N$ 2-cocycle $B^\se_2 = B_2^\se[X_1,X_2]$ in terms of $X_2$ and the $\mathbb{Z}_p$ 2-cocycle $X_1^*(\zeta)$ as in (\ref{expression B2 special case}). The coupling has a parameter $\nu\in H^2(BG,\mathbb{Z}_N)$ that shifts $B^\se_2$ by the $\mathbb{Z}_N$ 2-cocycle $X_1^*(\nu)$.
We can also introduce additional coupling parameters $\lambda^{(i)}\in H^{d-1-2i}(BG,\mathbb{Z}_N)$ by turning on the background
\begin{equation}
B^\sm_{d-1}=\sum_{i=0}^{\left[ \frac{d-1}2 \right]} \big( B^\se_2[X_1,X_2] \big)^i \cup X_1^*(\lambda^{(i)})~.
\end{equation}
The 't Hooft anomaly for the external 2-group symmetry is given by substituting the backgrounds $B^\sm_{d-1}$, $B^\se_2$ into the mixed anomaly (\ref{eqn:ZNegDanom}).

Another particularly simple example, that does not involve 2-groups, is the following. Consider the $\mathbb{Z}_N$ gauge theory coupled to an ordinary symmetry $G$, where $G$ is a finite group. We set $B^\sm_{d-1} = 0$ and couple to the background field $X_1$ by setting $B^\se_2 = X_1^*(\eta)$ with $\eta \in H^2(BG, \bZ_N)$. The presence of the background $B^\se_2$ implies that the Wilson line $e^{i\oint b}$, charged under the 1-form symmetry, needs to be attached to a surface with $B^\se_2$ flux. From Section~\ref{1danomsec}, this means that the line carries a projective representation of $G$ described by the cocycle $\eta$. Since there is no 't~Hooft anomaly in this case, we can promote $X_1$ to be a dynamical field. The action of the theory reads
\be
S = \frac{N}{2\pi}\int \phi_{(d-2)} \, \big[ db + X_1^*(\eta) \big] \;.
\ee
The equation of motion of $\phi_{(d-2)}$ no longer implies that $b$ is a $\bZ_N$ gauge field, but rather that $b$ and $X_1$ together constitute the gauge field for the group extension $\wh G$:
\begin{equation}
1\rightarrow \mathbb{Z}_N\rightarrow \wh G\rightarrow G\rightarrow 1 \;,
\end{equation}
where $G$ acts trivially on $\mathbb{Z}_N$ and the extension $\wh G$ is specified by $\eta$. The discussion can be generalized to the case that $G$ acts on $\mathbb{Z}_N$ by a general homomorphism $\rho:G\rightarrow \text{Aut}(\mathbb{Z}_N)$.

Finally, we consider the $\mathbb{Z}_2$ gauge theory coupled to various symmetries (with trivial Postnikov class) in various dimensions:
\begin{itemize}
\item $d=2$. We couple the theory to time-reversal and to an $SO(3)$ symmetry (with gauge field $X_1$) by the backgrounds
\begin{equation}
B^\sm_1 =\pi \, w_1 \;,\qquad\qquad B^\se_2 = \pi \, X_1^*w_2\big( SO(3) \big) \;,
\end{equation}
where $w_1$ is the first Stiefel-Whitney class of the spacetime manifold.
Substituting into (\ref{eqn:ZNegDanom}), we find the mixed anomaly
\begin{equation}
S_\text{anom} = \pi\int_X w_1 \cup X_1^*w_2 \big( SO(3) \big) \;,
\end{equation}
where $X$ is a closed non-orientable bulk 3-manifold with Stiefel-Whitney class $w_1$. Thus the 2d $\mathbb{Z}_2$ gauge theory can be the surface state of such a bulk term. In particular, the presence of the background $B^\se_2$ implies that the line $e^{i\oint b}$, charged under the 1-form symmetry, carries a projective representation of $SO(3)$ \ie{} it has half-integer isospin. Similarly, since the 0-form gauge transformation of $B^\sm_1$ is identified with a gauge transformation of $w_1$, the point operator $e^{i\phi}$ is odd under time-reversal symmetry.

\item $d=3$. We couple the theory to time-reversal symmetry. There are two linearly independent 1-form symmetries, with 2-form backgrounds $B^\sm_2$ and $B^\se_2$. Consider the coupling
\begin{equation}
\label{eqn:ZNegbgthree}
B^\sm_2 = B^\se_2 = \pi \, w_1^2 \;.
\end{equation}
The presence of the 2-form backgrounds implies that the lines $e^{i\oint \phi_{(1)}}$, $e^{i\oint b}$ charged under the two $\mathbb{Z}_2$ 1-form symmetries, respectively, carry projective representations of time-reversal symmetry ${\cal T}$ specified by the cocycle (\ref{eqn:ZNegbgthree}), \ie{} they obey ${\cal T}^2=-1$.
The background-coupled theory with such anyons is also called the $\se T\sm T$ state \cite{Vishwanath:2012tq}.
The mixed anomaly (\ref{eqn:ZNegDanom}) implies that time-reversal symmetry has anomaly
\begin{equation}
S_\text{anom} = \pi\int_X w_1^4 \;,
\end{equation}
where $X$ is a closed non-orientable bulk 4-manifold. This reproduces the time-reversal anomaly of the $\se T\sm T$ state  \cite{Kapustin:2014tfa}.

\item $d=4$. We couple the theory to the spacetime manifold using the following backgrounds for the 1-form and 2-form symmetries:
\begin{equation}
B^\sm_3 = \pi \, w_3 \;,\qquad\qquad B^\se_2 = \pi \, w_2 \;.
\end{equation}
Such a $B^\se_2$ background implies that the line $e^{i\oint b}$ describes a fermionic particle. Similarly, the $B^\sm_3$ background implies that the surface $e^{i\oint \phi_{(2)}}$ charged under the 2-form symmetry, is attached to a volume with $B^\sm_3 = \pi\, w_3$ flux. Such a surface is referred to as a fermionic string \cite{Thorngren:2014pza}.
Then (\ref{eqn:ZNegDanom}) implies that the $\mathbb{Z}_2$ gauge theory with such couplings has gravitational anomaly
\begin{equation}
S_\text{anom} = \pi\int_X w_2 \cup w_3 \;,
\end{equation}
where $X$ is a closed 5-manifold. This reproduces the result in \cite{Kapustin:2014tfa}.
\end{itemize}
A similar discussion for the examples in $d=3$ and $d=4$ can be found in \cite{Kapustin:2014tfa}, but here we propose the new interpretation that these couplings are realized by higher-form symmetries, and the anomalies for ordinary symmetries in those examples come from the anomaly of the higher-form symmetries.



\section*{Acknowledgments} 

We are grateful to M. Barkeshli, D. Ben-Zvi, M. Cheng, T. Dumitrescu, K. Intriligator, N.~Seiberg, S.-H. Shao, K. Ohmori, Y. Tachikawa, J. Wang and E. Witten for enlightening conversations, suggestions or correspondence. F.B. is supported in part by the MIUR-SIR grant RBSI1471GJ ``Quantum Field Theories at Strong Coupling: Exact Computations and Applications'', and by the IBM Einstein Fellowship at the Institute for Advanced Study.  C.C. is supported by the Marvin L. Goldberger Membership at the Institute for Advanced Study, and DOE grant \mbox{DE--SC000%
9988}.  The work of P.-S.H. is supported by the Department of Physics at Princeton University.

\appendix


\section{Singular Cohomology and Group Cohomology}
\label{app: cohomology}

To construct bundles for finite groups, as well as flat bundles for continuous groups, on a manifold $X$ we use simplicial calculus. First we triangulate $X$ with simplices: vertices are 0-simplices, lines (or edges) are 1-simplices, faces are 2-simplices, and so on up to $d$-simplices where $d = \dim X$. We indicates the set of vertices as $\{i\}$ and choose an arbitrary ordering.

The analog of an $n$-form is a simplicial $n$-cochain $f \in C^n(X,\cA)$, which is a function on $n$-simplices taking values in an Abelian group $\cA$ (we use additive notation for Abelian groups). This is a collection of elements $f_{i_0 \dots i_n} \in \cA$ for all $n$-simplices in $X$. There is one element for each $n$-simplex, and so we assume that $\{i_0, \dots, i_n\}$ are ordered: $i_0 < \ldots < i_n$.

The analog of the exterior differential of forms in this context is the simplicial differential $d: C^n(X, \cA) \to C^{n+1}(X,\cA)$ defined as 
\be
(df)_{i_0 \dots i_{n+1}} = \sum_{j=0}^{n+1} (-1)^j f_{i_0 \dots \wh{\imath_j} \dots i_{n+1}}
\ee
for ordered vertices $\{i_j\}$, where the hatted index is omitted. One uses the fact that, given an $n$-simplex, any possible subset of its indices forms a simplex. The differential is nilpotent, $d^2=0$. This allows us to define the groups $Z^n(X,\cA)$ of closed cochains, or cocycles; the groups $B^n(X,\cA)$ of exact cochains, or coboundaries; and then the cohomology groups \mbox{$H^n = Z^n / B^n$}.

Given a (possibly non-Abelian) group $G$, a flat $G$-bundle on $X$ is described by a 1-cocycle $A \in Z^1(X, G)$, \ie{} by elements $A_{ij} \in G$ associated to the edges of the triangulation, such that $A_{ij} A_{jk} = A_{ik}$ for ordered vertices $\{i,j,k\}$ of a face (we use multiplicative notation for non-Abelian groups). Given a group homomorphism $\rho:G \to \Aut(\cA)$, \ie{} an action of $G$ on $\cA$, we can construct a twisted differential $d_A$:
\be
(d_Af)_{i_0 \dots i_{n+1}} = \rho(A_{i_0 i_1})\, f_{i_1 \dots i_{n+1}} + \sum_{j=1}^{n+1} (-1)^j f_{i_0 \dots \wh{\imath_j} \dots i_{n+1}}
\ee
for ordered vertices $\{i_j\}$. This differential is nilpotent as well, $d_A^2 = 0$. Thus it leads to twisted cocycles, twisted coboundaries and twisted cohomology classes.

Let $\cA, \cA', \cA''$ be three Abelian groups with a given bilinear pairing
\be
\langle\;,\,\rangle: \cA \times \cA' \,\to\, \cA'' \;.
\ee
We allow for an action of $G$ on both $\cA$, $\cA'$, $\cA''$ (for simplicity we call both actions $\rho$), and demand that the pairing be covariant with respect to the $G$-action:
\be
\langle \rho_h f, \rho_h g\rangle = \rho_h \langle f, g \rangle \qquad\qquad \forall\; f \in \cA,\; g \in \cA',\; h \in G \;.
\ee
In the main text we will be mainly interested in the case that $\cA'' = \bR/\bZ$, there is no $G$-action on $\cA''$ and thus the pairing is $G$-invariant, but we will be general in this Section. Then we define a (twisted) cup product. Let $f \in C^p(X, \cA)$ and $g \in C^q(X, \cA')$, then the product $\langle f,  \cup\, g \rangle \in C^{p+q}(X, \cA'')$ is defined as
\be
\langle f, \cup\, g \rangle_{i_0 \dots i_{p+q}} = \big\langle f_{i_0 \dots i_p},\, \rho(A_{i_0 i_p}) g_{i_p \dots i_{p+1}} \big\rangle
\ee
for ordered vertices $\{i_j\}$. This product reduces to the standard cup product if there is no $G$-action $\rho$.

The twisted differential satisfies the Liebnitz rule when acting on the cup product, namely
\be
d_A \langle f, \cup\, g \rangle = \langle d_A f, \cup\, g \rangle + (-1)^p \langle f,\cup\, d_A g \rangle \;.
\ee
Notice that in each term, the action $\rho$ contained in the twisted differential is the one that pertains to the corresponding Abelian group. To verify the formula we first compute
\begin{multline}
\langle d_Af, \cup\, g \rangle_{i_0 \dots i_{p+q+1}} = \rho(A_{i_0i_1}) \big\langle f_{i_1 \dots i_{p+1}}, \rho(A_{i_1 i_{p+1}}) g_{i_{p+1} \dots i_{p+q+1}} \big\rangle + {} \\
{} + \sum_{j=1}^{p+1} (-1)^j \langle f_{i_0 \dots \wh{\imath_j} \dots i_{p+1}}, \rho(A_{i_0 i_{p+1}}) g_{i_{p+1} \dots i_{p+q+1}} \big\rangle
\end{multline}
where we used that $A$ is closed and the pairing is covariant, and
\begin{multline}
(-1)^p \, \langle f, \cup\, d_Ag \rangle_{i_0 \dots i_{p+q+1}} = (-1)^p \, \big\langle f_{i_0 \dots i_p}, \rho(A_{i_0 i_{p+1}}) g_{i_{p+1} \dots i_{p+q+1}} \big\rangle + {} \\
+ \sum_{j=p+1}^{p+q+1} (-1)^j \, \big\langle f_{i_0 \dots i_p}, \rho(A_{i_0 i_p}) g_{i_p \dots \wh{\imath_j} \dots i_{p+q+1}} \big\rangle \;.
\end{multline}
Then
\bea
& d_A \langle f, \cup\, g \rangle \big|_{i_0 \dots i_{p+q+1}} = 
\rho(A_{i_0i_1}) \big\langle f_{i_1 \dots i_{p+1}}, \rho(A_{i_1 \dots i_{p+1}}) g_{i_{p+1} \dots i_{p+q+1}} \big\rangle + {} \\
&\quad + \sum_{j=1}^p (-1)^j \big\langle f_{i_0 \dots \wh{\imath_j} \dots i_{p+1}} , \rho(A_{i_0 i_{p+1}}) g_{i_{p+1} \dots i_{p+q+1}} \big\rangle + \sum_{j=p+1}^{p+q+1} (-1)^j \big\langle f_{i_0 \dots i_p}, \rho(A_{i_0i_p}) g_{i_p \dots \wh{\imath_j} \dots i_{p+q+1}} \big\rangle \\
&= \Big[ \langle d_Af, \cup\, g \rangle_{i_0 \dots i_{p+q+1}} - (-1)^{p+1} \big\langle f_{i_0 \dots i_p}, \rho(A_{i_0 i_{p+1}}) g_{i_{p+1} \dots i_{p+q+1}} \big\rangle \Big] + {} \\
&\quad + \Big[ (-1)^p \langle f, \cup \, d_A g \rangle_{i_0 \dots i_{p+q+1}} - (-1)^p \big\langle f_{i_0 \dots i_p}, \rho(A_{i_0 i_{p+1}}) g_{i_{p+1} \dots i_{p+q+1}} \big\rangle \Big] \;.
\eea
The two extra terms after the last equality cancel out.

A particularly interesting case is when the second Abelian group is $\wh\cA$, the Pontryagin dual to $\cA$, namely the Abelian group of linear functions $\lambda: \cA \to \bR/\bZ$. Then there is a natural pairing
\bea
\langle \;,\,\rangle: \wh \cA \times \cA &\to \bR/\bZ \\ \langle \lambda, f \rangle &\mapsto \lambda(f) \;.
\eea
Given an action $\rho$ of $G$ on $\cA$, we can define an action on $\wh\cA$ such that $\langle \;,\,\rangle$ is invariant:
\be
\rho_g \lambda \quad\text{ such that }\quad (\rho_g \lambda)(f) = \lambda(\rho_g^{-1} f) \qquad\forall\; g \in G,\; f \in \cA \;.
\ee
It follows $\langle \rho_g \lambda, \rho_g a\rangle = (\rho_g\lambda)(\rho_g a) = \lambda (\rho_g^{-1} \rho_g a) = \langle \lambda, a \rangle$. Therefore we can construct a twisted cup product and twisted differentials, that satisfy the Liebnitz rule.

\subsection{Group Cohomology}
\label{app: group cohomology}

The group cohomology of $G$ consists of the cohomology classes $H^n_\rho(BG, \cA)$, where $BG$ is the classifying (or Eilenberg-Mac Lane) space of $G$ \cite{Eilenberg:1953}. When $\cA$ is finite, one can safely take the singular cohomology of $BG$. When $\cA$ is continuous, on the other hand, the topology of $\cA$ should be taken into account, and the correct language to use is sheaf cohomology. In particular, when we consider $\cA = \bR/\bZ \cong U(1)$ we use the standard (as opposed to the discrete) topology. This guarantees that
\be
H^n(BG, \bR/\bZ) \cong H^{n+1}(BG, \bZ)
\ee
in all cases (see \eg{} \cite{Segal:1970, Brylinski:2000}).

When $G$ is finite, the cohomology groups have an algebraic description \cite{Milnor:1956} (see also \cite{Hatcher:book}) as cohomology groups of functions
\be
f: G^n \to \cA \;,
\ee
``normalized'' such that $f(\dots, \unit, \dots) = 0$. The twisted differential is defined as
\bea
(d_\rho f)(g_1, \dots, g_{n+1}) &= \rho_{g_1} f(g_2, \dots, g_{n+1}) + \sum_{j=1}^n (-1)^j f(g_1, \dots, g_j g_{j+1}, \dots, g_{n+1}) \\
&\quad + (-1)^{n+1} f(g_1, \dots, g_n) \;.
\eea
Such differential is nilpotent, $d_\rho^2 = 0$, and this allows us to construct the (twisted) cohomology groups $H^n_\rho$. For finite groups, the notation $H^n_\rho(G,\cA) \equiv H^n_\rho(BG,\cA)$ is often used in the literature.

Given Abelian groups $\cA, \cA', \cA''$ and a pairing $\langle\;,\,\rangle: \cA \times \cA' \to \cA''$, covariant with respect to the $G$-action $\rho$, we construct the twisted cup product of $f\in C^p(BG, \cA)$ and $b \in C^q(BG, \cA')$:
\be
\langle f, \cup\, b\rangle (g_1, \dots, g_{p+q}) = \big\langle f(g_1, \dots, g_p), \rho_{g_1 \cdots g_p} b(g_{p+1}, \dots, g_{p+q}) \big\rangle
\ee
which is an element of $C^{p+q}(BG, \cA'')$. The twisted differential satisfies the Liebnitz rule when acting on such a product:
\be
d_\rho \langle f, \cup\, b \rangle = \big\langle d_\rho f, \cup\, b \big\rangle + (-1)^p \big\langle f, \cup\, d_\rho b \big\rangle \;.
\ee
The proof is essentially the same as before.

In fact, if we regard the cocycle $A \in Z^2(X,G)$ as a homotopy class of maps $A: X \to BG$, we realize that
\be
A^* d_\rho = d_A A^* \;,
\ee
where the pull-back $A^*$ corresponds to substituting $g_j \to A_{i_{j-1} i_j}$ for all $n$-simplices $\{i_j\}$.

\subsection{Steenrod's Cup Products}

Following Steenrod \cite{Steenrod:1947}, we can introduce higher generalizations of the cup product. As before, we let $\cA$, $\cA'$, $\cA''$ be Abelian groups, $\langle\;,\,\rangle: \cA \times \cA' \to \cA''$ a bilinear pairing, and $\rho:G \to \Aut(\cA^{(0,1,2)})$ three actions of $G$ on $\cA,\cA',\cA''$ (we use the same symbol $\rho$ for both). In order not to clutter our formulas, here we will keep the twist and the pairing implicit.

As discussed before, the differential satisfies the Liebnitz rule when acting on the cup product ($f \in C^p(X,\cA)$ and $g \in C^q(X,\cA')$):
\be
d \, ( f\cup g ) = df \cup g + (-1)^p f \cup dg \;.
\ee
On the other hand, as opposed to the case of differential forms, the cup product is not graded-commutative:
\be
\label{cup 1}
f \cup g - (-1)^{pq} g \cup f = (-1)^{p+q-1} \big[ d \,(f \cup_1 g) - df \cup_1 g - (-1)^p f \cup_1 dg \big] \;.
\ee
In the second term on the left we have made a slight abuse of notation (unless $\cA = \cA'$ and the pairing is symmetric): when we write $g \cup f$ we still pair $f$ with $g$ in the correct order; what we commute is the assignment of indices. The object $\cup_1$ on the right is a (twisted) bilinear cup product from $C^p(X,\cA) \times C^q(X,\cA')$ to $C^{p+q-1}(X,\cA'')$ defined as
\be
(f\cup_1 g)_{i_0 \dots i_{p+q-1}} = \sum_{j=0}^{p-1} (-1)^{(p-j)(q+1)} \; f_{i_0 \dots i_j i_{j+q} \dots i_{p+q-1}} \; \rho(A_{i_j i_{j+q}}) g_{i_j \dots i_{j+q}}
\ee
for ordered $\{i_j\}$. This is the pull-back of its version in group cohomology:
\begin{align}
& (f \cup_1 b)(g_1,\dots, g_{p+q-1}) = \\
&\quad \sum_{j=0}^{p-1} (-1)^{(p-j)(q+1)} \; f(\underbrace{g_1,\ldots, g_j}_j,\, \underbrace{g_{j+1}\cdots g_{j+q}}_1,\, \underbrace{g_{j+q+1}, \ldots, g_{p+q-1}}_{p-j-1}) \; \rho_{g_1\cdots g_j} b(g_{j+1}, \dots, g_{j+q}) \;. \nn
\end{align}
To facilitate the reading, we have indicated the number of entries with braces.

The cup product $\cup_1$ does not satisfies the Liebnitz rule, and equation (\ref{cup 1}) describes its failure. It is also not graded commutative:
\be
f \cup_1 g + (-1)^{pq} g \cup_1 f = (-1)^{p+q} \big[ d\,(f\cup_2g) - df \cup_2 g - (-1)^p f \cup_2 dg \big] \;,
\ee
where $\cup_2$ is a bilinear cup product from $C^p(X,\cA) \times C^q(X,\cA')$ to $C^{p+q-2}(X,\cA'')$ (its definition can be found in \cite{Steenrod:1947} or \cite{Kapustin:2014gua}). The structure continues, the general formula being
\be
f \cup_i g - (-1)^{pq - i} g \cup_i f = (-1)^{p+q- i -1} \big[ d \, (f \cup_{i+1} g) - df \cup_{i+1} g - (-1)^p f \cup_{i+1} dg \big]
\ee
for $i\geq 0$. The formula can be rewritten as a modification of the Liebnitz rule:
\be
d \, (f \cup_i g) = df \cup_i g + (-1)^p f \cup_i dg + (-1)^{p+q-i} f \cup_{i-1} g + (-1)^{pq+p+q} g \cup_{i-1} f
\ee
for $i\geq 1$. At some point the structure stops, because $f\cup_i g$ vanishes if $i> \min(p,q)$.


\section{Bockstein Homomorphism}
\label{app: Bockstein}

Suppose we have a short exact sequence of Abelian groups
\be
1 \;\to\; \cB \;\xrightarrow{i}\; \cA \;\xrightarrow{p}\; \cC \equiv \cA/\cB \;\to\; 1 \;,
\ee
where $i$ is inclusion and $p$ is projection to the quotient. If we consider cohomology groups with values in those Abelian groups, the short exact sequence induces a long exact sequence
\be
\ldots \;\to\; H^n(X,\cB) \;\xrightarrow{i}\; H^n(X, \cA) \;\xrightarrow{p}\; H^n(X,\cA/\cB) \;\xrightarrow \Bock\; H^{n+1}(X,\cB) \;\to\; \ldots \;.
\ee
The map on the right is called the Bockstein homomorphism (see \eg{} \cite{Hatcher:book}),
\be
\Bock:\; H^n(X,\cA/\cB) \;\to\; H^{n+1}(X,\cB) \;.
\ee
It is the obstruction to lifting a class in $H^n(X,\cA/\cB)$ to a class in $H^n(X,\cA)$, because the image of $p$ is the kernel of $\Bock$.

The map is constructed as follows. Take a representative $\omega \in Z^n(X,\cA/\cB)$ of the class $[\omega]$. Lift it to an element $\wt\omega \in C^n(X,\cA)$, namely
\be
p(\wt\omega) = \omega \;.
\ee
In general $\wt\omega$ will not be closed, but $p(d\wt\omega)=0$. Because of the short exact sequence, $d\wt\omega$ is in the image of $i$ and thus
\be
d\wt\omega = i(\nu) \qquad\text{for some } \nu \in C^{n+1}(X,\cB) \;.
\ee
Since $i$ is injective, $d\nu = 0$ and thus $\nu$ defines a class $[\nu] \in H^{n+1}(X,\cB)$. One can check that this class only depends on the class $[\omega]$ and not on the other choices we made.


\section{Pontryagin Square and Affine Generalization}
\label{app: Pontryagin}

In this Appendix we give more details on the construction of the four-dimensional action that describes the 1-form anomaly in 3D, and in particular on the Pontryagin square and its affine generalization.

Consider first the case of a pure 1-form symmetry $\cA$. The background it couples to is described by $B \in H^2(X,\cA)$, which can be thought of as a homotopy class of maps $B: X \to B^2\cA$. Four-dimensional actions \cite{Kapustin:2013qsa} are written in terms of $\cL \in H^4(B^2\cA, \bR/\bZ)$ as
\be
\label{4D action for 1-form symm}
S = 2\pi \int_X B^* \cL \;.
\ee
We can rewrite the action in a more convenient way. First we use that \cite{Eilenberg:1954, Kapustin:2013qsa}
\be
\label{equality for H4 B2A}
H^4(B^2\cA, \bR/\bZ) = \Hom\big( H_4(B^2\cA, \bZ), \bR/\bZ \big) = \Hom\big( \Gamma(\cA), \bR/\bZ \big) = \wh{\Gamma(\cA)}
\ee
where $\Gamma(\cA)$ is the universal quadratic group of $\cA$. We review the definition of this group in Section \ref{app: quadratic group}. Thus we can think of $\cL$ as an element $\tilde q \in \wh{\Gamma(\cA)}$. Then we use a canonical map $\fP: H^2(X,\cA) \to H^4\big( X, \Gamma(\cA)\big)$, called the Pontryagin square \cite{Whitehead:1949} and reviewed in Section \ref{app: pontryagin square}, to rewrite the action as
\be
S = 2\pi \int_X \tilde q(\fP B) \;.
\ee
This is the notation used in the main text.

In the case of a 2-group symmetry, $B \in C^2(X,\cA)$ is not closed. Rather, it has fixed differential $d_AB = A^*\beta$ (for fixed 0-form background $A$). It is still the case that gauge transformations shift it by a coboundary, therefore $B$ defines what we call an affine cohomology class. In this case we need an extension of the Pontryagin square $\fP$ to affine cohomology classes, that we will present in Section \ref{sec: affine Pontryagin}.

\subsection{Universal Quadratic Group}
\label{app: quadratic group}

Let $\cA$, $\cB$ be any two Abelian groups. A map $q:\cA \to \cB$ is a quadratic function if%
\footnote{In particular $\langle x, x \rangle_q = 2q(x)$. This does not fix $q(x)$ completely, indeed $q$ is also called a quadratic refinement. From $0 = \langle 0,0\rangle_q = -q(0)$ one finds $q(0)=0$. From $\langle x, x\rangle_q = - \langle x, -x \rangle_q$ and writing both sides in terms of $q$, one finds $q(2x) = 4q(x)$. Then from $\langle x, (n-1)x\rangle_q = (n-1) \langle x, x\rangle_q$, writing both sides in terms of $q$ and using induction, one finds $q(nx) = n^2 q(x)$.}
\begin{enumerate}
\item $q(x) = q(-x)$;
\item $\langle x, y \rangle_q \equiv q(x+y) - q(x) - q(y)$ is a bilinear form $\langle \;,\; \rangle: \cA \times \cA \to \cB$.
\end{enumerate}
It turns out \cite{Whitehead:1950} that there exist a unique Abelian group $\Gamma(\cA)$ and a (not unique) quadratic function $\gamma: \cA \to \Gamma(\cA)$ such that, for any other Abelian group $\cB$ and quadratic function $q:\cA \to \cB$, we can write
\be
q = \tilde q \circ \gamma \qquad\text{ for some unique }\qquad \tilde q \in \Hom\big( \Gamma(\cA), \cB \big) \;.
\ee
Consider the following cases.
\begin{itemize}
\item If $\cA = \bZ_r$ with $r$ even, then $\Gamma(\cA) = \bZ_{2r}$ and $\gamma(1) = 1$. This fixes $\gamma(x) = x^2$ and thus $q(x) = \tilde q(x^2)$. We find the relation
\be
\langle x, y \rangle_q = q(x+y) - q(x) - q(y) = \tilde q\big( (x+y)^2 - x^2 - y^2 \big) = \tilde q(2xy) \;.
\ee
\item If $\cA = \bZ_r$ with $r$ odd, then $\Gamma(\cA) = \bZ_r$ and $\gamma(1) = 1$. The same relation as above holds.
\item A general finite Abelian group is $\cA = \oplus_i A_i$ where each $A_i$ is a cyclic group, then%
\footnote{In this notation $\bZ_p \oplus \bZ_q$ is the group of pairs $(a,b)$, usually denoted as $\bZ_p \times \bZ_q$, while $\bZ_p \otimes \bZ_q$ is the group constructed out of the elements $ab$ and which turns out to be equal to $\bZ_{\gcd(p,q)}$.}
\be
\Gamma(\oplus_i A_i) = \bigoplus_i \Gamma(A_i) \oplus \bigoplus_{i<j} A_i \otimes A_j \;.
\ee
If we call $e_i$ the generator of $A_i$ and $e_{ij}$ the generator of $A_i \otimes A_j$, we have
\be
\gamma(e_i) = e_i \;,\qquad \gamma(e_i + e_j) = e_i + e_j + e_{ij} \;.
\ee
\end{itemize}
We will consider finite groups.

\subsection{Pontryagin Square}
\label{app: pontryagin square}

The Pontryagin square \cite{Whitehead:1949} is a map from $H^n(X,\cA)$ to $H^{2n}\big( X, \Gamma(\cA)\big)$. Let us give an explicit construction of the Pontryagin square, specializing to the case $n=2$ which is relevant to this paper, in other words we construct a representative of the class in $H^4\big(X, \Gamma(\cA)\big)$.

Consider first the case that $\cA = \bZ_r$ with $r$ odd. We let the 2-cocycle $f\in Z^2(X, \bZ_r)$ be a representative of the cohomology class $[f]$. Then a representative of its Pontryagin square, that with some abuse of notation we indicate as $\fP f$, is simply $f\cup f \in Z^4(X, \bZ_r)$.

Next consider the case that $\cA = \bZ_r$ with $r$ even, and let $f \in Z^2(X,\bZ_r)$ be a representative. We take an integer lift of $f$, namely $\tilde f \in C^2(X, \bZ)$ such that $\tilde f = f \pmod{r}$. Such a lift will satisfy $d\tilde f = ru$ for some $u \in B^3(X,\bZ)$. We construct
\be
\fP f \,\equiv\, \tilde f \cup \tilde f - \tilde f \cup_1 d\tilde f \;.
\ee
We ask whether this is well-defined modulo $2r$. Suppose we chose another lift $\tilde f' = \tilde f + rw$ for some $w \in C^2(X, \bZ)$. Then, discarding multiples of $2r$, we find
\be
\Big( \tilde f' \cup \tilde f' - \tilde f' \cup_1 d\tilde f' \Big) - \Big( \tilde f \cup \tilde f - \tilde f \cup_1 d\tilde f \Big) = - r \, d ( \tilde f \cup_1 w ) \pmod{2r} \;.
\ee
Thus $\fP f$ is not uniquely defined as a cochain, but it is uniquely defined modulo exact terms. The action of $\tilde q \in \Hom(\bZ_{2r}, \bR/\bZ)$ does not change this fact, since $\tilde q(d\omega) = d \, \tilde q(\omega)$.

Then we verify that $\fP f$ is closed modulo $2r$:
\be
d \Big( \tilde f \cup \tilde f - \tilde f \cup_1 d\tilde f \Big) = 2r\, u \cup \tilde f - r^2 \, u \cup_1 u \;.
\ee
Thus $d\, \fP f = 0 \pmod{2r}$ and $d\, \tilde q(\fP f) = 0$. We see that $\fP f$ defines an element of $H^4(X, \bZ_{2r})$, and also a closed representative although the specific representative depends on the lift. Besides, $\tilde q(\fP f)$ defines an element of $H^4(X, \bR/\bZ)$.

Finally, we check that the cohomology class $[\fP f]$ only depends on the class $[f]$: if we choose a different representative $\tilde f'' = \tilde f + dv$ for some $v \in C^1(X,\bZ)$, we find
\be
\Big( \tilde f'' \cup \tilde f'' - \tilde f'' \cup_1 d\tilde f'' \Big) - \Big( \tilde f \cup \tilde f - \tilde f \cup_1 d\tilde f \Big) = d \big( 2 \tilde f \cup v + v \cup dv - dv \cup_1 \tilde f \big) - 2r\, u \cup v \;.
\ee
We can construct the integral $\int_X \fP f \in \bZ_{2r}$, or more importantly the action
\be
\int_X \tilde q(\fP f) \,\in\, \bR/\bZ \;.
\ee
Notice that, in this case, $\tilde q$ is multiplication by $1/2r$ times an integer. The integral is a well-defined function of $[f] \in H^2(X, \bZ_r)$.

For general $\cA = \oplus_i A_i$, we decompose $f = \sum_i f_i$ where each component is valued in $A_i$, then $\fP f = \sum_i \fP f_i + \sum_{i<j} f_i \cup f_j$. Notice that in the second summation $f_i$ and $f_j$ commute in cohomology.

\bigskip

We can construct other similar objects using the higher cup products. For instance, let $\cA = \bZ_r$ with $r$ even and consider $\Omega \in Z^3(X,\bZ_r)$. We let $\tilde\Omega \in C^3(X,\bZ)$ be an integer lift, namely $\tilde\Omega = \Omega \pmod{r}$ and then $d\tilde\Omega = r u$ for some $u\in B^4(X,\bZ)$. We construct
\be
\fP_1 \Omega \,\equiv\, \tilde\Omega \cup_1 \tilde\Omega - \tilde\Omega \cup_2 d\tilde\Omega \;.
\ee
Choosing a different lift $\tilde\Omega' = \tilde\Omega + r w$ for some $w \in C^3(X,\bZ)$, we find
\be
\Big( \tilde\Omega' \cup_1 \tilde\Omega' - \tilde\Omega' \cup_2 d\tilde\Omega' \Big) - \Big( \tilde\Omega \cup_1 \tilde\Omega - \tilde\Omega \cup_2 d\tilde\Omega \Big) = r\, d( \tilde\Omega \cup_2 w) \pmod{2r} \;,
\ee
therefore $\fP_1 \Omega$ is well-defined modulo $2r$ and modulo exact terms. On the other hand it is not quite closed:
\be
d \Big( \tilde\Omega \cup_1 \tilde\Omega - \tilde\Omega \cup_2 d\tilde\Omega \Big) = -2 \, \tilde\Omega \cup \tilde\Omega + 2r\, u \cup_1 \tilde\Omega - r^2 \, u \cup_2 u \;.
\ee
Notice that the first term is well-defined modulo $2r$. Thus
\be
d\, \fP_1 \Omega = -2\, \Omega \cup \Omega \pmod{2r}
\ee
and it is a well-defined element of an affine cohomology class defined modulo exact terms.

\subsection{Affine Pontryagin Square}
\label{sec: affine Pontryagin}

Now we would like to construct an object similar to the Pontryagin square, but for $df = \Omega$ namely in the case that $f$ is not closed. Again, we focus on the two cases that $\cA = \bZ_r$ with $r$ odd or even.

If $f\in C^2(X, \bZ_r)$ and $\Omega \in B^3(X, \bZ_r)$ with $r$ odd, we construct $f\cup f - f \cup_1\Omega$. It satisfies a shifted cocycle condition
\be
d\Big( f \cup f - f \cup_1 \Omega \Big) = 2\, \Omega \cup f - \Omega \cup_1 \Omega \;.
\ee
This condition will be very important later on.

More involved is that case that $r$ is even. Let $f \in C^2(X,\bZ_r)$ and $\Omega \in B^3(X,\bZ_r)$. We take lifts $\tilde f \in C^2(X, \bZ)$, $\tilde\Omega \in C^3(X,\bZ)$ with $\tilde f = f \pmod{r}$, $\tilde \Omega = \Omega \pmod{r}$ and
\be
d\tilde f = \tilde\Omega + ru
\ee
for some $u \in C^3(X,\bZ)$. We construct
\be
\fP f \,\equiv\, \tilde f \cup \tilde f - \tilde f \cup_1 d\tilde f + \tilde\Omega \cup_2(d\tilde f - \tilde\Omega) \;.
\ee
We verify that this is a well-defined quantity in $C^4(X, \bZ_{2r})$ that does not depend on the particular choice of the lift $\tilde f$, as long as we mod out by exact terms: setting $\tilde f' = \tilde f + rw$ we find
\begin{multline}
\Big( \tilde f' \cup \tilde f' - \tilde f' \cup_1 d\tilde f' + \tilde\Omega \cup_2(d\tilde f' - \tilde\Omega) \Big) - \Big( \tilde f \cup \tilde f - \tilde f \cup_1 d\tilde f + \tilde\Omega \cup_2(d\tilde f - \tilde\Omega) \Big) = \\
= -r\, d\big( \tilde f \cup_1 w + d\tilde f \cup_2 w \big) \pmod{2r} \;.
\end{multline}
In other words, $\fP f$ is a well-defined element of an affine cohomology class with values in $\bZ_{2r}$.%
\footnote{Notice that there is a dependence on the lift $\tilde \Omega$: setting $\tilde\Omega' = \tilde\Omega + r\chi$ we find
\be
\Big( \tilde f \cup \tilde f - \tilde f \cup_1 d\tilde f + \tilde\Omega' \cup_2(d\tilde f - \tilde\Omega') \Big) - \Big( \tilde f \cup \tilde f - \tilde f \cup_1 d\tilde f + \tilde\Omega \cup_2(d\tilde f - \tilde\Omega) \Big) = - r\, \tilde\Omega \cup_2 \chi \pmod{2r} \;.
\ee
This has to do with the fact that the differential of $\fP f$ contains $- \fP_1\Omega$ and the latter shifts by an exact term, see (\ref{shifted cocycle condition of Pf}). In the physical context of 't~Hooft anomalies, such a dependence is reabsorbed into the 0-form anomaly $\omega$ as discussed in Section \ref{sec: 3D anomaly}.}

We compute the differential of $\fP f$:
\begin{multline}
d\Big( \tilde f \cup \tilde f - \tilde f \cup_1 d\tilde f + \tilde\Omega \cup_2(d\tilde f - \tilde\Omega) \Big) = \\
= 2 \, \tilde\Omega \cup \tilde f - \tilde\Omega \cup_1 \tilde\Omega + \tilde\Omega \cup_2 d\tilde\Omega + \big[ 2r\, u \cup \tilde f - r^2\, u \cup_1 u - 2r\, u \cup_1 \tilde\Omega - r^2\, du\cup_2 u \big] \;.
\end{multline}
Thus we can write
\be
\label{shifted cocycle condition of Pf}
d\, \fP f = 2\, \Omega \cup f - \fP_1 \Omega \;,
\ee
and both sides are defined in $\bZ_{2r}$.

Finally, if we shift $f$ by an exact term, $f'' = f + dv$, we find
\begin{multline}
\Big( \tilde f'' \cup \tilde f'' - \tilde f'' \cup_1 d\tilde f'' + \tilde\Omega \cup_2(d\tilde f'' - \tilde\Omega) \Big) - \Big( \tilde f \cup \tilde f - \tilde f \cup_1 d\tilde f + \tilde\Omega \cup_2(d\tilde f - \tilde\Omega) \Big) = \\
= d\big( 2 \tilde f \cup v + v\cup dv - dv \cup_1 \tilde f \big) -2r\, u \cup v - 2\, \tilde\Omega \cup v \;.
\end{multline}
Modulo exact terms and modulo $2r$, under $f \to f + dv$ we have
\be
\fP f \,\to\, \fP f - 2 \, \Omega \cup v \;.
\ee
In other words, the affine Pontryagin square is not closed and is not gauge invariant under $f \to f + dv$. Both problems are cured by writing the full anomaly action.%
\footnote{Notice that $\fP f$ is also not invariant under a change of representative $\Omega \to \Omega + d\theta$. This is not surprise, as in general all our expressions depend on the representative $\beta$ of the class $[\beta]$. Such a dependence, though, is reabsorbed in a shift of $B$ (here $f$).}

\subsection{Full Anomaly}

The 3d anomaly discussed in Section \ref{sec: 3D anomaly} is controlled by the following integral, here written in the variables used in this Appendix:
\be
\int_X \bigg[ \tilde q\big( \fP f \big) - \langle \lambda, \cup f \rangle + \omega \bigg]
\ee
where:
\bea
f &\in C^2(X,\cA) \qquad\qquad\qquad\qquad\qquad & \lambda &\in C^2(X,\wh\cA) \\
df &= \Omega \in B^3(X,\cA) & d\lambda &= \langle \Omega, \star \rangle_q \\
\tilde q &\in \Hom\big( \Gamma(\cA), \bR/\bZ \big) = \wh{\Gamma(\cA)} & \omega &\in C^4(X, \bR/\bZ) \\
\langle \,\;,\; \rangle_q &: \cA \times \cA \to \bR/\bZ \text{ bilinear} & d\omega &= \langle \lambda, \cup\Omega\rangle + \tilde q(\fP_1\Omega) \;.
\eea
The integral takes values in $\bR/\bZ$.

First, we check that the integrand is closed:
\bea
d\Big[ \tilde q\big( \fP f \big) - \langle \lambda, \cup f \rangle + \omega \Big] &= \tilde q\big( 2\, \Omega \cup f - \fP_1\Omega\big) - \langle d\lambda, \cup f\rangle - \langle \lambda, \cup\Omega\rangle + d\omega \\
&= \langle \Omega, \cup f\rangle_q - \tilde q(\fP_1\Omega) - \langle \Omega, \cup f\rangle_q - \langle \lambda, \cup\Omega\rangle + d\omega = 0 \;.
\eea
Therefore the action is topological (invariant under change of triangulation). Next we check that the action is gauge invariant under $f \to f + dv$, namely that the integrand changes by exact terms under the gauge variation:
\be
\Delta\Big[ \tilde q(\fP f) - \langle \lambda, \cup f \rangle \Big] = - \langle \Omega, \cup v \rangle_q - \langle \lambda, \cup dv \rangle + \text{exact} = \text{exact} \;.
\ee
We conclude that it is a good action.

\section{Trivial Anyon Permutation Symmetry in 3d TQFT}
\label{app:trivialpermutation}

It has been conjectured in \cite{Barkeshli:2014cna} that for ordinary global symmetry that does not permute the anyons in 3d TQFT, the symmetry action on the three-punctured spheres (which form a basis of the Hilbert space) must be trivial.
In this Appendix we will prove the simple case for 3d TQFT that has fusion multiplicity $N^x_{y,z} \in \{0,1\}$, namely the symmetry action takes the following form on three-punctured spheres
\begin{equation}\label{naturalisofac}
U(x,y,z)|x,y;z\rangle=\gamma(x) \, \gamma(y) \, \gamma(z)^{-1} \, |x,y;z\rangle~,
\end{equation}
where $x,y,z$ are anyons in the TQFT such that $N^{x,y}_z\neq 0$, and $\gamma$ is some $U(1)$ function.

The symmetry action on the spheres must preserve the $F$- and $R$-symbols of the 3d TQFT. For fusion multiplicity either zero or one the condition is: \cite{Barkeshli:2014cna}
\begin{align}
&U(y,x;z)R^{x,y}_zU(x,y,z)^{-1} = R^{x',y'}_{z'}\cr
&U(x,y;u)U(u,z;w)(F^{x,y,z}_w)_{u,v} U(y,z;v)^{-1}U(x,v;w)^{-1} = (F^{x',y',z'}_{w'})_{u',v'}~,
\end{align}
where the lower case letters are anyons, and the anyon with prime denotes the permutation acting on the anyon.
For symmetry that does not permute the anyons they become
\begin{equation}\label{eqn:condUnonperm}
U(y,x;z) = U(x,y,z),\quad U(x,y;u)U(u,z;w)= U(y,z;v)U(x,v;w)~.
\end{equation}
Applying (\ref{eqn:condUnonperm}) for the invariance of $(F^{x,y,z}_{w})_{u_i,v_k},(F^{x,y,z}_{w})_{u_j,v_k},(F^{x,y,z}_{w})_{u_j,v_l},\cdots$ with all possible anyons $u_i,v_j$ appear in the fusion channels generalizes the second equation to%
\footnote{The fusion rules are associative:
\begin{align}
\sum\nolimits_u N^{x,y}_u N^{u,z}_w = \sum\nolimits_v N^{y,z}_v N^{v,x}_w~,
\end{align}
thus if there exists an $u=u_i$ such that the left-hand-side is nonzero, there must also exist some $v=v_i$ to make the right-hand-side  nonzero, and vice versa. 
}
\begin{align}\label{eqn:condUnonpermp}
&U(x,y;u_i)U(u_i,z;w) = U(x,y;u_j)U(u_j,z;w)=\cdots\cr
&=U(y,z;v_k)U(x,v_k;w)=U(y,z;v_l)U(x,v_l;w)=\cdots~.
\end{align}

\medskip

We begin by showing $U(x,y;u)$ takes the following form 
\begin{equation}\label{eqn:auxzero}
U(x,y;u)=f(x,y) h(u)^{-1}~.
\end{equation}
Note the equation holds automatically if the fusion channel of $x,y$ is unique, {\it e.g.} when at least one of $x,y$ is Abelian.
\begin{itemize}
\item If there does not exist another pair $x',y'\neq x,y$ that can fuse into $u$ {\it i.e.} $u$ appears only in the fusion channel of $x,y$, then the equation (\ref{eqn:auxzero}) holds with $f=1$ and $h(u)=U(x,y;u)^{-1}$.

\item If $u$ appears in the fusion channel of the pairs $x,y$ and $x',y'$ but there does not exist another $u'$ that appear in both fusion channels, then (\ref{eqn:auxzero}) is still true by choosing some $(x,y)=(x_*,y_*)$ and taking $f(x_*,y_*)=1$, $h(u)=f(x_*,y_*)U(x_*,y_*;u)^{-1}$. For any other pair $(x',y')$ we can then take $f(x',y')=f(x_*,y_*)U(x_*,y_*;u)^{-1}U(x',y';u)$.

\item
To see the possible problem if both $u,u'$ appear in the fusion channels of more than one pairs, say $x,y$ and $x',y'$, 
note demanding $U(x,y,u')$ and $U(x',y';u')$ to be of the same form (\ref{eqn:auxzero}) for an additional $u'$ requires the following constraint
\begin{equation}
U(x',y';u)U(x,y;u)^{-1}=U(x',y';u')U(x,y;u')^{-1}~,
\end{equation}
which might not be true if $U$ were an arbitrary function.
Instead, we will explore the constraints (\ref{eqn:condUnonperm}) that $U$ is the symmetry action on the three-puncture spheres.

Consider the fusion tree of fusing $x',y',\overline{x}$ into $y$ by the fusion channels $u,v$:
\begin{align}
&(x'y')\overline{x}\supset u\overline{x}\supset y\cr
&x'(y'\overline{x})\supset x' v\supset y~,
\end{align}
where there exists $v$ satisfies $v\cdot x\supset y'$ and $v\cdot x'\supset y$ by the associativity of the fusion rules.
The condition (\ref{eqn:condUnonperm}) implies
\begin{equation}\label{eqn:auxone}
U(x',y';u)U(u,\overline{x};y)=U(y',\overline{x};v)U(x',v;y)~.
\end{equation}
Next, consider the fusion tree of fusing $y,x,\overline{x}$ into $y$ by the fusion channels $u,1$:
\begin{align}\label{eqn:treetwo}
&(yx)\overline{x}\supset  u\overline{x}\supset y\cr
&y(x\overline{x})\supset  y\cdot 1=y~.
\end{align}
The condition (\ref{eqn:condUnonperm}) and $U(1,y;y)=1$ implies
\begin{equation}\label{eqn:auxtwo}
U(x,y;u)U(u,\overline{x};y)=U(x,\overline{x};1)~.
\end{equation}
Comparing the equations (\ref{eqn:auxone}) and (\ref{eqn:auxtwo}) gives
\begin{equation}
U(x',y';u)U(x,y;u)^{-1}=U(y',\overline{x};v)U(x',v;y)U(x,\overline{x};1)^{-1}~.
\end{equation}
The right hand side is independent of which $u$ appear in both the fusion channels of $x,y$ and $x',y'$, 
and therefore from the left hand side we again establish (\ref{eqn:auxzero}).

\end{itemize}

\medskip

Next, applying the condition (\ref{eqn:condUnonpermp}) on the fusion trees (\ref{eqn:treetwo}) for $x\cdot y=u+u'+\cdots$ gives
\begin{equation}
U(x,y;u)U(u,\overline{x};y)=U(x,y;u')U(u',\overline{x};y)~.
\end{equation}
Multiplying both sides by $U(u,\overline{x};y)^{-1}U(x,y;u')^{-1}$ and substituting (\ref{eqn:auxzero}) gives
\begin{equation}
h(u)h(u')^{-1}=U(u',\overline{x};y)U(u,\overline{x};y)^{-1}~.
\end{equation}
The left hand side is independent of $x,y$, thus we find $U$ also has the form
\begin{equation}\label{eqn:auxfour}
U(u,\overline{x};y)=p(u)g(\overline{x},y)~.
\end{equation}
Note if the fusion of $x,y$ gives a unique anyon $u$ so $u'=u$, the above equation holds automatically.
If there is a unique pair $x,y$ that has both $u,u'$ in their fusion channel, then by a similar discussion from the first two points under (\ref{eqn:auxzero}) we again obtain (\ref{eqn:auxfour}).

The equations (\ref{eqn:auxzero}),(\ref{eqn:auxfour}) and (\ref{eqn:condUnonperm}) together imply
\begin{equation}
U(u,\overline{x};y) = p(u)p(\overline{x})q(y)^{-1}~
\end{equation}
for some function $q$.
Substituting it into (\ref{eqn:auxtwo}) gives $q(y)=l\, p(y)$ with some constant $l^2=1$ independent of the anyons $y$.
By rescaling $p(b)=l\,\gamma(b)$ and renaming, we can rewrite it as
\begin{equation}
U(r,s;t) = \gamma(r)\gamma(s)\gamma(t)^{-1}~,
\end{equation}
for any three-punctured spheres with anyons $r,s$ fusing into $t$.

\section{'t Hooft Anomaly and Hall Conductivity}
\label{sec:hall}

In this appendix we use the method of Section~\ref{sec: general coupling} to show that for 3d TQFTs coupled to a $U(1)$ symmetry that does not permute the anyons, the fractional Hall conductivity (\ie{} the response to varying the $U(1)$ background gauge field) can be computed from the 't~Hooft anomaly of the 1-form symmetry. This provides a method for computing the fractional $U(1)$ Hall conductivity in a TQFT without relying on a dynamical Abelian gauge theory description.

Denote the $U(1)$ background gauge field by $A$. From the discussion in Section~\ref{sec: general coupling},
the theory can couple to the $U(1)$ symmetry by activating a background $B_2$ for some $\mathbb{Z}_N$ 1-form symmetry, with $B_2$ expressed in terms of $A$ (the normalization is $\oint B_2\in \frac{2\pi}{N}\mathbb{Z}$):
\begin{equation}\label{eqn:hallbg}
B_2 = \frac{dA}{N}~.
\end{equation}
The 1-form gauge transformation changes the background $A$ as $B_2\rightarrow B_2+d\lambda$, $A\rightarrow A+N\lambda$ with $U(1)$ gauge field $\lambda$, and it does not correspond to a $U(1)$ gauge transformation of $A$, and thus the 't Hooft anomaly of the 1-form symmetry does not imply any inconsistency of the theory coupled to $A$.

If we perform a singular 1-form gauge transformation $\lambda=-A/N$ to remove the background (\ref{eqn:hallbg}) that couples the theory to $A$, 
the anomaly of the 1-form symmetry produces additional Chern-Simons contact term of $A$, which accounts for the $U(1)$ Hall conductivity \cite{Closset:2012vg}.
The anomaly of the 1-form symmetry can be computed from the spin of the generating lines \cite{Gaiotto:2014kfa,Gomis:unpublish}. If the line that generates the $\mathbb{Z}_N$ 1-form symmetry has spin $\frac{L}{2N}$ mod 1 for some integer $L$, the anomaly is
\begin{equation}\label{eqn:anomoneform}
\frac{NL}{4\pi}\int_Y B_2 \, B_2 \quad \text{mod }2\pi\mathbb{Z}~,
\end{equation}
where $Y$ is a closed 4-manifold. In particular, if the coefficient $NL$ is odd in the anomaly (\ref{eqn:anomoneform}), the theory must be a spin theory for the anomaly to be well-defined. 
The Hall conductivity $j=\kappa \, dA / 2\pi$ can then be computed from the anomaly as
\begin{equation}\label{eqn:hall}
\kappa = -\frac{L}{N}~,
\end{equation}
which is meaningful modulo an integer for a spin theory, and modulo an even integer for a non-spin theory, since 
we can add a 3d local counterterm $\frac{k}{4\pi}AdA$ for some integer $k$ ($k$ needs to be even if the theory is non-spin).
If we change the coupling (\ref{eqn:hallbg}) to the $U(1)$ symmetry by replacing the coefficient $1/N$ with $q/N$ for some $q\in\mathbb{Z}$, then the anomaly (\ref{eqn:anomoneform}) implies the Hall conductivity (\ref{eqn:hall}) changes by multiplying with $q^2$.
The discussion can be generalized to multiple $U(1)$ symmetries that do not permute the anyons, and again the 't~Hooft anomaly of the 1-form symmetry gives the Hall conductivity.


\bibliographystyle{ytphys}
\bibliography{Non_SUSY_dualities}

\providecommand{\href}[2]{#2}\begingroup\raggedright\begin{thebibliography}{10}

\bibitem{Gaiotto:2014kfa}
D.~Gaiotto, A.~Kapustin, N.~Seiberg, and B.~Willett, ``{Generalized Global
  Symmetries},'' \href{http://dx.doi.org/10.1007/JHEP02(2015)172}{{\em JHEP}
  {\bfseries 02} (2015) 172},
\href{http://arxiv.org/abs/1412.5148}{{\ttfamily arXiv:1412.5148 [hep-th]}}.

\bibitem{Bhardwaj:2017xup}
L.~Bhardwaj and Y.~Tachikawa, ``{On finite symmetries and their gauging in two
  dimensions},'' \href{http://dx.doi.org/10.1007/JHEP03(2018)189}{{\em JHEP}
  {\bfseries 03} (2018) 189},
\href{http://arxiv.org/abs/1704.02330}{{\ttfamily arXiv:1704.02330 [hep-th]}}.

\bibitem{Chang:2018iay}
C.-M. Chang, Y.-H. Lin, S.-H. Shao, Y.~Wang, and X.~Yin, ``{Topological Defect
  Lines and Renormalization Group Flows in Two Dimensions},''
  \href{http://dx.doi.org/10.1007/JHEP01(2019)026}{{\em JHEP} {\bfseries 01}
  (2019) 026},
\href{http://arxiv.org/abs/1802.04445}{{\ttfamily arXiv:1802.04445 [hep-th]}}.

\bibitem{baez2004higher}
J.~C. Baez and A.~D. Lauda, ``{Higher-Dimensional Algebra V: 2-Groups},'' {\em
  Theory Appl. Categ.} {\bfseries 12} (2004) 423--491,
  \href{http://arxiv.org/abs/math/0307200}{{\ttfamily arXiv:math/0307200
  [math.QA]}}.

\bibitem{Baez:2004in}
J.~Baez and U.~Schreiber, ``{Higher gauge theory: 2-connections on
  2-bundles},''
\href{http://arxiv.org/abs/hep-th/0412325}{{\ttfamily arXiv:hep-th/0412325
  [hep-th]}}.

\bibitem{Baez:2005qu}
J.~C. Baez and U.~Schreiber,
  \href{http://dx.doi.org/10.1090/conm/431}{``{Higher gauge theory},''} in {\em
  Categories in Algebra, Geometry and Mathematical Physics}, A.~Davydov {\em et
  al.}, eds., vol.~431 of {\em Contemp. Math.}, pp.~7--30.
\newblock AMS, Providence, 2007.
\newblock
\href{http://arxiv.org/abs/math/0511710}{{\ttfamily arXiv:math/0511710
  [math-dg]}}.
\newblock

\bibitem{Schreiber:2008}
U.~Schreiber and K.~Waldorf, ``{Connections on non-Abelian Gerbes and their
  Holonomy},'' {\em Theory Appl. Categ.} {\bfseries 28} (2013) 476--540,
  \href{http://arxiv.org/abs/0808.1923}{{\ttfamily arXiv:0808.1923 [math.DG]}}.

\bibitem{Kapustin:2013uxa}
A.~Kapustin and R.~Thorngren, ``{Higher symmetry and gapped phases of gauge
  theories},''
\href{http://arxiv.org/abs/1309.4721}{{\ttfamily arXiv:1309.4721 [hep-th]}}.

\bibitem{Gukov:2013zka}
S.~Gukov and A.~Kapustin, ``{Topological Quantum Field Theory, Nonlocal
  Operators, and Gapped Phases of Gauge Theories},''
\href{http://arxiv.org/abs/1307.4793}{{\ttfamily arXiv:1307.4793 [hep-th]}}.

\bibitem{Kapustin:2013qsa}
A.~Kapustin and R.~Thorngren, ``{Topological Field Theory on a Lattice,
  Discrete Theta-Angles and Confinement},''
  \href{http://dx.doi.org/10.4310/ATMP.2014.v18.n5.a4}{{\em Adv. Theor. Math.
  Phys.} {\bfseries 18} (2014) 1233--1247},
\href{http://arxiv.org/abs/1308.2926}{{\ttfamily arXiv:1308.2926 [hep-th]}}.

\bibitem{Thorngren:2015gtw}
R.~Thorngren and C.~von Keyserlingk, ``{Higher SPT's and a generalization of
  anomaly in-flow},''
\href{http://arxiv.org/abs/1511.02929}{{\ttfamily arXiv:1511.02929
  [cond-mat.str-el]}}.

\bibitem{Gaiotto:2017zba}
D.~Gaiotto and T.~Johnson-Freyd, ``{Symmetry Protected Topological phases and
  Generalized Cohomology},''
\href{http://arxiv.org/abs/1712.07950}{{\ttfamily arXiv:1712.07950 [hep-th]}}.

\bibitem{Bhardwaj:2016clt}
L.~Bhardwaj, D.~Gaiotto, and A.~Kapustin, ``{State sum constructions of
  spin-TFTs and string net constructions of fermionic phases of matter},''
  \href{http://dx.doi.org/10.1007/JHEP04(2017)096}{{\em JHEP} {\bfseries 04}
  (2017) 096},
\href{http://arxiv.org/abs/1605.01640}{{\ttfamily arXiv:1605.01640
  [cond-mat.str-el]}}.

\bibitem{Tachikawa:2017gyf}
Y.~Tachikawa, ``{On gauging finite subgroups},''
\href{http://arxiv.org/abs/1712.09542}{{\ttfamily arXiv:1712.09542 [hep-th]}}.

\bibitem{Delcamp:2018wlb}
C.~Delcamp and A.~Tiwari, ``{From gauge to higher gauge models of topological
  phases},'' \href{http://dx.doi.org/10.1007/JHEP10(2018)049}{{\em JHEP}
  {\bfseries 10} (2018) 049},
\href{http://arxiv.org/abs/1802.10104}{{\ttfamily arXiv:1802.10104
  [cond-mat.str-el]}}.

\bibitem{Cordova:2018cvg}
C.~C\'{o}rdova, T.~T. Dumitrescu, and K.~Intriligator, ``{Exploring 2-Group
  Global Symmetries},'' \href{http://dx.doi.org/10.1007/JHEP02(2019)184}{{\em
  JHEP} {\bfseries 02} (2019) 184},
\href{http://arxiv.org/abs/1802.04790}{{\ttfamily arXiv:1802.04790 [hep-th]}}.

\bibitem{Sharpe:2015mja}
E.~Sharpe, ``{Notes on generalized global symmetries in QFT},''
  \href{http://dx.doi.org/10.1002/prop.201500048}{{\em Fortsch. Phys.}
  {\bfseries 63} (2015) 659--682},
\href{http://arxiv.org/abs/1508.04770}{{\ttfamily arXiv:1508.04770 [hep-th]}}.

\bibitem{Bauer:2004nh}
M.~Bauer, G.~Girardi, R.~Stora, and F.~Thuillier, ``{A Class of topological
  actions},'' \href{http://dx.doi.org/10.1088/1126-6708/2005/08/027}{{\em JHEP}
  {\bfseries 08} (2005) 027},
\href{http://arxiv.org/abs/hep-th/0406221}{{\ttfamily arXiv:hep-th/0406221
  [hep-th]}}.

\bibitem{Kapustin:2014zva}
A.~Kapustin and R.~Thorngren, ``{Anomalies of discrete symmetries in various
  dimensions and group cohomology},''
\href{http://arxiv.org/abs/1404.3230}{{\ttfamily arXiv:1404.3230 [hep-th]}}.

\bibitem{Green:1984sg}
M.~B. Green and J.~H. Schwarz, ``{Anomaly Cancellation in Supersymmetric
  $D{=}10$ Gauge Theory and Superstring Theory},''
\href{http://dx.doi.org/10.1016/0370-2693(84)91565-X}{{\em Phys. Lett.}
  {\bfseries 149B} (1984) 117--122}.

\bibitem{Eilenberg:1953}
S.~Eilenberg and S.~Mac~Lane, ``{On the groups $H(\Pi,n)$, I},''
  \href{http://dx.doi.org/10.2307/1969820}{{\em Ann. Math.} {\bfseries 58}
  (1953) 55--106}.

\bibitem{Barkeshli:2014cna}
M.~Barkeshli, P.~Bonderson, M.~Cheng, and Z.~Wang, ``{Symmetry, Defects, and
  Gauging of Topological Phases},''
\href{http://arxiv.org/abs/1410.4540}{{\ttfamily arXiv:1410.4540
  [cond-mat.str-el]}}.

\bibitem{Callan:1984sa}
C.~G. Callan, Jr. and J.~A. Harvey, ``{Anomalies and Fermion Zero Modes on
  Strings and Domain Walls},''
\href{http://dx.doi.org/10.1016/0550-3213(85)90489-4}{{\em Nucl. Phys.}
  {\bfseries B250} (1985) 427--436}.

\bibitem{tHooft:1979rat}
G.~'t~Hooft, ``{Naturalness, chiral symmetry, and spontaneous chiral symmetry
  breaking},''
\href{http://dx.doi.org/10.1007/978-1-4684-7571-5_9}{{\em NATO Sci. Ser. B}
  {\bfseries 59} (1980) 135--157}.

\bibitem{Kapustin:2014lwa}
A.~Kapustin and R.~Thorngren, ``{Anomalies of discrete symmetries in three
  dimensions and group cohomology},''
  \href{http://dx.doi.org/10.1103/PhysRevLett.112.231602}{{\em Phys. Rev.
  Lett.} {\bfseries 112} (2014) 231602},
\href{http://arxiv.org/abs/1403.0617}{{\ttfamily arXiv:1403.0617 [hep-th]}}.

\bibitem{Kitaev:2013talk}
A.~Kitaev, ``{On the classification of short-range entangled states}.'' Seminar
  at {S}imons {C}enter for {G}eometry and {P}hysics, June 2013.
\newblock \url{http://scgp.stonybrook.edu/video/video.php?id=2010}.

\bibitem{Kapustin:2014tfa}
A.~Kapustin, ``{Symmetry Protected Topological Phases, Anomalies, and
  Cobordisms: Beyond Group Cohomology},''
\href{http://arxiv.org/abs/1403.1467}{{\ttfamily arXiv:1403.1467
  [cond-mat.str-el]}}.

\bibitem{Freed:2016rqq}
D.~S. Freed and M.~J. Hopkins, ``{Reflection positivity and invertible
  topological phases},''
\href{http://arxiv.org/abs/1604.06527}{{\ttfamily arXiv:1604.06527 [hep-th]}}.

\bibitem{EtingofNOM2009}
P.~Etingof, D.~Nikshych, V.~Ostrik, and E.~Meir, ``{Fusion Categories and
  Homotopy Theory},'' \href{http://dx.doi.org/10.4171/QT/6}{{\em Quantum
  Topology} {\bfseries 1} (2010) 209--273},
  \href{http://arxiv.org/abs/0909.3140}{{\ttfamily arXiv:0909.3140 [math.QA]}}.

\bibitem{Teo:2015xla}
J.~C.~Y. Teo, T.~L. Hughes, and E.~Fradkin, ``{Theory of Twist Liquids: Gauging
  an Anyonic Symmetry},''
  \href{http://dx.doi.org/10.1016/j.aop.2015.05.012}{{\em Annals Phys.}
  {\bfseries 360} (2015) 349--445},
\href{http://arxiv.org/abs/1503.06812}{{\ttfamily arXiv:1503.06812
  [cond-mat.str-el]}}.

\bibitem{Barkeshli:2017rzd}
M.~Barkeshli and M.~Cheng, ``{Time-reversal and spatial reflection symmetry
  localization anomalies in (2+1)-dimensional topological phases of matter},''
  \href{http://dx.doi.org/10.1103/PhysRevB.98.115129}{{\em Phys. Rev.}
  {\bfseries B98} (2018) 115129},
\href{http://arxiv.org/abs/1706.09464}{{\ttfamily arXiv:1706.09464
  [cond-mat.str-el]}}.

\bibitem{Moore:1988qv}
G.~W. Moore and N.~Seiberg, ``{Classical and Quantum Conformal Field Theory},''
\href{http://dx.doi.org/10.1007/BF01238857}{{\em Commun. Math. Phys.}
  {\bfseries 123} (1989) 177}.

\bibitem{Moore:2006dw}
G.~W. Moore and G.~Segal, ``{D-branes and K-theory in 2D topological field
  theory},''
\href{http://arxiv.org/abs/hep-th/0609042}{{\ttfamily arXiv:hep-th/0609042
  [hep-th]}}.

\bibitem{Komargodski:2017dmc}
Z.~Komargodski, A.~Sharon, R.~Thorngren, and X.~Zhou, ``{Comments on Abelian
  Higgs Models and Persistent Order},''
  \href{http://dx.doi.org/10.21468/SciPostPhys.6.1.003}{{\em SciPost Phys.}
  {\bfseries 6} (2019) 003},
\href{http://arxiv.org/abs/1705.04786}{{\ttfamily arXiv:1705.04786 [hep-th]}}.

\bibitem{Kapustin:2014gua}
A.~Kapustin and N.~Seiberg, ``{Coupling a QFT to a TQFT and Duality},''
  \href{http://dx.doi.org/10.1007/JHEP04(2014)001}{{\em JHEP} {\bfseries 04}
  (2014) 001},
\href{http://arxiv.org/abs/1401.0740}{{\ttfamily arXiv:1401.0740 [hep-th]}}.

\bibitem{Moore:1988uz}
G.~W. Moore and N.~Seiberg, ``{Polynomial Equations for Rational Conformal
  Field Theories},''
\href{http://dx.doi.org/10.1016/0370-2693(88)91796-0}{{\em Phys. Lett.}
  {\bfseries B212} (1988) 451--460}.

\bibitem{Segal:1970}
G.~Segal, ``{Cohomology of topological groups},'' in {\em {Symposia
  Mathematica, Vol. IV (INDAM, Rome, 1968/69)}}, pp.~377--387.
\newblock Academic Press, London, 1970.

\bibitem{Brylinski:2000}
J.-L. Brylinski, ``{Differentiable Cohomology of Gauge Groups},''
  \href{http://arxiv.org/abs/math/0011069}{{\ttfamily arXiv:math/0011069
  [math.DG]}}.

\bibitem{Chen:2015bia}
X.~Chen, F.~J. Burnell, A.~Vishwanath, and L.~Fidkowski, ``{Anomalous Symmetry
  Fractionalization and Surface Topological Order},''
  \href{http://dx.doi.org/10.1103/PhysRevX.5.041013}{{\em Phys. Rev.}
  {\bfseries X5} (2015) 041013},
\href{http://arxiv.org/abs/1403.6491}{{\ttfamily arXiv:1403.6491
  [cond-mat.str-el]}}.

\bibitem{Gomis:unpublish}
J.~Gomis, Z.~Komargodski, and N.~Seiberg. Unpublished, 2017.

\bibitem{Kitaev:2005}
A.~Kitaev, ``Anyons in an exactly solved model and beyond,''
  \href{http://dx.doi.org/10.1016/j.aop.2005.10.005}{{\em Annals of Physics}
  {\bfseries 321} (2006) 2--111},
  \href{http://arxiv.org/abs/cond-mat/0506438}{{\ttfamily
  arXiv:cond-mat/0506438 [cond-mat]}}.

\bibitem{Whitehead:1949}
J.~H.~C. Whitehead, ``{On simply connected, 4-dimensional polyhedra},''
  \href{http://dx.doi.org/10.1007/BF02568048}{{\em Comm. Math. Helv.}
  {\bfseries 22} (1949) 48--92}.

\bibitem{Essin:2013rca}
A.~M. Essin and M.~Hermele, ``{Classifying fractionalization: Symmetry
  classification of gapped $\mathbb{Z}_2$ spin liquids in two dimensions},''
  \href{http://dx.doi.org/10.1103/PhysRevB.87.104406}{{\em Phys. Rev.}
  {\bfseries B87} (2013) 104406},
\href{http://arxiv.org/abs/1212.0593}{{\ttfamily arXiv:1212.0593
  [cond-mat.str-el]}}.

\bibitem{Tarantino:2016}
N.~Tarantino, N.~H. Lindner, and L.~Fidkowski, ``{Symmetry fractionalization
  and twist defects},''
  \href{http://dx.doi.org/10.1088/1367-2630/18/3/035006}{{\em New Journal of
  Physics} {\bfseries 18} (2016) 035006},
  \href{http://arxiv.org/abs/1506.06754}{{\ttfamily arXiv:1506.06754
  [cond-mat.str-el]}}.

\bibitem{bakalov2001lectures}
B.~Bakalov and A.~A. Kirillov, {\em {Lectures on Tensor Categories and Modular
  Functors}}, vol.~21 of {\em University Lecture Series}.
\newblock American Mathematical Soc., 2001.

\bibitem{Vafa:1988ag}
C.~Vafa, ``{Toward Classification of Conformal Theories},''
\href{http://dx.doi.org/10.1016/0370-2693(88)91603-6}{{\em Phys. Lett.}
  {\bfseries B206} (1988) 421--426}.

\bibitem{Bernevig:2015fqa}
A.~Bernevig and T.~Neupert, ``{Topological Superconductors and Category
  Theory},''
\href{http://arxiv.org/abs/1506.05805}{{\ttfamily arXiv:1506.05805
  [cond-mat.str-el]}}.

\bibitem{Maldacena:2001ss}
J.~M. Maldacena, G.~W. Moore, and N.~Seiberg, ``{D-brane charges in five-brane
  backgrounds},'' \href{http://dx.doi.org/10.1088/1126-6708/2001/10/005}{{\em
  JHEP} {\bfseries 10} (2001) 005},
\href{http://arxiv.org/abs/hep-th/0108152}{{\ttfamily arXiv:hep-th/0108152
  [hep-th]}}.

\bibitem{Banks:2010zn}
T.~Banks and N.~Seiberg, ``{Symmetries and Strings in Field Theory and
  Gravity},'' \href{http://dx.doi.org/10.1103/PhysRevD.83.084019}{{\em Phys.
  Rev.} {\bfseries D83} (2011) 084019},
\href{http://arxiv.org/abs/1011.5120}{{\ttfamily arXiv:1011.5120 [hep-th]}}.

\bibitem{Benini:2017dus}
F.~Benini, P.-S. Hsin, and N.~Seiberg, ``{Comments on global symmetries,
  anomalies, and duality in (2+1)d},''
  \href{http://dx.doi.org/10.1007/JHEP04(2017)135}{{\em JHEP} {\bfseries 04}
  (2017) 135},
\href{http://arxiv.org/abs/1702.07035}{{\ttfamily arXiv:1702.07035
  [cond-mat.str-el]}}.

\bibitem{Aharony:2016jvv}
O.~Aharony, F.~Benini, P.-S. Hsin, and N.~Seiberg, ``{Chern-Simons-matter
  dualities with $SO$ and $USp$ gauge groups},''
  \href{http://dx.doi.org/10.1007/JHEP02(2017)072}{{\em JHEP} {\bfseries 02}
  (2017) 072},
\href{http://arxiv.org/abs/1611.07874}{{\ttfamily arXiv:1611.07874
  [cond-mat.str-el]}}.

\bibitem{Komargodski:2017keh}
Z.~Komargodski and N.~Seiberg, ``{A symmetry breaking scenario for
  QCD$_{3}$},'' \href{http://dx.doi.org/10.1007/JHEP01(2018)109}{{\em JHEP}
  {\bfseries 01} (2018) 109},
\href{http://arxiv.org/abs/1706.08755}{{\ttfamily arXiv:1706.08755 [hep-th]}}.

\bibitem{Cordova:2017vab}
C.~C\'{o}rdova, P.-S. Hsin, and N.~Seiberg, ``{Global Symmetries, Counterterms,
  and Duality in Chern-Simons Matter Theories with Orthogonal Gauge Groups},''
  \href{http://dx.doi.org/10.21468/SciPostPhys.4.4.021}{{\em SciPost Phys.}
  {\bfseries 4} (2018) 021},
\href{http://arxiv.org/abs/1711.10008}{{\ttfamily arXiv:1711.10008 [hep-th]}}.

\bibitem{witten:unpublished}
E.~Witten. Unpublished, 2016.

\bibitem{Benini:2017aed}
F.~Benini, ``{Three-dimensional dualities with bosons and fermions},''
  \href{http://dx.doi.org/10.1007/JHEP02(2018)068}{{\em JHEP} {\bfseries 02}
  (2018) 068},
\href{http://arxiv.org/abs/1712.00020}{{\ttfamily arXiv:1712.00020 [hep-th]}}.

\bibitem{Cordova:2017kue}
C.~C\'{o}rdova, P.-S. Hsin, and N.~Seiberg, ``{Time-Reversal Symmetry,
  Anomalies, and Dualities in (2+1)$d$},''
  \href{http://dx.doi.org/10.21468/SciPostPhys.5.1.006}{{\em SciPost Phys.}
  {\bfseries 5} (2018) 006},
\href{http://arxiv.org/abs/1712.08639}{{\ttfamily arXiv:1712.08639
  [cond-mat.str-el]}}.

\bibitem{Seiberg:2016rsg}
N.~Seiberg and E.~Witten, ``{Gapped Boundary Phases of Topological Insulators
  via Weak Coupling},'' \href{http://dx.doi.org/10.1093/ptep/ptw083}{{\em PTEP}
  {\bfseries 2016} (2016) 12C101},
\href{http://arxiv.org/abs/1602.04251}{{\ttfamily arXiv:1602.04251
  [cond-mat.str-el]}}.

\bibitem{hsin:unpublish}
P.-S. Hsin. Unpublished, 2017.

\bibitem{Wang:2016qkb}
C.~Wang and M.~Levin, ``{Anomaly indicators for time-reversal symmetric
  topological orders},''
  \href{http://dx.doi.org/10.1103/PhysRevLett.119.136801}{{\em Phys. Rev.
  Lett.} {\bfseries 119} (2017) 136801},
\href{http://arxiv.org/abs/1610.04624}{{\ttfamily arXiv:1610.04624
  [cond-mat.str-el]}}.

\bibitem{Barkeshli:2016mew}
M.~Barkeshli, P.~Bonderson, C.-M. Jian, M.~Cheng, and K.~Walker, ``{Reflection
  and time reversal symmetry enriched topological phases of matter: path
  integrals, non-orientable manifolds, and anomalies},''
\href{http://arxiv.org/abs/1612.07792}{{\ttfamily arXiv:1612.07792
  [cond-mat.str-el]}}.

\bibitem{Vishwanath:2012tq}
A.~Vishwanath and T.~Senthil, ``{Physics of three dimensional bosonic
  topological insulators: Surface Deconfined Criticality and Quantized
  Magnetoelectric Effect},''
  \href{http://dx.doi.org/10.1103/PhysRevX.3.011016}{{\em Phys. Rev.}
  {\bfseries X3} (2013) 011016},
\href{http://arxiv.org/abs/1209.3058}{{\ttfamily arXiv:1209.3058
  [cond-mat.str-el]}}.

\bibitem{Thorngren:2014pza}
R.~Thorngren, ``{Framed Wilson Operators, Fermionic Strings, and Gravitational
  Anomaly in 4d},'' \href{http://dx.doi.org/10.1007/JHEP02(2015)152}{{\em JHEP}
  {\bfseries 02} (2015) 152},
\href{http://arxiv.org/abs/1404.4385}{{\ttfamily arXiv:1404.4385 [hep-th]}}.

\bibitem{Milnor:1956}
J.~Milnor, ``{Construction of Universal Bundles, II},''
  \href{http://dx.doi.org/10.2307/1970012}{{\em Ann. Math} {\bfseries 63}
  (1965) 430--436}.

\bibitem{Hatcher:book}
A.~Hatcher, {\em {Algebraic topology}}.
\newblock Cambridge University Press, Cambridge, 2002.

\bibitem{Steenrod:1947}
N.~E. Steenrod, ``{Products of cocycles and Extensions of Mappings},''
  \href{http://dx.doi.org/10.2307/1969172}{{\em Ann. Math} {\bfseries 48}
  (1947) 290--320}.

\bibitem{Eilenberg:1954}
S.~Eilenberg and S.~Mac~Lane, ``{On the Groups $H(\Pi,n)$, II: Methods of
  Computation},'' \href{http://dx.doi.org/10.2307/1969702}{{\em Ann. Math.}
  {\bfseries 60} (1954) 49--139}.

\bibitem{Whitehead:1950}
J.~H.~C. Whitehead, ``{A Certain Exact Sequence},''
  \href{http://dx.doi.org/10.2307/1969511}{{\em Ann. Math.} {\bfseries 52}
  (1950) 51--110}.

\bibitem{Closset:2012vg}
C.~Closset, T.~T. Dumitrescu, G.~Festuccia, Z.~Komargodski, and N.~Seiberg,
  ``{Contact Terms, Unitarity, and $F$-Maximization in Three-Dimensional
  Superconformal Theories},''
  \href{http://dx.doi.org/10.1007/JHEP10(2012)053}{{\em JHEP} {\bfseries 10}
  (2012) 053},
\href{http://arxiv.org/abs/1205.4142}{{\ttfamily arXiv:1205.4142 [hep-th]}}.

\end{thebibliography}\endgroup
\end{document}